\documentclass[fleqn,usenatbib]{mnras}

\usepackage{newtxtext,newtxmath}

\usepackage[T1]{fontenc}

\DeclareRobustCommand{\VAN}[3]{#2}
\let\VANthebibliography\thebibliography
\def\thebibliography{\DeclareRobustCommand{\VAN}[3]{##3}\VANthebibliography}

\usepackage{graphicx}	
\usepackage{amsmath}	
\usepackage{subfigure}
\usepackage{natbib}
\usepackage{xcolor}
\usepackage{tabularx}
\usepackage{longtable}
\usepackage{bm}
\usepackage{threeparttable}
\usepackage[normalem]{ulem}
\usepackage{verbatim}

\newcommand{\tess}{{\it TESS}}
\newcommand{\nres}{{\it NRES}}
\newcommand{\chiron}{{\it CHIRON}}
\newcommand{\coralie}{{\it CORALIE}}
\newcommand{\code}[1]{{\texttt{#1}}}

\newcommand{\lco}{{\it LCOGT}}
\newcommand{\gaia}{{\it Gaia}}
\newcommand{\wise}{{\it WISE}}

\newcommand{\sample}{{previous transiting BDs and low mass stars sample}}

\title[TOI-1608, 2336 and 2521]{Three low-mass companions around aged stars discovered by TESS}

\author[Z. Lin et al.]{Zitao Lin,$^{1}$\thanks{E-mail: lzt22@mails.tsinghua.edu.cn}
Tianjun Gan,$^{1}$\thanks{E-mail: gtj18@mails.tsinghua.edu.cn}
Sharon X. Wang,$^{1}$\thanks{E-mail: sharonw@mail.tsinghua.edu.cn}
Avi Shporer,$^{2}$ 
Markus Rabus,$^{3}$ 
George Zhou,$^{4}$ 
\newauthor
Angelica Psaridi,$^{5}$ 
Fran\c{c}ois Bouchy,$^{5}$ 
Allyson Bieryla,$^{6}$ 
David W. Latham,$^{3}$ 
Shude Mao,$^{1,7}$ 
Keivan G. Stassun,$^{8}$ 
\newauthor
Coel Hellier,$^{9}$ 
Steve B. Howell,$^{10}$ 
Carl Ziegler,$^{11}$ 
Douglas A. Caldwell,$^{10,12}$ 
Catherine A. Clark,$^{13,14}$ 
\newauthor
Karen A. Collins,$^{15}$
Jason L. Curtis,$^{16}$ 
Jacqueline K. Faherty,$^{17}$ 
Crystal L. Gnilka,$^{10}$ 
Samuel K. Grunblatt,$^{18}$  
\newauthor
Jon M. Jenkins,$^{10}$ 
Marshall C. Johnson,$^{19}$
Nicholas Law,$^{20}$ 
Monika Lendl,$^{5}$ 
Colin Littlefield,$^{10,21}$ 
\newauthor
Michael B. Lund,$^{22}$ 
Mikkel N. Lund,$^{23}$ 
Andrew W. Mann,$^{20}$ 
Scott McDermott,$^{24}$ 
Lokesh Mishra,$^{5}$ 
\newauthor
Dany Mounzer,$^{5}$ 
Martin Paegert,$^{25}$ 
Tyler Pritchard,$^{26}$ 
George R. Ricker,$^{2}$ 
Sara Seager,$^{2,27,28}$ 
Gregor Srdoc,$^{29}$ 
\newauthor
Qinghui Sun,$^{1}$ 
Jiaxin Tang,$^{1}$ 
Stéphane Udry,$^{5}$ 
Roland Vanderspek,$^{2}$ 
David Watanabe,$^{30}$ 
Joshua N. Winn,$^{31}$ 
\newauthor
and Jie Yu$^{32}$ 
\\
Affiliations are listed at the end of the paper
}

\date{Accepted XXX. Received YYY; in original form ZZZ}

\pubyear{2015}

\begin{document}
\label{firstpage}
\pagerange{\pageref{firstpage}--\pageref{lastpage}}
\maketitle

\begin{abstract}
We report the discovery of three transiting low-mass companions to aged stars: a brown dwarf (TOI-2336b) and two objects near the hydrogen burning mass limit (TOI-1608b and TOI-2521b). These three systems were first identified using data from the Transiting Exoplanet Survey Satellite (\tess). TOI-2336b has a radius of $1.05\pm 0.04\ R_J$, a mass of $69.9\pm 2.3\ M_J$ and an orbital period of 7.71 days. TOI-1608b has a radius of $1.21\pm 0.06\ R_J$, a mass of $90.7\pm 3.7\ M_J$ and an orbital period of 2.47 days. TOI-2521b has a radius of $1.01\pm 0.04\ R_J$, a mass of $77.5\pm 3.3\ M_J$ and an orbital period of 5.56 days. We found all these low-mass companions are inflated. We fitted a relation between radius, mass and incident flux using the sample of known transiting brown dwarfs and low-mass M dwarfs. We found a positive correlation between the flux and the radius for brown dwarfs and for low-mass stars that is weaker than the correlation observed for giant planets. We also found that TOI-1608 and TOI-2521 are very likely to be spin-orbit synchronized, leading to the unusually rapid rotation of the primary stars considering their evolutionary stages. Our estimates indicate that both systems have much shorter spin-orbit synchronization timescales compared to their ages. These systems provide valuable insights into the evolution of stellar systems with brown dwarf and low-mass stellar companions influenced by tidal effects.
\end{abstract}

\begin{keywords}
stars: brown dwarfs -- stars: low mass -- techniques: photometric -- techniques: radial velocities
\end{keywords}



\section{Introduction}

Brown dwarfs are objects with masses between those of giant planets and stars. Typically, the lower mass limit is taken to be $\sim 13\ M_J$ \citep{Burrows1997_BD_low_mass_range_2, Spiegel2011}, the minimum mass to ignite deuterium fusion, and the upper mass limit is taken to be $\sim 80\ M_J$ \citep{Baraffe2002}, the minimum mass to ignite ordinary hydrogen fusion. The term ``brown dwarf" was first used by \cite{Tarter1975}. Since the earliest brown dwarf discoveries such as Gliese 229B \citep{Nakajima1995_Gl229b1, Oppenheimer1995_Gl229b2} and Teide 1 \citep{Rebolo1995_Teide1_1, Rebolo1996_Teide1_2}, thousands of isolated brown dwarfs have been discovered \citep[e.g.,][]{Henry2018}. However, previous works using radial velocity data have shown that, on relatively short orbital periods, brown dwarf companions around solar-type stars are significantly rarer compared to planetary and stellar companions, which is known as the `brown dwarf desert’ \citep[e.g.,][]{Marcy2000, Grether2006}. Similarly, the known transiting brown dwarfs are outnumbered by hot Jupiters (e.g., the NASA Exoplanet Archive). Despite the similar radii of hot Jupiters and brown dwarfs \citep{Chen2017_m_r_relation}, only about 30 transiting brown dwarfs have been found, in contrast with the hundreds of transiting hot Jupiters. This desert may result from two different formation mechanisms, i.e., gravitational instability and cloud fragmentation, dominating different mass regimes for the formation of brown dwarfs \citep[see discussion in][]{Palle2021}. For example, \cite{Ma2014_BDgap} analyzed a catalog of 62 brown dwarf companions and found that they are most depleted at $30\ M_J<M\sin i<55\ M_J$ and $P<100$ days. Using $42.5\ M_J$ as a threshold, they found that the eccentricity distribution of lower-mass brown dwarfs is consistent with that of massive planets, while the eccentricity distribution of higher mass brown dwarfs is more consistent with that of stellar binaries. They suggested that the brown dwarfs below $42.5\ M_J$ may form through disc gravitational instability, while the brown dwarfs above $42.5\ M_J$ may form through molecular cloud fragmentation like stars. However, the scope and robustness of such statistical works is limited by the small size of the brown dwarf sample. Therefore, we need more brown dwarf companions with precisely determined parameters such as mass, radius, and age to analyze the origin of the `brown dwarf desert’ and their formation mechanism.

NASA’s Transiting Exoplanet Survey Satellite (\tess) mission \citep{Ricker2015} is an all-sky transit survey, searching for planets around nearby bright stars, which also provides an exciting opportunity to enlarge the sample of transiting brown dwarfs. TESS has completed its Primary Mission (July 2018 - July 2020), where it observed most of the sky, and its First Extended Mission (July 2020 - September 2022) where it observed most of the sky a second time, and it is now in its Second Extended Mission (started September 2022) where it continues to re-observe the sky. Up to late 2022, \tess \ has successfully found more than 10 transiting brown dwarfs \citep[e.g.,][]{Grieves2021, Carmichael2022, Psaridi2022}. Light curves from \tess \ can provide us with the radius of a transiting brown dwarf, its period, and its orbital inclination. However, transit data alone are not sufficient to determine the masses, which are important to distinguish brown dwarfs from planetary-mass companions having similar radii. In addition, mass is a fundamental parameter for brown dwarf population studies as in the works mentioned above. Therefore, we need to observe the spectra of the host stars of transiting brown dwarfs to extract their radial velocities (RVs) and derive the true mass (instead of the minimum mass $m\sin i$) and other orbital parameters such as the eccentricity. 

Furthermore, the stellar spectra are crucial for stellar characterization, which will also encompass photometry, parallax, and other potential data. Proper characterization of the host star is important in order to characterize its companions. For instance, to estimate the radius of a companion using transit, the radius of its host star is necessary. Ultimately, a wealth of physical information can be obtained for a transiting brown dwarf system, including the mass, radius, and orbit of the companion. Such systems are essentially ideal samples for further studies such as population statistics and investigations into their formation and evolution scenarios. In addition, only a limited number of field BDs have dynamic mass measurements, and their radii are too small for any ‘resolved’ measurement even with interferometry. Therefore, transiting brown dwarf systems are the only ideal means to constrain the masses and radii of brown dwarfs, which is essential for the study of these objects.

As different evolution models are developed for brown dwarfs and low-mass stars \citep[e.g.,][]{Baraffe2003, Phillips2020_ATMO2020, Marley2021}, the transiting brown dwarf sample can be used to test these models \citep[e.g.,][]{Carmichael2021}. Compared with field brown dwarfs, the transiting ones tend to be better characterized in terms of radius, mass, age and even elemental abundances, thanks to the wealth of information provided by their host stars, for example, assuming they were formed together and co-evolved. 

A special stage in the evolution of brown dwarfs in binary systems is when their host stars evolve off the main sequence. Low-mass objects orbiting evolved stars are poorly understood due to the challenges in their detection and characterization largely caused by the large radii and stellar RV jitter of the host stars \citep{Yu2018, Tayar2019}. It has been suggested that planets around evolved stars should be different from planets around main-sequence stars in many aspects due to their dynamical interactions with the evolved host \citep{Veras2016}. For example, the increasing tidal interaction may speed up the orbital decay and planet engulfment \citep{Hut1981, Jackson2008_Tidal}, which may induce a different eccentricity distribution in the observed population of surviving hot Jupiters \citep{Villaver2014, Grunblatt2018}. Additionally, due to the increasing luminosity of the evolved host stars, close-in planets would receive more flux, which makes such systems ideal for studies of planet inflation \citep[e.g.,][]{Saunders2022_Jup_inflate, Grunblatt2022} or re-inflation \citep[e.g.,][]{Grunblatt2016_Jup_reinflate}. In contrast to the numerous works on planets around evolved stars, few works discuss brown dwarfs or very low-mass stars around evolved stars, probably due to the small sample size of such systems. Do low-mass stars and brown dwarfs around evolved stars undergo orbital or radius changes in a similar way as planets? Gathering a sample of well-characterized transiting brown dwarfs would be the crucial first step in addressing this question.

The tidal effect plays a crucial role in the evolution of binary systems. For instance, it has been observed that the rotation of close binary stars tends to synchronize with their orbital motion due to tidal interactions \cite[e.g.,][]{Levato1974, Giuricin1984}. However, for planetary systems, the rotations of the host stars are rarely synchronized with the orbits \citep[though suspected, e.g., in][]{Donati2008_tau_Boo}. This is primarily because the angular momentum of planetary orbits is often insufficient to significantly influence the rotation of the host stars. Therefore, brown dwarf companions, which serve as transitional objects between planets and stars, can offer ideal examples to study tidal effects under different companion masses. Furthermore, brown dwarf companions can also provide valuable insights into the theories about the conditions for tidal synchronization and tidal equilibrium \cite[e.g.,][]{Hut1980}.

Here we report the discovery of a transiting brown dwarf, TOI-2336b, and two transiting objects near the hydrogen burning mass limit, TOI-1608b and TOI-2521b, around three aged stars using \tess\ data and ground-based follow-up observations.\footnote{These three old stars have slightly evolved off the main sequence and are not all into the subgiant phase yet, so we refer to them as ``aged" instead of evolved stars to be precise.} In Section \ref{observations} we describe the data we used in this work. Section \ref{stellar_properties} presents the analyses to determine the properties of the host stars of these three systems. Section \ref{analysis} presents the joint analysis of the light curves and the RV data and how we obtained the orbital parameters of the companions. In Section \ref{discussion}, we analyze the radius inflation of brown dwarfs and low mass stars, and we discuss the tidal evolution of these systems. We conclude our findings in Section \ref{summary}.

\section{Observations}\label{observations}
\subsection{\tess\ photometry}\label{tess_data}

TOI-1608 (TIC 138017750) was observed with the 2-minute cadence exposures in \tess \ Sector 18 from 3rd November 2019 to 27th November 2019. The photometry were extracted by the Science Processing Operations Center \citep[SPOC;][]{Jenkins2016_SPOC} pipeline. SPOC produces two kinds of light curves, using Simple Aperture Photometry (SAP) or the Presearch Data Conditioning SAP \citep[PDCSAP;][]{Smith2012_PDC2, Stumpe2012_PDC1, Stumpe2014_PDC3}. PDCSAP is corrected for the instrumental systematic effects and dilution, and we used PDCSAP light curves in our analysis. Eight transit signals were identified in this light curve with a period of 2.47 days. SPOC DV reports \citep{Twicken:DVdiagnostics2018, Li:DVmodelFit2019} presents the difference image centroiding results of TOI-1608, which indicates that the source of the transit signal is located within $1.3\pm 2.5$ arcsec of the target star, complementing the results of the high-resolution imaging presented in Section \ref{subsect:high_resolution_image}.

TOI-2336 (TIC 88902249) was observed with the 30-minute cadence exposures in \tess \ Sector 11 from 23rd April 2019 to 20th May 2019, and it was re-observed with the 10-minute cadence exposures in \tess \ Sector 38 from 29th April 2021 to 26th May 2021 during the \tess\ First Extended Mission. Both light curves were extracted by the TESS-SPOC pipeline \citep{Caldwell2020} and we used the PDCSAP in our analysis. Two transit signals were identified in the 30-minute data and four transit signals were identified in the 10-minute data with a period of 7.71 days. The TESS-SPOC DV reports of Sector 38 presents the difference image centroiding results of TOI-2336, which indicates that the source of the transit signal is located within $3.8\pm 2.5$ arcsec of the target star, complementing the results of the high-resolution imaging presented in Section \ref{subsect:high_resolution_image}.

TOI-2521 (TIC 72556406) was observed with the 30-minute cadence exposures in \tess \ Sector 6 from 15th December 2018 to 6th January 2019, and it was re-observed with the 10-minute cadence exposures in \tess \ Sector 33 from 18th December 2020 to 13th January 2021 during the \tess\ First Extended Mission. Again we used the PDCSAP  light curves extracted by the TESS-SPOC pipeline \citep{Caldwell2020} in our analysis. Four transit signals were identified in the 30-minute data and five transit signals were identified in the 10-minute data with a period of 5.56 days.

\begin{figure*}
\centering
\subfigure[TOI-1608]{
\begin{minipage}{\textwidth}
    \includegraphics[width=0.5\textwidth]{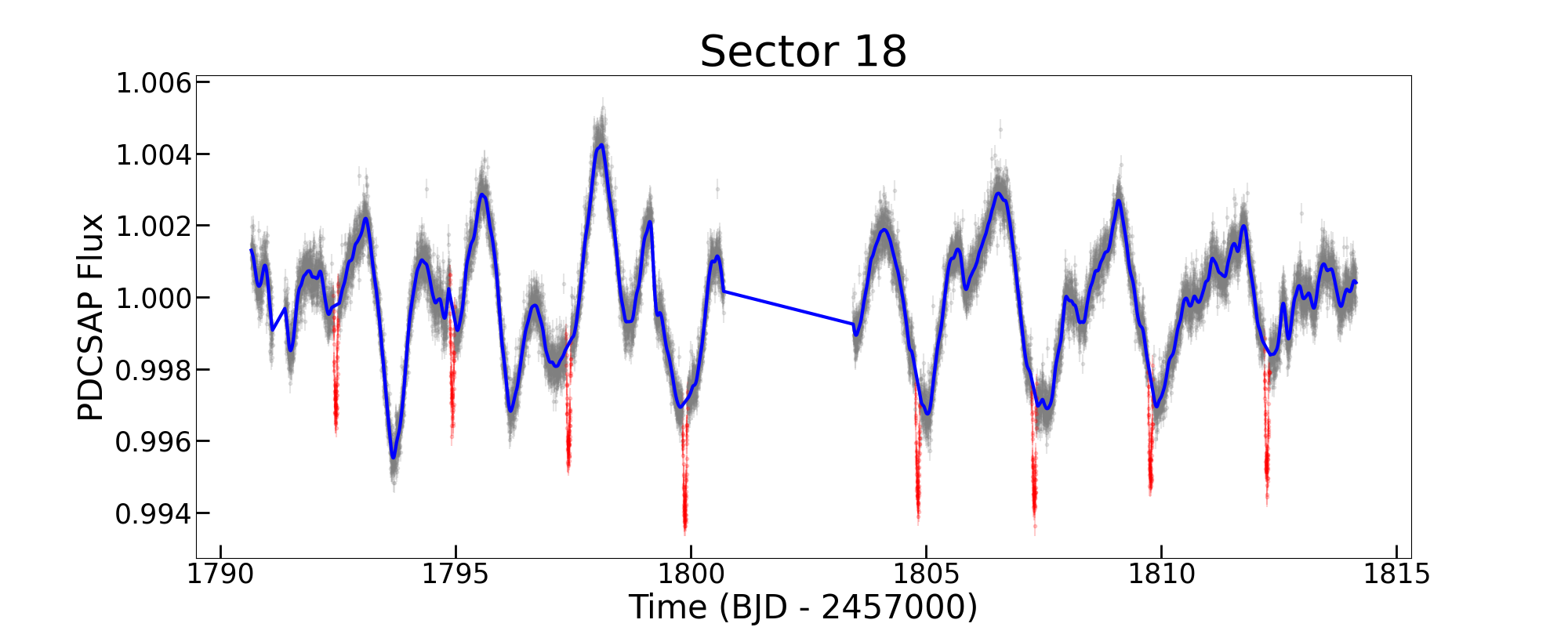}
    \includegraphics[width=0.5\textwidth]{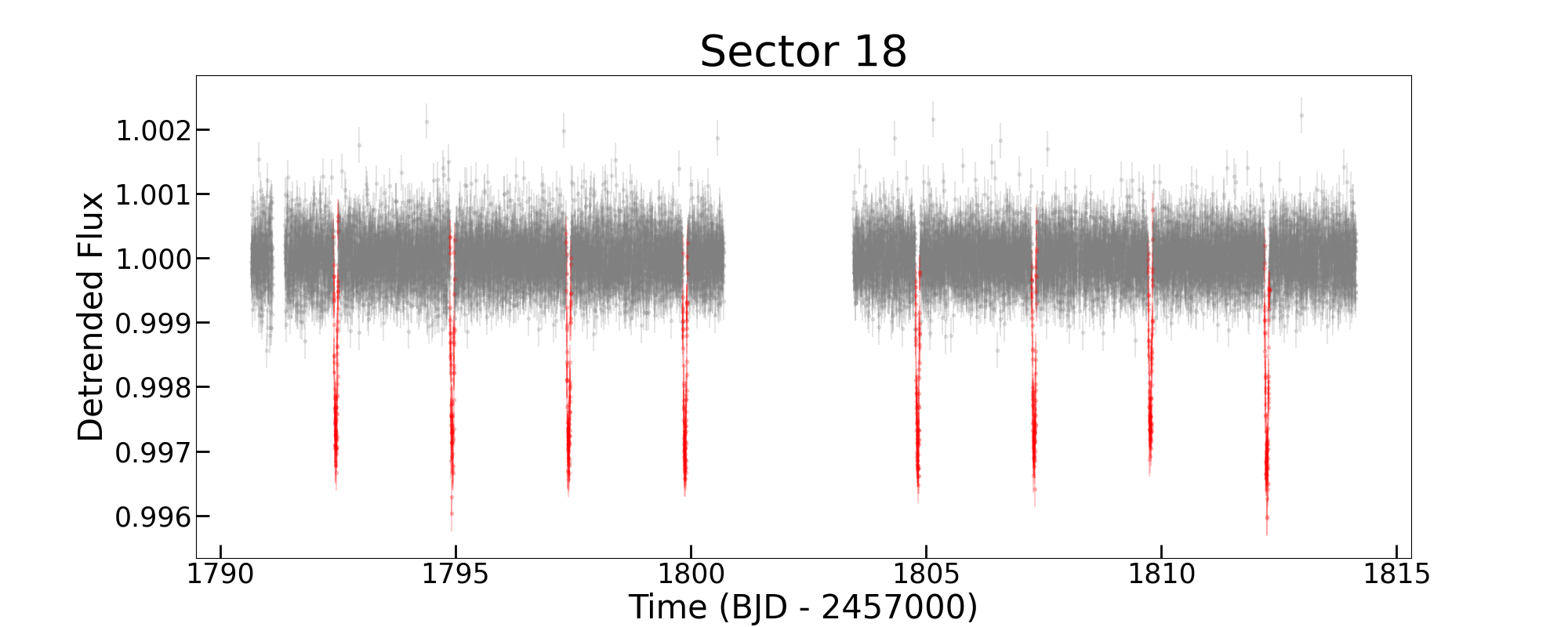}
\end{minipage}
}
\subfigure[TOI-2336]{
\begin{minipage}{\textwidth}
    \includegraphics[width=0.5\textwidth]{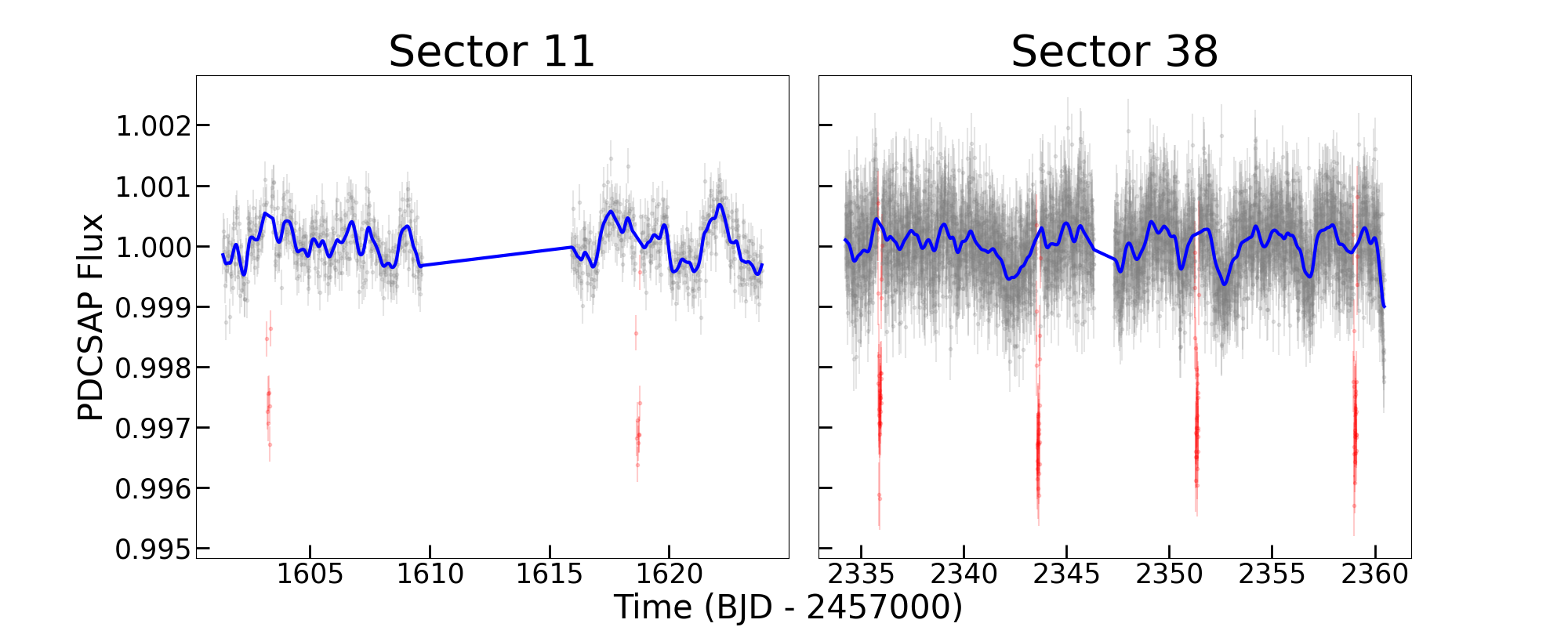}
    \includegraphics[width=0.5\textwidth]{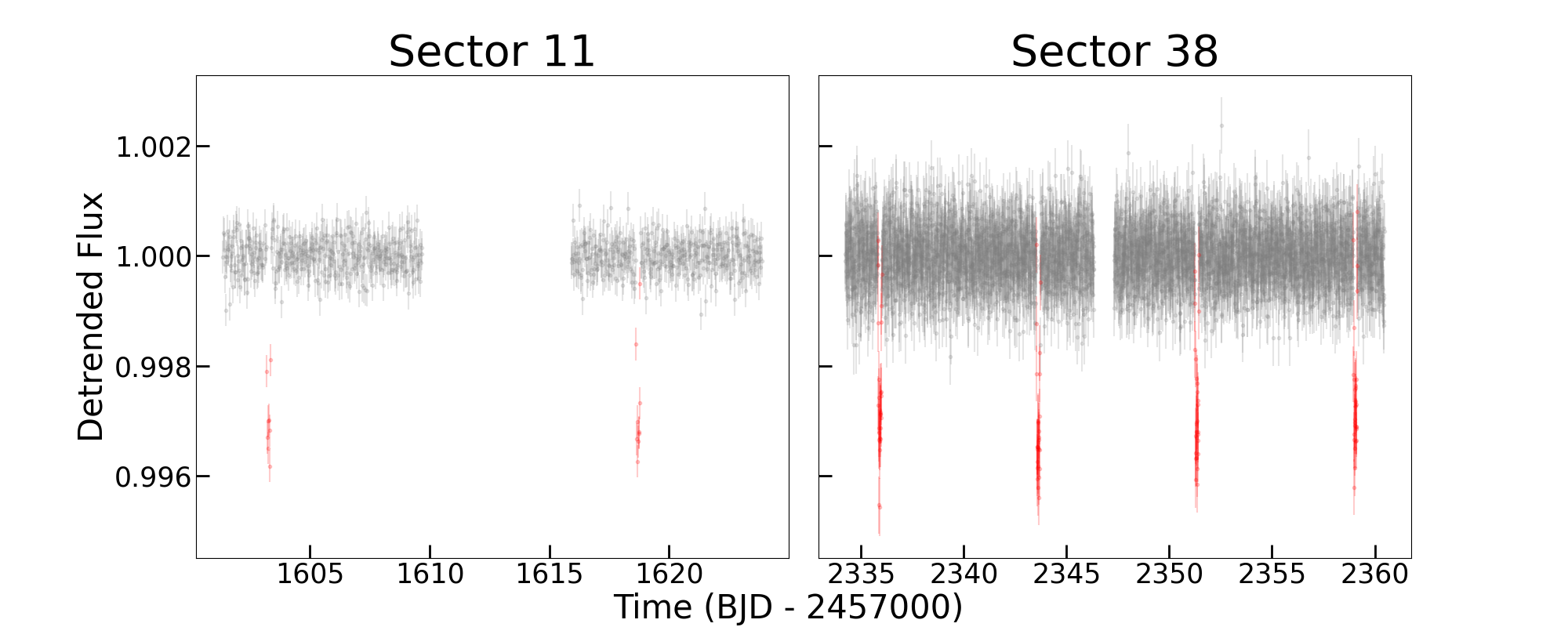}
\end{minipage}
}
\subfigure[TOI-2521]{
\begin{minipage}{\textwidth}
    \includegraphics[width=0.5\textwidth]{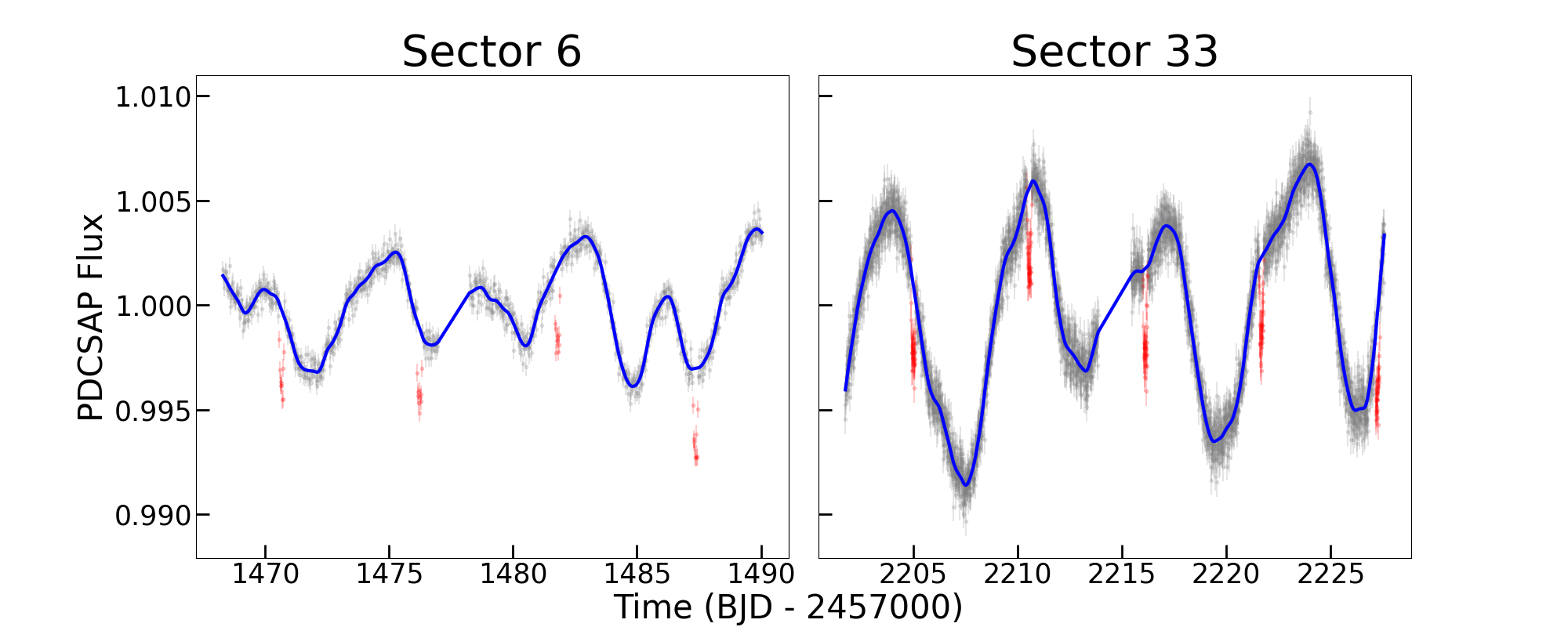}
    \includegraphics[width=0.5\textwidth]{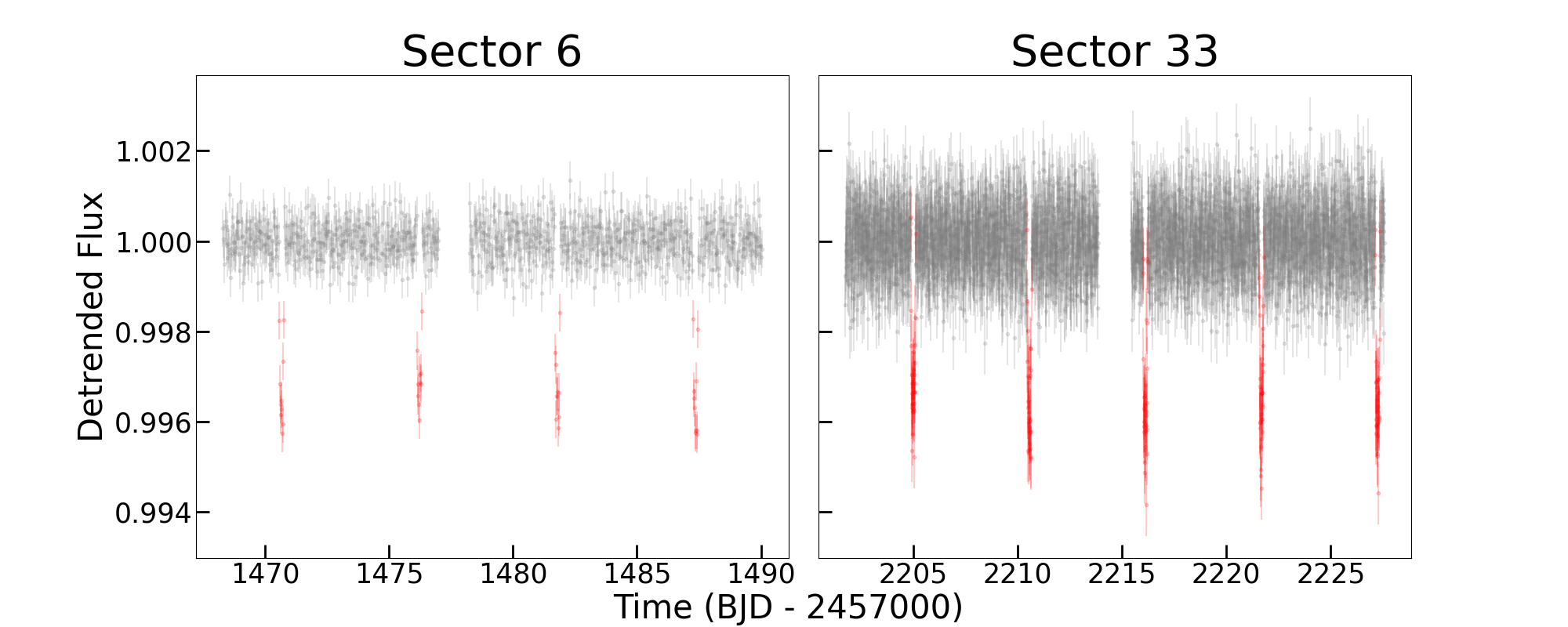}
\end{minipage}
}

\caption{\tess \ light curves of TOI-1608, TOI-2336 and TOI-2521. {\it Left panels:} The dots are the \tess \ PDCSAP data and the red dots highlight the transit signals. The blue lines are the best-fit GP models for detrending. The sector 18 for TOI-1608 was observed with the 2-minute cadence exposures. The sector 11 for TOI-2336 and the sector 6 for TOI-2521 were observed with the 30-minute cadence exposures. The sector 38 for TOI-2336 and the sector 33 for TOI-2521 were observed with the 10-minute cadence exposures.{\it Right panels:} The final detrended \tess \ PDCSAP light curves.} 
\label{fig_tess_PDCSAP}
\end{figure*}

We obtained all \tess \ data from the Mikulski Archive for Space Telescopes (MAST) Portal\footnote{\url{https://mast.stsci.edu/portal/Mashup/Clients/Mast/Portal.html}}. We then detrended the \tess \ light curve to remove the remaining systematic trends by fitting a Gaussian Process (GP) model after masking out the in-transit data using the \code{juliet} package \citep{Espinoza2019_Juliet}, which employs the package \code{celerite} \citep{Foreman-Mackey2017_celerite} to build the GP model. The PDCSAP light curves with the best-fit GP models are shown in Figure \ref{fig_tess_PDCSAP}.

\subsection{Ground-Based photometry}\label{gbp}
 
\subsubsection{LCOGT}\label{LCOGT}

We conducted ground-based follow-up observations of TOI-2336 using the Las Cumbres Observatory Global Telescope\ (LCOGT\footnote{\url{https://lco.global/}}) network \citep{Brown2013} on 1$^{\rm st}$ July 2021. We used the {\tt TESS Transit Finder}, which is a customized version of the \code{Tapir} software package \citep{Jensen2013}, to schedule the transit observation. The observation was taken using the 1.0-m telescopes located at Cerro Tololo Interamerican Observatory\ (CTIO). The images were acquired with the Sinistro cameras in the Pan-STARRS $z$-short band ($z_{s}$) with an exposure time of 30 s. The photometric analysis was then carried out using \code{AstroImageJ} \citep{Collins2017}. We show the ground light curve in Figure \ref{fig_2336_lco}.

\subsubsection{WASP}\label{WASP}
WASP-South has an array of 8 cameras forming the South African station of the WASP transit-search survey \citep{2006PASP..118.1407P}. The field of TOI-2336 was observed during 2006, 2007 and 2008, and then again in 2011 and 2012. Within each year, observations of the field every $\sim$\,15 min were obtained on each clear night, over a span of $\sim$\,150 nights.  During that time, WASP-South was equipped with 200~mm, f/1.8 lenses, observing with a 400--700 nm bandpass, and with a photometric extraction aperture of 48 arcsecs.  A total of 32\,000 photometric data points were obtained. 

TOI-2336 was not selected as a WASP candidate, but with hindsight we find that the standard WASP search algorithm \citep{2007MNRAS.380.1230C} produces a match to the \tess \ transit detection with a period of 7.712 days.

We also searched the WASP data for signs of a stellar rotational modulation, using the methods from \citet{2011PASP..123..547M}, but found only a 95\%\ upper limit of 1 mmag on any modulation in the range 1 d to 100 d.

\subsection{Spectroscopic Observations}

\subsubsection{NRES}

We acquired 16 spectra for TOI-1608 from 3$^{\rm rd}$ December 2021 to 9$^{\rm th}$ January 2022 at the McDonald Observatory, and 21 spectra for TOI-2336 between 22$^{\rm nd}$ April 2021 and 19$^{\rm th}$ July 2021 at South African Astronomical Observatory (SAAO) using the LCOGT Network of Robotic Echelle Spectrographs \citep[NRES;][]{Siverd2018}. NRES has four sets of identical echelle spectograph units at four LCOGT sites.  It has a resolving power of $R\sim53,000$, covering a wavelength range from 3900 to 8600 \AA. We reduced the spectra and extracted the RVs using the \texttt{CERES} pipeline \citep{Brahm2017}.

\subsubsection{CHIRON} \label{subsect_CHIRON}

We observed TOI-2336 and TOI-2521 with the CHIRON fiber-fed echelle spectrograph at the SMARTS 1.5-m telescope located at Cerro Tololo Inter-American Observatory, Chile \citep{Tokovinin2013}. CHIRON is a high-resolution echelle spectrograph with a spectral coverage of 4100 to 8700\,\AA{}. Both targets were observed in the \emph{fiber} mode, with a spectral resolving power of $R\sim$ 28,000. 
For TOI-2336 we obtained 12 spectra from May 7 to August 5 2021 with an exposure time of 1200~s. For TOI-2521 we gathered 8 RVs, from March 20 to March 30 2021, using an exposure time of 1800~s. 
We used the spectra extracted via the official CHIRON pipeline as per \citet{Paredes2021}. Radial velocities were derived from a least-squares deconvolution between the observation and a nonrotating synthetic template, generated using the ATLAS9 atmosphere models \citep{Castelli2004} at the spectral parameters of the targets. The derived broadening profile is fitted with a kernel accounting for the effects of rotational, macroturbulent, instrumental broadening, and radial velocity shift. Spectroscopic atmosphere parameters were derived by matching the observed CHIRON spectra against a library of 10,000 observed spectra previously classified via the Stellar Parameter Classification \citep{Buchhave2012}. The library is convolved to the instrument resolution of CHIRON and interpolated via a gradient-boosting regressor. Uncertainties on the spectroscopic parameters are determined by the spectrum-to-spectrum scatter of these values. 

\subsubsection{CORALIE}
We used the high-resolution spectrograph CORALIE to obtain spectra for TOI-2336. CORALIE is installed at the Swiss 1.2-m Leonhard Euler Telescope at La Silla Observatory, Chile \citep{Queloz2001}, and it has a resolution of R = 60, 000 and is fed by a 2\arcsec~fibre. We acquired 16 RV measurements from April 14 to June 08 2021 with exposure times ranging between 1200 s and 2400 s and S/N between 15 and 33. The RV measurement of each epoch was extracted by cross-correlating the spectrum with a binary 
G2 mask \citep{Pepe2002}. The bisector-span (BIS), the contrast (depth) and the full width at half-maximum (FWHM) were recorded using the standard CORALIE data reduction pipeline.

\subsubsection{TRES} \label{subsect_TRES}

We observed TOI-1608 and TOI-2521 with the Tillinghast Reflector Echelle Spectrograph \citep[TRES;][]{tres}. TRES is a fiber-fed echelle spectrograph mounted on the 1.5m telescope at the Fred Lawrence Whipple Observatory (FLWO) in Arizona. TRES covers a wavelength range of 3900 to 9100\,\AA{} and has a resolving power of $R\sim$ 44,000. Two reconnaissance spectra of TOI-1608 were obtained on December 31 2019 and January 4 2020 and similarly two reconnaissance spectra of TOI-2521 were obtained on March 11 and March 19 2021. The spectra were extracted as described in \citet{buchhave2010} and were then used to derive stellar parameters using the Stellar Parameter Classification \citep[SPC;][]{Buchhave2012} tool. SPC cross correlates the observed spectrum against a grid of synthetic spectra derived from the Kurucz atmospheric models \citep{kurucz1992}. SPC was used to derive parameters $T_{\rm eff}$, $\rm [m/H]$, and $v\sin i_{\ast}$ by fixing the $\log g_{\ast}$ from isochrone fitting as described in section \ref{subsect_isochrone}. We did not use the TRES spectra for RV measurements as they include only two epochs for each of TOI-1608 and TOI-2521.

\subsection{High Angular Resolution Imaging} \label{subsect:high_resolution_image}

The presence of a close companion star can be the cause of a transit-like event (i.e., a false positive), can cause incorrect stellar and exoplanet parameters to be determined \citep{FH1, FH2}, or can lead to non-detections of small planets residing with the same exoplanetary system \citep{Lester}. Given that nearly one-half of FGK stars are in binary or multiple star systems \citep{Matson}, high-resolution imaging provides crucial information toward our understanding of exoplanetary formation, dynamics and evolution \citep{Howell2021}.  We conducted high-resolution imaging observations on the three systems and summarize our imaging observations and results in this section.

\subsubsection{Gemini} \label{subsect:Gemini}

TOI-1608 was observed on 2022 February 10 UT using the ‘Alopeke speckle instrument on the Gemini North 8-m telescope, TOI-2336 was observed on 2022 May 05 UT using the Zorro instrument on the Gemini South 8-m telescope \citep{SH,HF}. TOI-2521 was observed on 2023 January 8 UT using the Zorro speckle instrument on the Gemini South 8-m telescope.

‘Alopeke and Zorro both provide simultaneous speckle imaging in two bands (562~nm and 832~nm) with output data products including a reconstructed image with robust contrast limits on companion detections \citep[e.g.,][]{Howell2016}.

\begin{figure}
\centering
\subfigure{\includegraphics[width=0.49\textwidth]{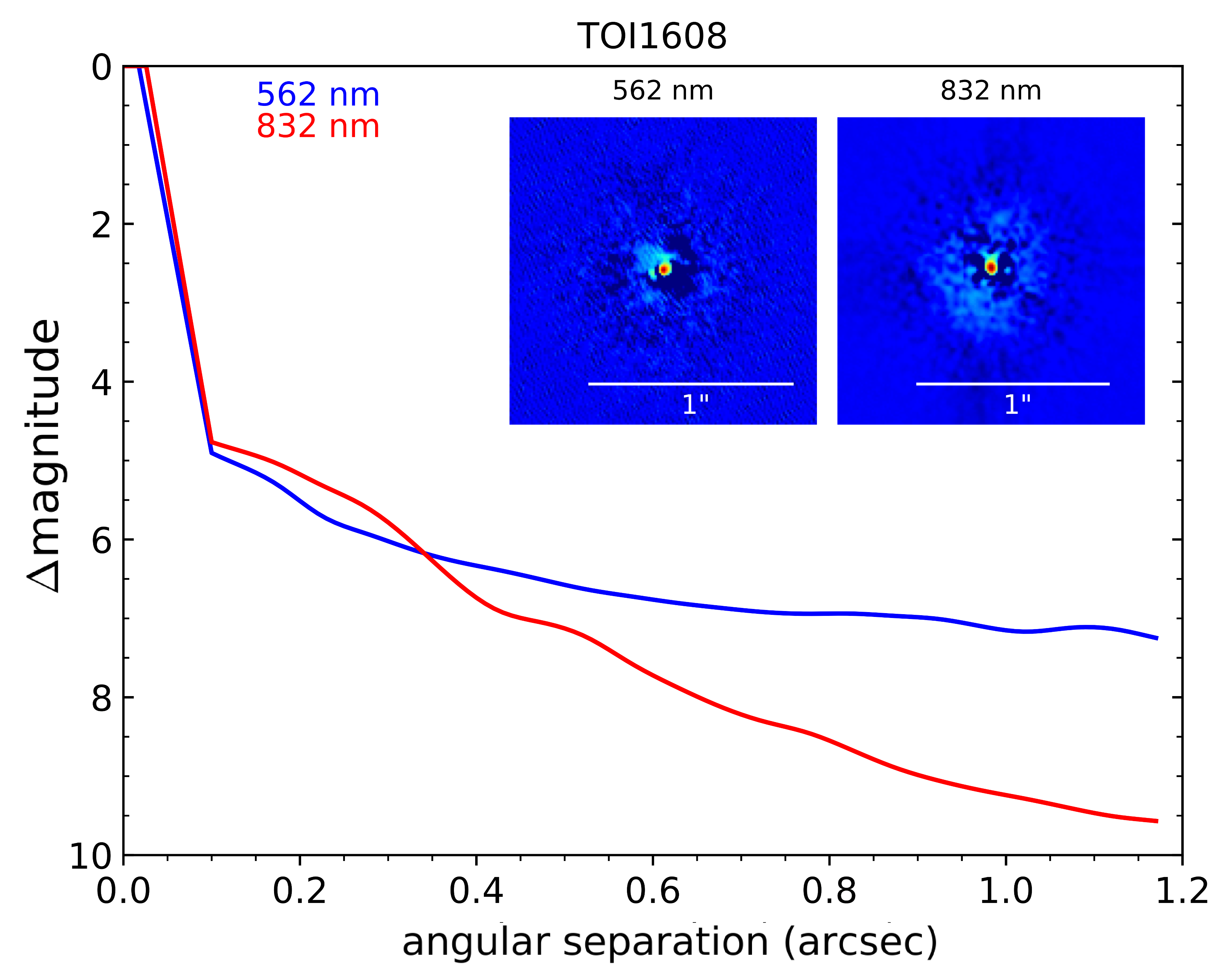}}

\subfigure{\includegraphics[width=0.49\textwidth]{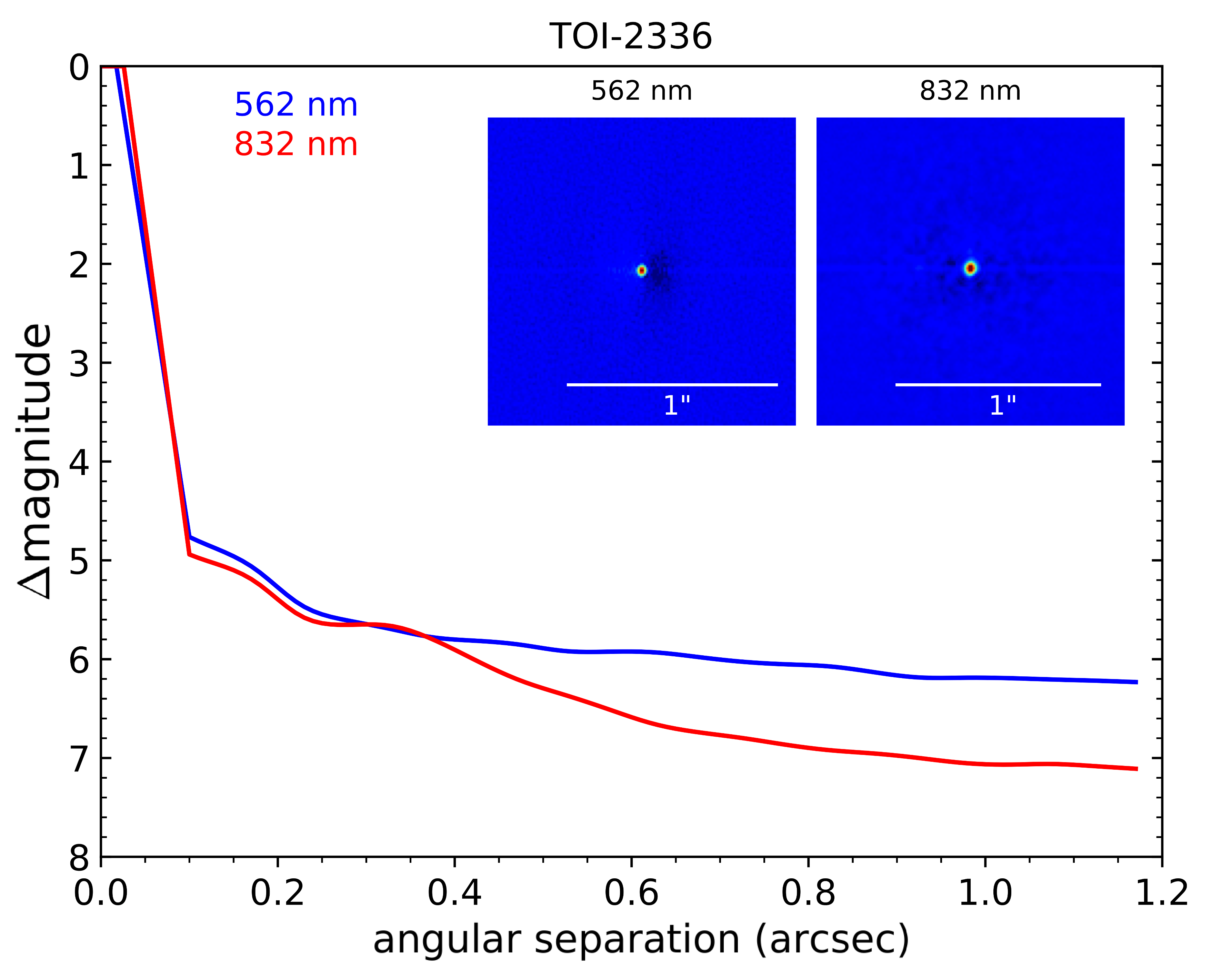}}

\subfigure{\includegraphics[width=0.49\textwidth]{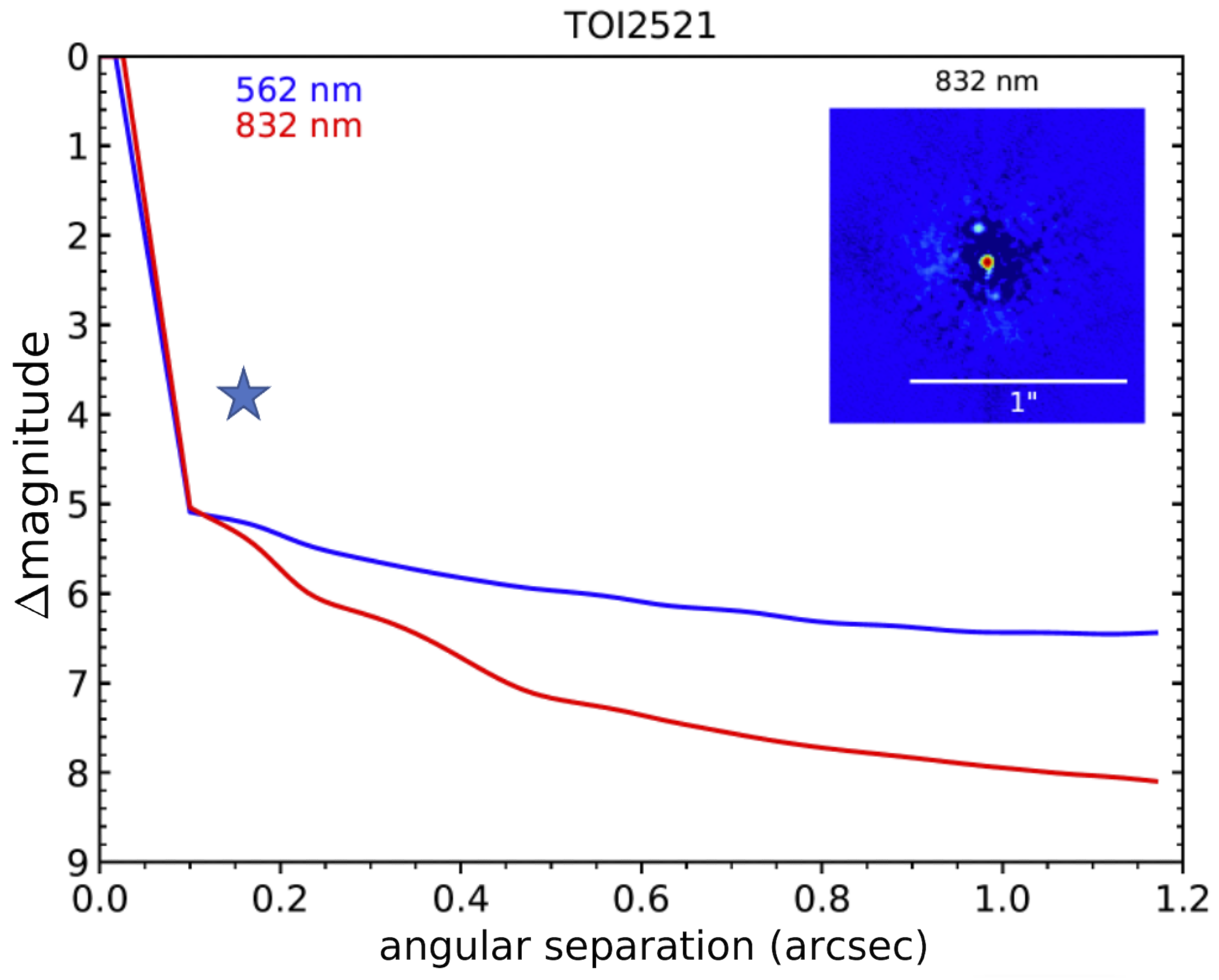}}
\caption{We show the $5\sigma$ speckle imaging contrast curves in both filters as a function of the angular separation from the diffraction limit out to 1.2 arcsec. The inset shows the reconstructed 832 nm image with a 1 arcsec scale bar. For TOI-1608 and TOI-2336, our results rule out companions and background objects that are potentially problematic. For TOI-2521, the blue star symbol shows its faint close companion. With a magnitude of 3.94 fainter, the companion will not influence our analysis.}
\label{fig_Gemini}
\end{figure}

For TOI-1608 and TOI-2336, three sets of 1000 x 0.06 second images were obtained and processed in our standard reduction pipeline \citep{Howell2011}. Figure \ref{fig_Gemini} shows our final contrast curves and the 832 nm reconstructed speckle image. We find that both TOI-1608 and TOI-2336 have no stellar companions brighter than 5-9 magnitudes below that of the target star from the 8-m telescope diffraction limit (20 mas) out to 1.2”. At the distance of TOI-1608 (d=101 pc) these angular limits correspond to (sky-projected) spatial limits of 2  to 120 AU and for TOI-2336 (d=297 pc) the angular limits correspond to 6 to 358 AU. Objects that are 5-9 magnitudes fainter than TOI-1608 (an F5 star) or TOI-2336 (an F4 star) will be late K-dwarfs or smaller objects \citep{Pecaut2012_Mamajek_Table1, Pecaut2013_Mamajek_Table2}, and will not have any significant impact on our analysis.

For TOI-2521, five sets of 1000 x 0.06 second images were obtained during a time of good seeing (0.47 arcsec) and processed in our standard reduction pipeline. Figure \ref{fig_Gemini} shows our achieved 5-$\sigma$ magnitude contrast curves and the 832 nm reconstructed image. The resulting high-resolution imaging revealed that TOI-2521 has a very close companion star. The companion, only detected in the 832 nm band, lies 0.162 arcsec away at a position angle of 16.44 degrees.  The close companion is 3.94 magnitudes fainter than the (G8) primary star, suggesting that it is near spectral type M6, in agreement with its red-only detection. \cite{Ciardi2015} showed that ``third-light” contamination from a close companion will cause the determined exoplanet radius to be underestimated, however, at a magnitude difference of 3.94, the contamination due to the faint companion would cause a $<$1\% error in the transit derived exoplanet size. At a separation of 0.162 arcsec, it is 99\% probable that the companion is bound \citep{Matson2018}. Our SOAR observation did not detect this companion, which is consistent with the precision of the instrument. No additional close companions were detected near TOI-2521 to within the contrast limits of 5-8 magnitudes over the angular limits of the diffraction limit out to 1.2 arcsec. At the distance of TOI-2521 (d=334 pc), these angular limits correspond to spatial limits of 6.8 to 409 AU.

\subsubsection{SOAR}

\begin{figure}
\centering
\subfigure[TOI-2336]{\includegraphics[width=0.49\textwidth]{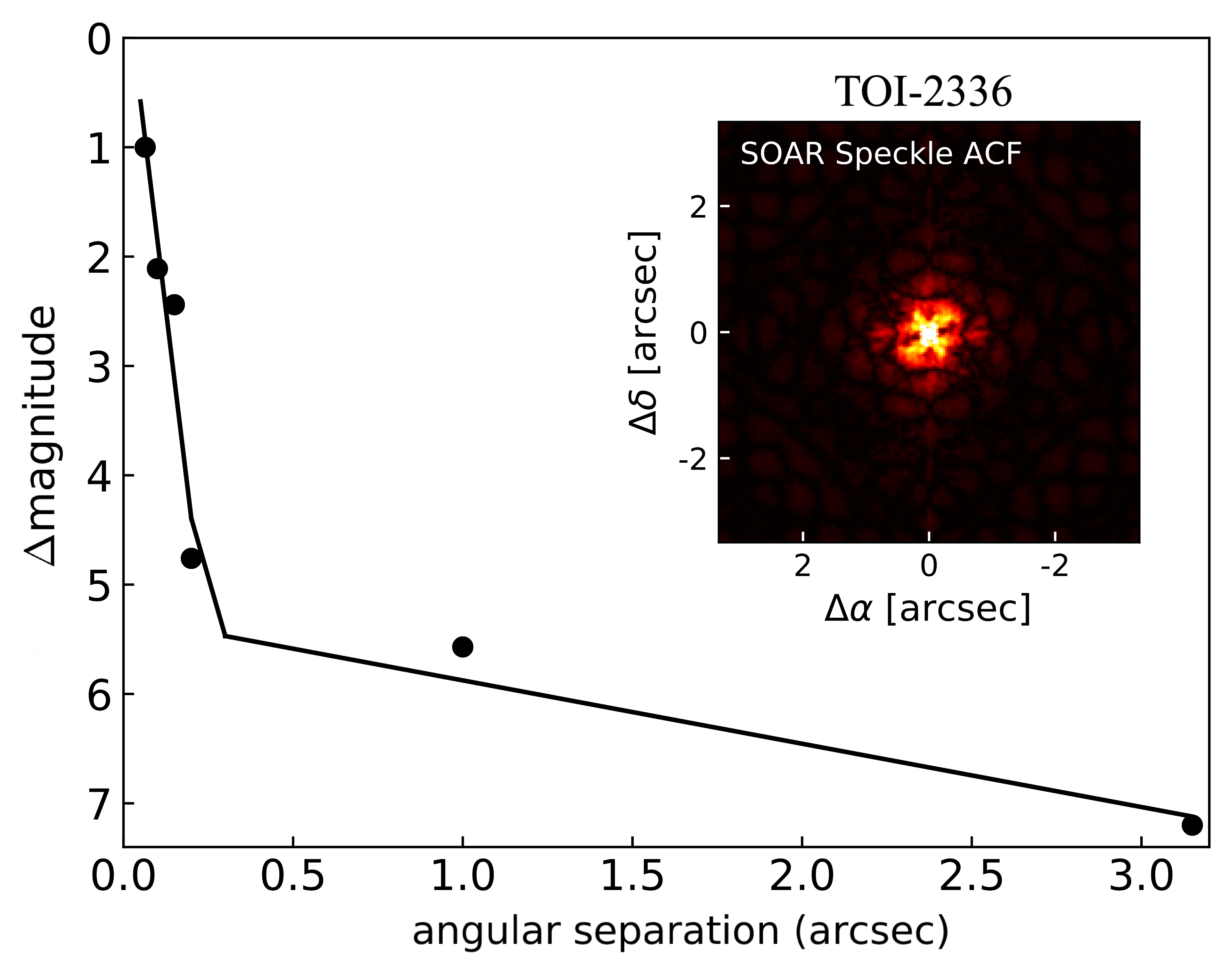}}

\subfigure[TOI-2521]{\includegraphics[width=0.49\textwidth]{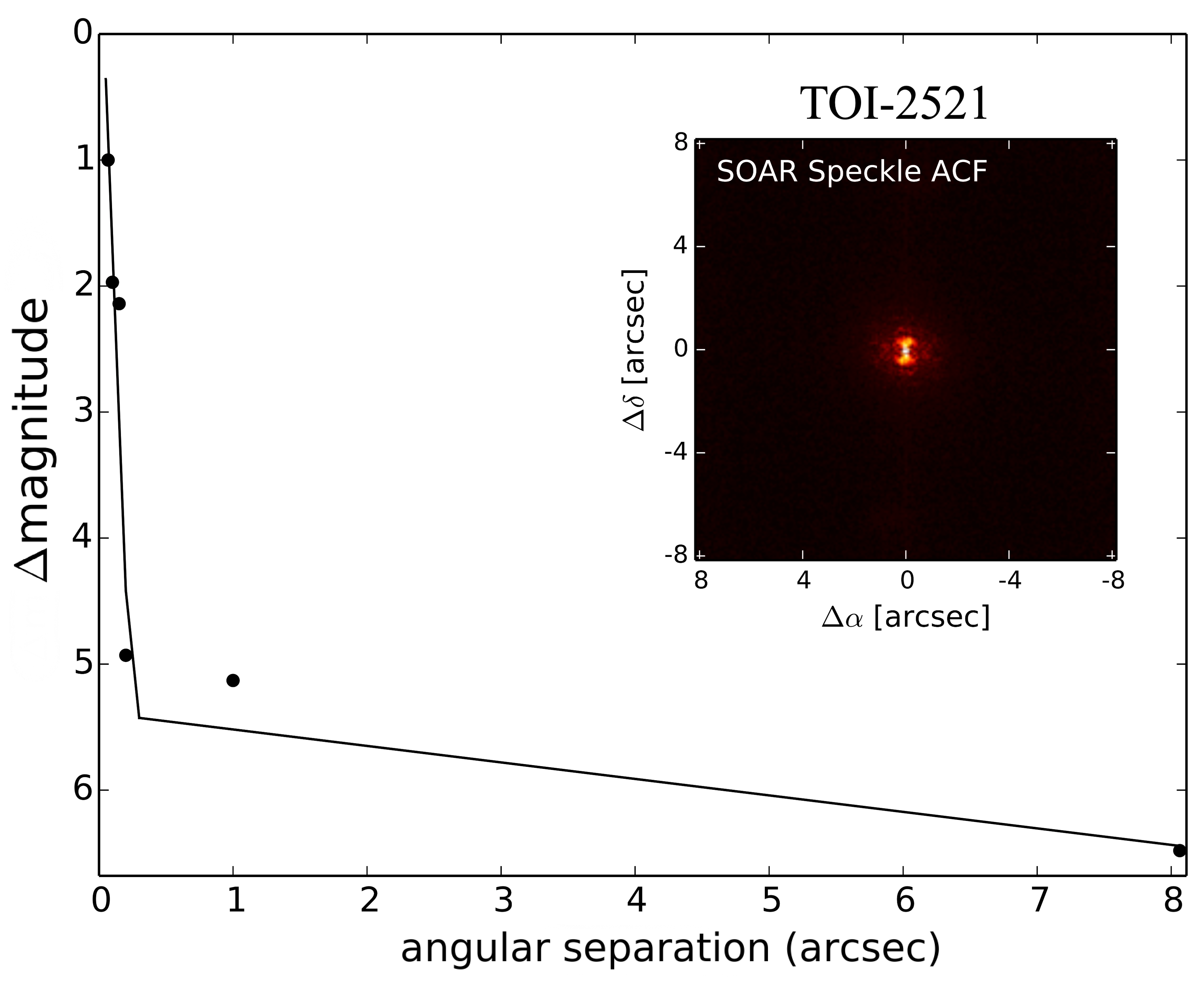}}
\caption{Speckle auto-correlation functions for TOI-2336 and TOI-2521 obtained in the $I$ band using SOAR. The $5\sigma$ contrast curves are shown as the black points. The black solid line shows the linear fit of the data.} 
\label{fig_SOAR}
\end{figure}

We searched for stellar companions to TOI-2336 and TOI-2521 with speckle imaging on the 4.1-m Southern Astrophysical Research (SOAR) telescope \citep{Tokovinin2018} on 7 February 2021 UT and 1 October 2021, respectively. The observations were performed in Cousins I-band, a similar visible bandpass as \tess. Both observations were sensitive to an approximately 5-magnitude fainter star at an angular distance of 0.25 arcsec from the target. More details of the observations within the SOAR \tess\ survey are available in \citet{Ziegler2020}. The 5$\sigma$ detection sensitivity and speckle auto-correlation functions from the observations are shown in Figure \ref{fig_SOAR}. No nearby stars were detected within 3\arcsec of either TOI-2336 or TOI-2521 in the SOAR observations.

\section{Stellar Properties}\label{stellar_properties}

In this section, we present the characterization of the host stars of the three targets. Briefly, we first performed an analysis of the broadband spectral energy distribution (SED) in Section \ref{SED}. Then in Section \ref{subsect_isochrone} we perform isochrone fitting utilizing the $T_{\rm eff}$, $[{\rm Fe/H}]$ and $\log g$ from SED fitting, the photometry of various bands, and the density derived from the transit model. Additionally, we performed independent spectral analyses in Section \ref{subsect_star_spectra} to validate the results from the isochrone fitting. Finally, we adopted the results from the isochrone fitting as the stellar parameters. Section \ref{subsect_star_rotation}-\ref{subsect:asteroseismology} are further discussion regarding the stellar rotation, age, and asteroseismology, respectively. All of the adopted stellar parameters are listed in Table \ref{tab:stellar_param}.

\begin{table*}
    \centering
    \caption{\textbf{Adopted stellar parameters} for TOI-1608, TOI-2336, and TOI-2521}
    \begin{tabular}{lcccl}
    \hline\hline
   Parameter &  TOI-1608   &   TOI-2336   &  TOI-2521  &  Source\\ \hline
   \it{Main identifiers}\\
   TIC     &138017750     &88902249    &72556406\\
   \gaia ID    &125548289270234880     &6206457922905803392    &3005862518856922752\\
   
   \it{Equatorial Coordinates} \\
   RA  & 3:23:12.28	& 15:20:55.30	& 6:14:06.99   &\gaia \ DR3$^{[1]}$\\
   DEC     & 33:04:41.25	& -33:41:32.87	& -8:03:44.31 &\gaia \ DR3\\
   
   \it{Photometric properties} \\
   \tess \ (mag) &$7.42812\pm 0.0064$	&$10.1757\pm 0.0061$	&$10.8218\pm 0.0065$ &TICv8$^{[2]}$ \\
   \gaia \ (mag) &$7.8280\pm 0.0004$	&$10.5905\pm 0.0002$	&$11.2692\pm 0.0008$ &\gaia \ DR3\\
   \gaia \ BP (mag) &$8.1097\pm 0.0016$	&$10.8992\pm 0.0006$	&$11.6116\pm 0.0024$ &\gaia \ DR3\\
   \gaia \ RP (mag) &$7.3789\pm 0.0012$	&$10.1146\pm 0.0004$	&$10.7518\pm 0.0018$ &\gaia \ DR3\\
   $B$ (mag) &$8.519\pm 0.027$	&$11.134\pm 0.119$	&$12.076\pm 0.231$  & Hipparcos\\
   $V$ (mag) &$7.96\pm 0.03$	&$10.756\pm 0.009$	&$11.263\pm 0.017$  & Hipparcos\\
   $J$ (mag) &$6.885\pm 0.027$	&$9.61\pm 0.024$	&$10.161\pm 0.026$  &2MASS$^{[3]}$\\
   $H$ (mag) &$6.636\pm 0.017$	&$9.386\pm 0.026$	&$9.839\pm 0.023$  &2MASS\\
   $K$ (mag) &$6.584\pm 0.029$	&$9.283\pm 0.019$	&$9.743\pm 0.019$  &2MASS\\
   \wise 1 (mag)    &$7.571\pm 0.402$	&$9.247\pm 0.023$	&$9.707\pm 0.023$  &\wise $^{[4]}$\\
   \wise 2 (mag)    &$6.554\pm 0.024$	&$9.247\pm 0.02$	&$9.726\pm 0.021$  &\wise\\
   \wise 3 (mag)    &$6.604\pm 0.015$	&$9.195\pm 0.039$	&$9.688\pm 0.043$  &\wise\\
   \wise 4 (mag)    &$6.52\pm 0.072$	&$8.49\pm 0.331$	&$9.193^{(a)}$  &\wise\\
   
   \it{Astrometric properties} \\
   $\varpi$ (mas)   &$9.87\pm 0.04$	&$3.37\pm 0.02$	&$2.99\pm 0.02$     &\gaia \ DR3\\
   $\mu_{\rm \alpha}\ ({\rm mas~yr^{-1}})$  &$109.30\pm 0.04$	&$6.22\pm 0.02$	&$-7.68\pm 0.03$    &\gaia \ DR3\\
   $\mu_{\rm \delta}\ ({\rm mas~yr^{-1}})$  &$-58.30\pm 0.03$	&$-18.97\pm 0.02$	&$-0.37\pm 0.03$    &\gaia \ DR3\\
   RV\ (km~s$^{-1}$)    &$44.5\pm 2.8$  &$35.6\pm 1.5$  &$-32.5\pm 2.0$    &\gaia \ DR3\\
   
   \it{Derived parameters} \\
   Distance (pc)    & $101.3\pm 0.4$    & $296.6\pm 1.7$    & $334.4\pm 2.3$    & Section \ref{subsect_isochrone}\\
   $M_{\ast}$  ($M_\odot$)      &  $1.31 \pm 0.07$    &   $1.40 \pm 0.07$   &   $0.95 \pm 0.06$    & Section \ref{subsect_isochrone}\\
   $R_{\ast}$  ($R_\odot$)      &  $2.16 \pm 0.08$    &  $1.82 \pm 0.06$   &   $1.74 \pm 0.06$     & Section \ref{subsect_isochrone}\\
   $\rho_\ast$ ($\rm g~cm^{-3}$)    & $0.18\pm 0.03$    & $0.33\pm 0.04$    & $0.25\pm 0.03$    & Section \ref{subsect_isochrone}\\
   $\log g_{\ast}$     &   $3.89 \pm 0.04$    &  $4.06 \pm 0.04$    &  $3.93 \pm 0.04$  & Section \ref{subsect_isochrone}\\
   $L_{\ast}$  (L$_\odot$)  &   $5.56 \pm 0.26$   &  $5.10 \pm 0.21$    &   $2.74 \pm 0.12$   & Section \ref{subsect_isochrone}\\
   $T_{\rm eff}$ (K)     &  $6028 \pm 82$  &  $6433 \pm 84$  & $5625 \pm 74$ & Section \ref{subsect_isochrone}\\
   $\rm [Fe/H]$ (dex)   & $0.09 \pm 0.12$   & $0.09 \pm 0.11$   & $-0.28 \pm 0.13$   & Section \ref{subsect_isochrone}\\
   Age (Gyr)    & $4.0 \pm 0.8$     & $2.4 \pm 0.5$     & $10.1 \pm 1.1$     & Section \ref{subsect_isochrone}\\
   $P_{\rm rot}$ (days)    & $2.7\pm 0.2$  &$4.2\pm 0.5$   &$6.9\pm 0.9$   &Section \ref{subsect_star_rotation} \\
    \hline\hline 
    \end{tabular}
    \begin{tablenotes}
    \item[1]  [1]\ \cite{Gaia2021_DR3}, [2]\ \cite{Stassun2017_TIC, Stassun2019_TICv8}, [3]\ \cite{Cutri2003_2MASS_JKH1}, [4]\ \cite{Wright2010_WISE}.
    \item[2] (a) The WISE catalog did not report an uncertainty for this $W4$ magnitude for TOI-2521. Therefore, we did not use this magnitude in our analysis.
    \end{tablenotes}
    \label{tab:stellar_param}
\end{table*}

\subsection{Spectral Energy Distribution}\label{SED}

\begin{figure}
    \centering
    \subfigure[TOI-1608]{\includegraphics[width=\linewidth,trim=90 70 90 70,clip]{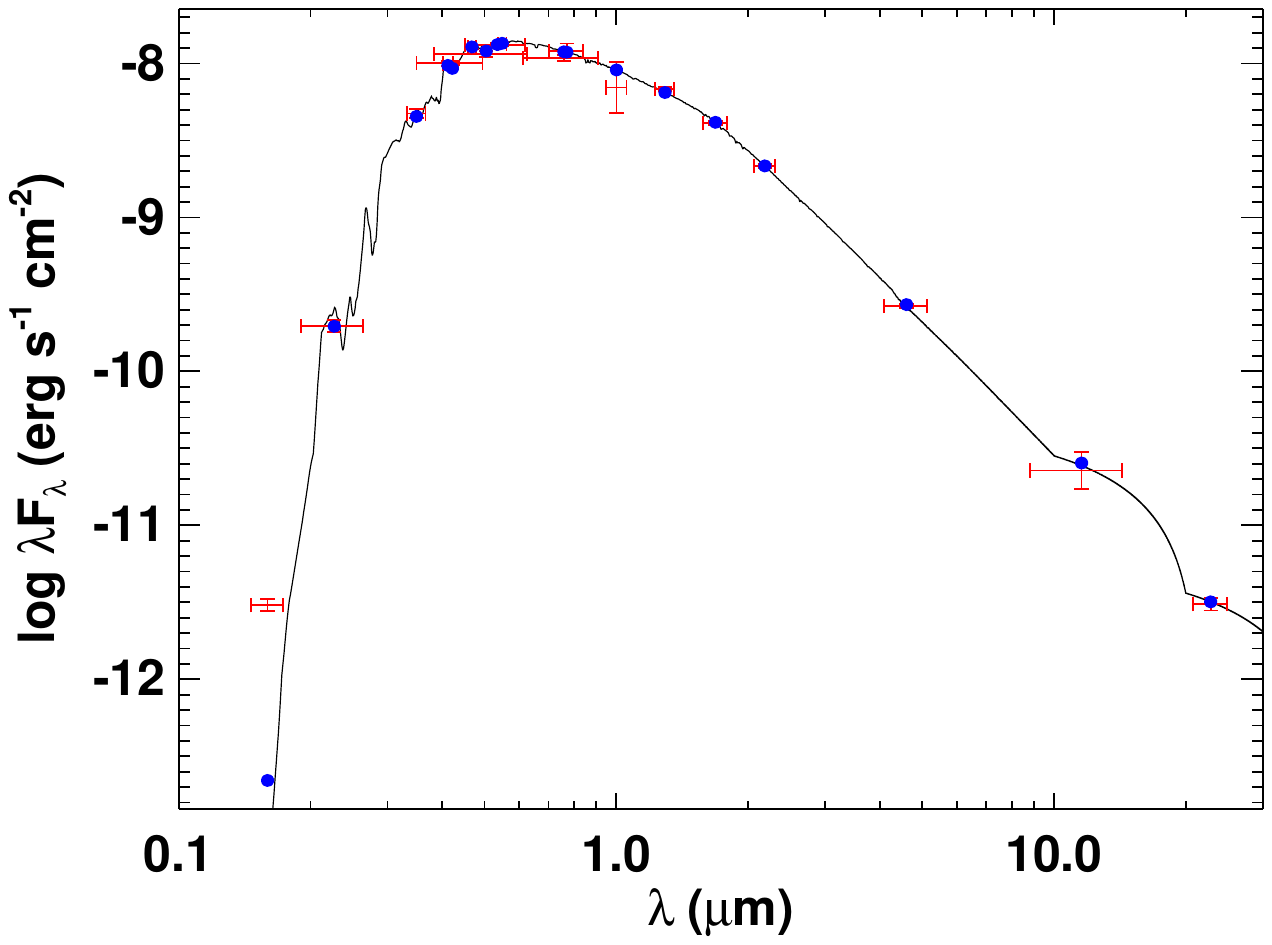}}
    \subfigure[TOI-2336]{\includegraphics[width=\linewidth,trim=90 70 90 70,clip]{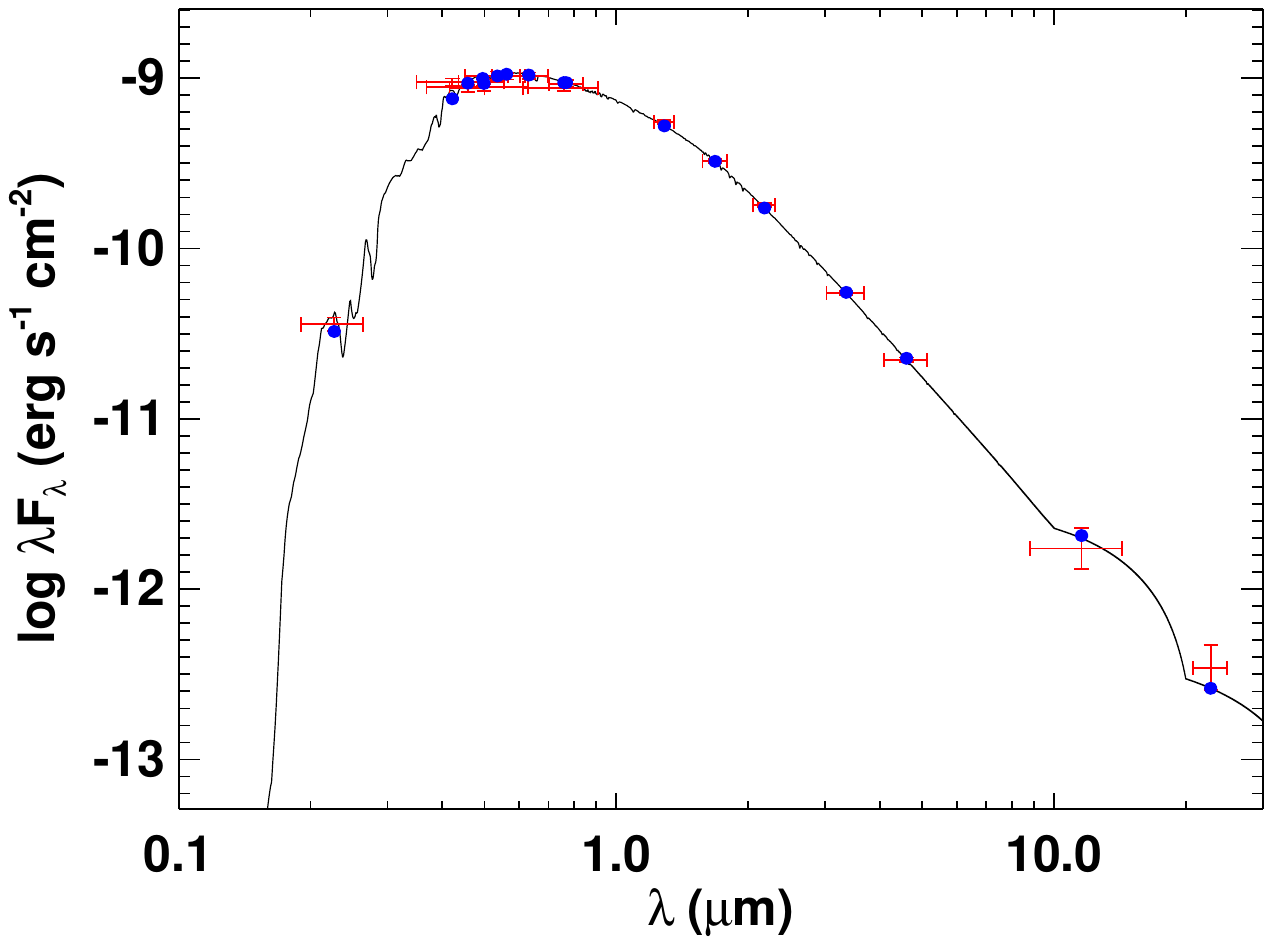}}
    \subfigure[TOI-2521]{\includegraphics[width=0.745\linewidth,trim=70 70 90 90,clip,angle=90]{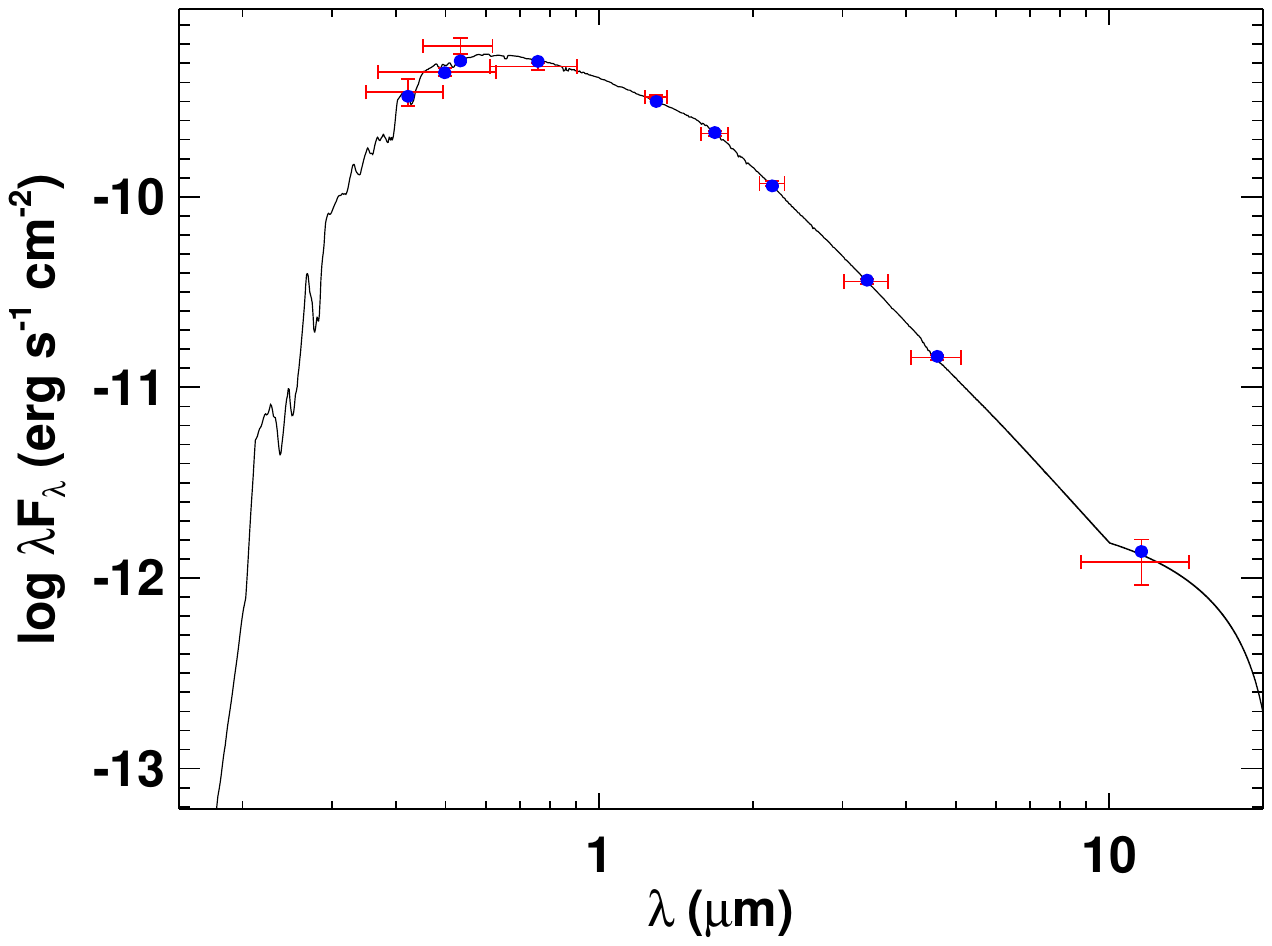}}
    \caption{Spectral energy distribution of TOI-1608 (top), TOI-2336 (middle), and TOI-2521 (bottom). Red symbols represent the observed photometric measurements, where the horizontal bars represent the effective width of the passband. Blue symbols are the model fluxes from the best-fit Kurucz atmosphere model (black).}
    \label{fig:sed}
\end{figure}

As an independent determination of the basic stellar parameters, we performed an analysis of the broadband spectral energy distribution (SED) of each star together with the {\it Gaia\/} EDR3 parallax \citep[with no systematic offset applied; see, e.g.,][]{StassunTorres:2021}, in order to determine an empirical measurement of the stellar radius, following the procedures described in \citet{Stassun:2016,Stassun:2017,Stassun:2018}. Depending on the photometry available for each source, we pulled the $B_T V_T$ magnitudes from {\it Tycho-2}, the $BVgri$ magnitudes from {\it APASS}, the Str\"omgren $uvby$ magnitudes from \citet{Paunzen:2015}, the $JHK_S$ magnitudes from {\it 2MASS}, the W1--W4 magnitudes from {\it WISE}, the $G G_{\rm BP} G_{\rm RP}$ magnitudes from {\it Gaia}, and the FUV and/or NUV fluxes from {\it GALEX}. All these magnitudes are shown in Table \ref{tab:stellar_param}. Together, the available photometry spans the stellar SED over the approximate wavelength range 0.2--22~$\mu$m (see Figure~\ref{fig:sed}).  

We performed fits to the photometry using Kurucz stellar atmosphere models, with the free parameters being the effective temperature ($T_{\rm eff}$), metallicity ([Fe/H]), and surface gravity ($\log g_{\ast}$). We also fit for the extinction, $A_V$, limited to the full line-of-sight value from the Galactic dust maps of \citet{Schlegel:1998}. The resulting fits shown in Figure~\ref{fig:sed} have a reduced $\chi^2$ ranging from 1.2 to 1.5, and the best-fit parameters are summarized in Table~\ref{tab:sed}. 

Integrating the model SED gives the bolometric flux at Earth, $F_{\rm bol}$. Taking the $F_{\rm bol}$ together with the {\it Gaia\/} parallax directly gives the luminosity, $L_{\rm bol}$. Similarly, the $F_{\rm bol}$ together with the $T_{\rm eff}$ and the parallax gives the stellar radius, $R_{\ast}$. Moreover, the stellar mass, $M_{\ast}$, can be estimated from the empirical eclipsing-binary based relations of \citet{Torres:2010}. When available, the {\it GALEX} photometry allows the activity index, $\log R'_{\rm HK}$ to be estimated from the empirical relations of \citet{Findeisen:2011}. All quantities are summarized in Table~\ref{tab:sed}. 

\begin{table}
    \centering
    \caption{Summary of derived and empirical parameters determined from fits to the stellar spectral energy distributions in Section \ref{SED}. $T_{\rm eff}$, $[{\rm Fe/H}]$ and $\log g_{\ast}$ here are used as the input of isochrone fitting in Section \ref{subsect_isochrone}}
    \begin{tabular}{lccc}
    \hline\hline
   Parameter &  TOI-1608   &   TOI-2336   &  TOI-2521  \\ \hline
    $A_V$     &  $0.01 \pm 0.01$   & $0.44 \pm 0.03$  &  $0.07 \pm 0.07$ \\
    $T_{\rm eff}$ [K]     &  $5950 \pm 100$  &  $6550 \pm 100$  & $5600 \pm 100$   \\
    $[{\rm Fe/H}]$       &   $0.1 \pm 0.3$    &  $0.0 \pm 0.3$   &  $-0.3 \pm 0.3$   \\
    $\log g_{\ast}$     &   $4.0 \pm 0.5$    &  $4.2 \pm 0.3$    &  $4.0 \pm 0.3$  \\
    $F_{\rm bol}$  [10$^{-9}$ erg s$^{-1}$ cm$^{-2}$]  &  $17.40 \pm 0.20$    &  $1.918 \pm 0.045$   &  $0.793 \pm 0.018$   \\ 
    $L_{\rm bol}$  [L$_\odot$]  &   $5.564 \pm 0.068$   &  $5.25 \pm 0.13$    &   $2.771 \pm 0.068$   \\
    $R_{\ast}$  [R$_\odot$]      &  $2.222 \pm 0.076$    &  $1.781 \pm 0.059$   &   $1.770 \pm 0.068$     \\
    $M_{\ast}$  [M$_\odot$]      &  $1.38 \pm 0.08$    &   $1.41 \pm 0.08$   &   $1.10 \pm 0.07$    \\
    $\log R'_{\rm HK}$       &   $-4.52 \pm 0.05$     &  ...  &  ...    \\
    \hline\hline 
    \end{tabular}
    \label{tab:sed}
\end{table}

\subsection{Isochrone fitting} \label{subsect_isochrone}

In order to obtain more precise stellar parameters, we performed isochrone fitting with the package \code{isochrones} \citep{Morton2015_isochrones} based on the results of the SED fitting and the photometry. We pulled $T_{\rm eff}$, $[{\rm Fe/H}]$ and $\log g$ from SED fitting, the \tess \ magnitudes from TICv8, the $JHK_S$ magnitudes from {\it 2MASS}, the W1--W3 magnitudes from {\it WISE} and the $G G_{\rm BP} G_{\rm RP}$ magnitudes from {\it Gaia} DR3. We also used the {\it Gaia\/} DR3 parallax as well as the stellar density from the transit model, which is described in Section \ref{analysis}. 

In the results of TOI-1608, we find bimodality in the posterior distribution of $M_{\ast}$, $\rho_{\ast}$, $\log g_{\ast}$ and age, which is shown in Figure \ref{fig_1608_iso_post}. We compared the best-fit values of these two peaks with the SED fitting results and the \gaia\ DR3 Final Luminosity Age Mass Estimator (FLAME) results \citep{Gaia2016_Gaia_1, Gaia2021_DR3}, and we found that the results of the posterior peak with a lower mass estimate are more consistent with the SED fitting and \gaia\ DR3. Therefore, we adopted the set of posterior distributions with a lower mass to derive the final stellar parameters and uncertainties from the isochrone fitting for TOI-1608. 

Similarly for TOI-2521, we checked for consistency and find that the mass estimates are different in these three sets of results. The mass values are $1.10\pm 0.07\ M_{\ast}$, $0.95\pm 0.06\ M_{\ast}$ and $1.05\pm 0.04\ M_{\ast}$ from the SED fitting, isochrone fitting and \gaia \ DR3, respectively. Relatively speaking, the mass derived from the SED fitting using the empirical relation from \citet{Torres:2010} is not as accurate as the other two estimates because the \citet{Torres:2010} relation probably does not work very well for aged stars due to the limit of their stellar sample. We consider the mass from our isochrone fitting as the more reliable one, as we incorporated the results from the SED fitting and also used more photometric data than \gaia\ DR3. Therefore, we adopted the results from our isochrone fitting as the final stellar parameters for TOI-2521.

For TOI-2336, our results from the isochrone fit are consistent with the SED fitting and \gaia\ DR3, so we adopted the isochrone fitting results as the final stellar parameters to stay consistent with the other two targets. The final stellar parameters we adopted in the following analyses are summarized in Table \ref{tab:stellar_param}.

\subsection{Spectral analysis} \label{subsect_star_spectra}

We analyzed CHIRON and TRES spectra to derive the stellar parameters. The observations and analyses are described in Section \ref{subsect_CHIRON} and \ref{subsect_TRES}, and the results are shown in Table \ref{tab_spec_parameter}. $T_{\rm eff}$ from both CHIRON and TRES and $[{\rm Fe/H}]$ from CHIRON are consistent with the results from other methods. $[{\rm M/H}]$ from TRES is obtained by using all of the lines in the wavelength region from $\sim 505\ nm$ to $\sim 536\ nm$. As iron lines dominate in this range, we compare these results with $[{\rm Fe/H}]$ from other methods and found they agree with each other. However, $\log g_{\ast}$ for TOI-2521 from CHIRON spectra is not consistent with isochrone fitting because TOI-2521 is a fast rotating star, where the spectra would have a relatively poor constraint on $\log g_{\ast}$. Thus, we only use the spectroscopic results as a check and adopt the results from isochrone fitting for further analyses.

\begin{table}
    \centering
    \caption{Summary of derived parameters determined from CHIRON and TRES spectra analysis.}
    \begin{tabular}{lccc}
    \hline\hline
   Parameter &  TOI-1608   &   TOI-2336   &  TOI-2521  \\ \hline
   \it{CHIRON} \\
    $T_{\rm eff}$ (K)    &  ...  &  $6551 \pm 100$  & $5526 \pm 50$   \\
    $[{\rm Fe/H}]$ (dex)      &   ...    &  $-0.03 \pm 0.10$   &  $-0.41 \pm 0.10$   \\
    $\log g_{\ast}$     &   ...    &  $4.16 \pm 0.10$    &  $3.70 \pm 0.10$  \\
    $v\sin i$\ (km~s$^{-1}$)    &  ...    &  $29.7 \pm 0.5$   &  $15.6 \pm 0.5$   \\ 
    \it{TRES} \\
    $T_{\rm eff}$ (K)    &  $6051 \pm 56$  &  ...  & $5604 \pm 50$   \\
    $[{\rm M/H}]$ (dex)      &   $0.10 \pm 0.08$    &  ...   &  $-0.21 \pm 0.08$   \\
    $v\sin i$\ (km~s$^{-1}$)    &  $48.3 \pm 0.5$    &  ...   &  $15.2 \pm 0.5$   \\
    \hline\hline 
    \end{tabular}
    \label{tab_spec_parameter}
\end{table}

\subsection{Stellar rotation period} \label{subsect_star_rotation}
\begin{figure}
    \centering
    \includegraphics[width=0.49\textwidth]{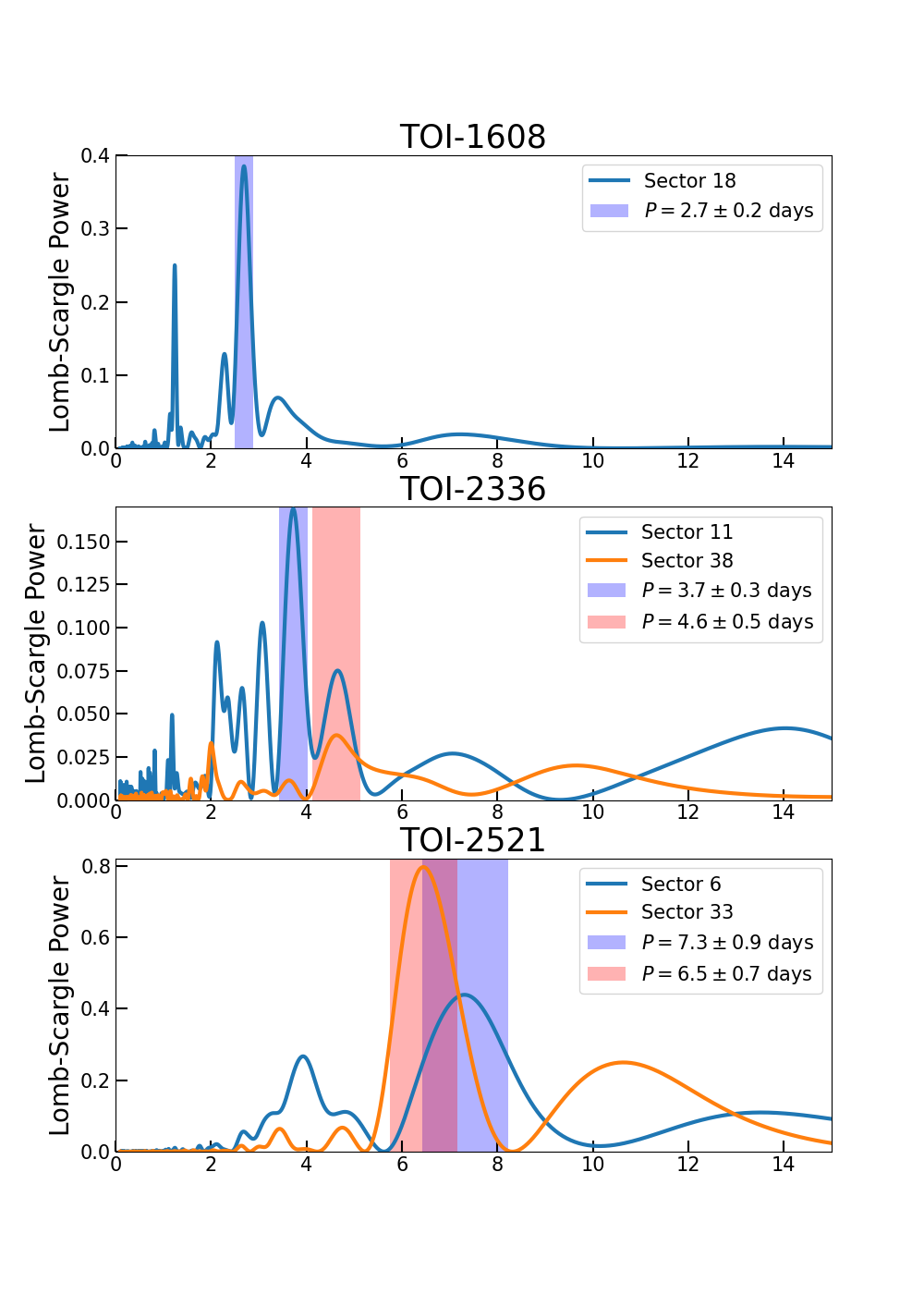}
    \caption{The generalized Lomb-Scargle periodogram from the \tess \ PDCSAP light curves of TOI-1608, TOI-2336, and TOI-2521. The periods of the highest peaks in each periodogram are marked as different color bars.} 
    \label{fig_GLS}
\end{figure}

We masked in-transit data points in the \tess\ PDCSAP light curves (shown in the left panels of Figure \ref{fig_tess_PDCSAP}) and used Generalized Lomb-Scargle periodograms to estimate the stellar rotation period of TOI-1608, TOI-2336, and TOI-2521. The Lomb-Scargle periodograms are shown in Figure \ref{fig_GLS}. We attributed the highest peak in the Lomb-Scargle periodogram to the stellar rotation period. For each peak, we used half of its FWHM as the uncertainty. For TOI-2336 and TOI-2521, we find the results obtained from different sectors are consistent. We used the average of different sectors as the final rotation period and use the higher uncertainty as the final error estimation. The final results are listed in Table \ref{tab:stellar_param}. Using these rotation period and $R_{\ast}$ derived in Section \ref{subsect_isochrone}, we can calculate the rotational velocities of TOI-1608, TOI2336, and TOI-2521 to be $40.5\pm 3.5$ km~s$^{-1}$, $21.9\pm 3.1$ km~s$^{-1}$, and $12.8\pm 1.9$ km~s$^{-1}$, respectively. Compared with $v\sin i$ derived from the spectra, the velocities of TOI-1608 and TOI-2336 are about $2\sigma$ smaller and the velocity of TOI-2521 is about $1\sigma$ smaller, which is unphysical (though still roughly consistent statistically speaking). This may result from the unreliable $v\sin i$ estimates from spectral analyses caused by the fast rotation and low $\log g_{\ast}$ of these three targets, or it may result from our estimates of the rotation period using the \tess\ light curves, which have a limited baseline. More accurate measurements on $v\sin i$ and the rotation period in the future would be able to put constraints on the projected stellar obliquity \citep{Masuda2020}.

\subsection{Stellar age} \label{subsect:stellar_age}

We determined the ages of these three stars from isochrone fitting in Section \ref{subsect_isochrone} and their rotational periods in Section \ref{subsect_star_rotation}. However, their rotations seem to be too fast for their ages. For example, according to Figure 7 in \cite{Curtis2020_Gyrochronology1}, the rotational period of TOI-2521 should be larger than 20 days as its age is larger than 3 Gyr, which is not consistent with its measured rotational period of $\sim 7$ days. Therefore, we adopted several methods to check whether these three stars are indeed old stars. 

\begin{figure}
\centering
\subfigure{\includegraphics[width=0.49\textwidth]{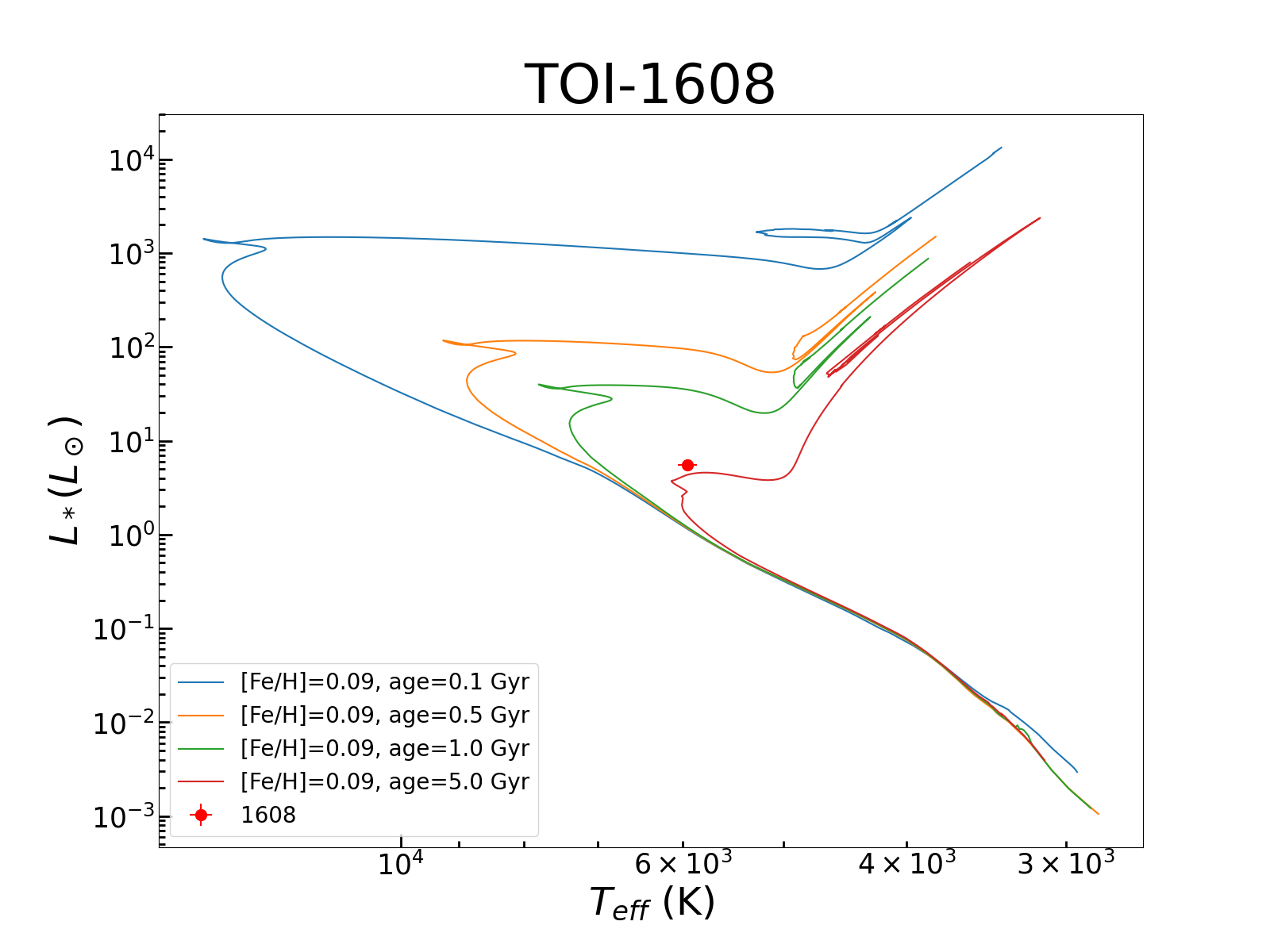}}

\subfigure{\includegraphics[width=0.49\textwidth]{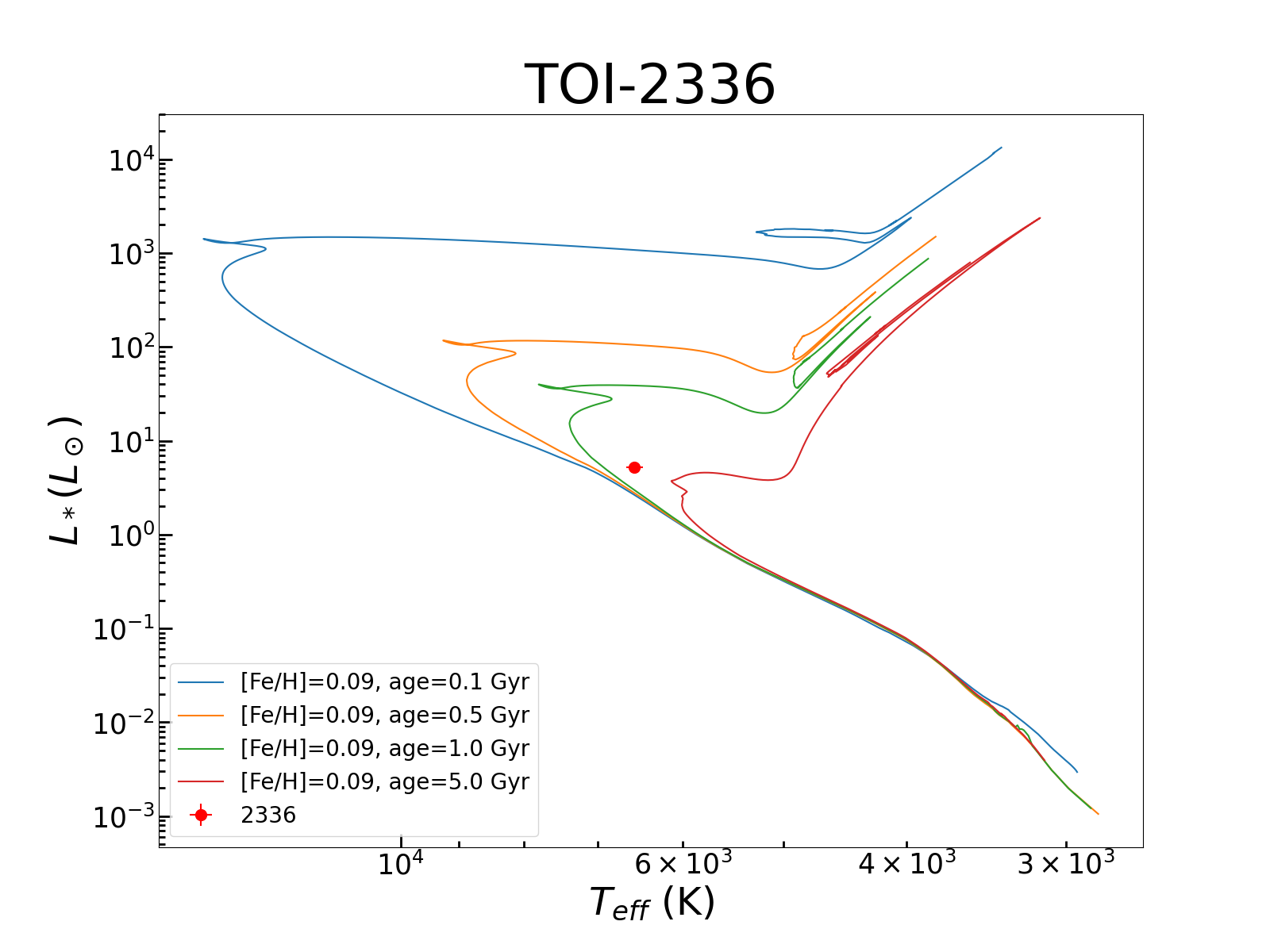}}

\subfigure{\includegraphics[width=0.49\textwidth]{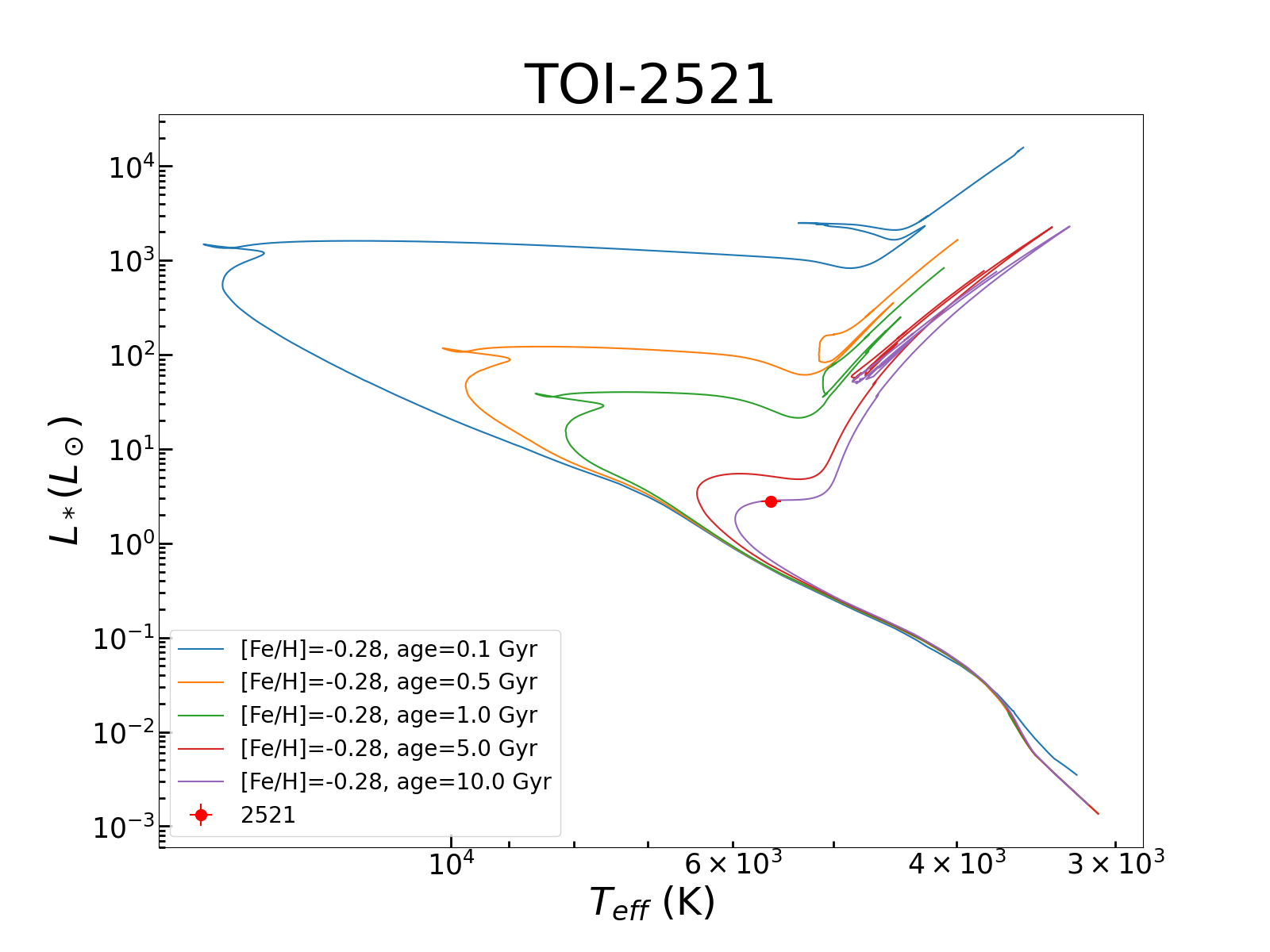}}

\caption{TOI-1608, TOI-2336, and TOI-2521 in HR diagrams. We also plot several isochrones of different ages as a comparison. These diagrams show that all of our targets have evolved a little off the main sequence and are consistent with being old stars.} 
\label{fig_HR-diagram}
\end{figure}

We compared these three targets with stellar evolution models in the Hertzsprung-Russell diagram (HR diagram; shown in Figure \ref{fig_HR-diagram}). We used the package \code{isochrones} which adopted MIST model \citep{Dotter2016_MIST} to plot the isochrones in Figure \ref{fig_HR-diagram}. As a result, we can see that all these three targets have evolved a little off the main sequence and are consistent with the old ages instead of young ages. In addition, we check the lithium absorption lines in the CHIRON and TRES spectra of these targets and found essentially no Li absorption, which also supports the old ages. Therefore, we conclude that all our targets are indeed old stars.

We also found explanations for the fast rotation of these three stars. For TOI-1608 and TOI-2521, we found that tidal spin-orbit synchronization can account for their fast rotation. We discuss this scenario in detail in Section \ref{subsubsect:Spin-orbit_synchronization}. For TOI-2336, we found that, given its effective temperature of $6433\pm 84$ K and mass of $1.40\pm 0.07\ M_{\ast}$, its fast rotation is not conflict with gyrochronology \citep[e.g.][]{Curtis2020_Gyrochronology1, Spada2020_Gyrochronology2}. Therefore, we believe our age estimations of all three targets are reliable.

In addition, we examined the kinematics of these three targets and determined that they are very likely thin disk stars. However, regarding TOI-2521, its age of $\sim 10$ Gry seems to conflict with the thin disk's age of $\sim 8$ Gyr. This discrepancy may indicate some inaccuracy in the isochrone fitting, and TOI-2521 may not be as old as 10 Gyr. However, it is important to note that stars also migrate and scatter, and there are old stars present in the thin disk as well. Therefore we cannot draw a strong conclusion for a single star.

\subsection{Asteroseismic analysis} \label{subsect:asteroseismology}

No unambiguous solar-like oscillations could be detected for these three stars. We calculated their expected frequencies of maximum power ($\nu_{\rm max}$) using the spectroscopic $T_{\rm eff}$ and $\log g_{\ast}$ from Section \ref{subsect_isochrone} with the seismic scaling relation \citep{brown1991,kb1995,lund2016}, yielding $\nu_{\rm max}\approx853\ {\rm \mu Hz}$ for TOI-1608, $\nu_{\rm max}\approx1221\ {\rm \mu Hz}$ for TOI-2336, and $\nu_{\rm max}\approx968\ {\rm \mu Hz}$ for TOI-2521. Since these values are all greater than the Nyquist frequencies for 30-min (${\sim}278\ {\rm \mu Hz}$) and 10-min (${\sim}833\ {\rm \mu Hz}$) cadence light curves, only 2-min cadence data are suited for the detection of solar-like oscillations -- which is only available for 1 sector for TOI-1608. We searched for oscillation signals using the \code{lightkurve} package \citep{Lightkurve2018} and from manual inspection of the power density spectra of the light curves with the transit signals masked out, but could not identify any solar-like oscillations. Considering the TESS magnitudes of the stars  \citep{Campante2016}, the fact that they were observed with TESS cameras 1 or 2 where background scattered light can be significant, and that clear rotational signals are evident in the photometry (at least for TOIs 1608 and 2521), indicating stellar surface activity that may suppress oscillation mode amplitudes \citep{Chaplin2011}, the non-detection is not entirely unexpected. Future \tess\ observations appended to the current coverage-limited light curves will be useful for searching for the oscillations of these stars.

\section{Analysis and results}\label{analysis}

\subsection{Joint RV and transit analysis}\label{Joint_fit}

We jointly fit the detrended \tess\ light curves and RVs by utilizing the \code{juliet} package \citep{Espinoza2019_Juliet}. For TOI-2336, we also include the \lco\ light curves in the joint fitting. \code{Juliet} is a modeling tool for exoplanetary systems. It employs the \code{batman} package \citep{Kreidberg2015_batman} to build transit models and employs \code{radvel} \citep{Fulton2018_radvel} to build RV models. It can fit both transit and RV data simultaneously by using bayesian inference with nested sampling with multiple choices of samplers. We used the sampler \code{dynesty} \citep{Speagle2020_DYNESTY} in the following analyses.

In the modeling for the three systems, we set the priors based in a similar fashion. We set wide uniform priors for the transit epoch ($T_{0, b}$) and the orbital period ($P_0$) around the values from ExoFOP \footnote{\url{https://exofop.ipac.caltech.edu/tess/}}, which are from the analysis by the \tess \ project. For the radius ratio ($p=R_b/R_*$) and the impact parameter ($b=a\cos i/R_*$), \code{juliet} uses the approach described in \cite{Espinoza2018_bp_sample} to fit the parametrizations $r_1$ and $r_2$ instead of fitting $b$ and $p$ directly. This parametrization allows us to efficiently sample the physically plausible zone in the (b,p) plane (i.e., b<1+p) by only simply uniformly sampling $r_1$ and $r_2$ between 0 and 1. Here we set uninformative uniform priors between 0 and 1 for $r_1$ and $r_2$. For the orbital eccentricity ($e$) and the argument of periapsis ($\omega_b$), we chose to fit for the parameters $e\sin \omega_b$ and $e\cos \omega_b$. The uninformative priors should be uniform priors between -1 and 1 for these two parameters and \code{juliet} will keep $e<1$ when sampling. However, we found the fitting would be stuck if we allow $e$ to be sampled very close to 1. Therefore, we set uniform priors between -0.7 and 0.7 for $e\sin \omega_b$ and $e\cos \omega_b$, which constrains $e<0.99$. As such priors are significantly wider than the resulting posteriors and changing the width down to (-0.1, 0.1) does not affect our final results, we think adopting such priors is essentially equivalent to using uninformative priors. 

For the \tess \ light curves, because the \tess\ PDCSAP light curves have already been corrected for the light dilution, we fixed the dilution factor D as 1. However, the dilution correction may not be optimal due to the complex TESS background flux like scattered light. To make sure these possible errors would not influence our results, we adopted a Gaussian prior with a mean of 1 and a variance of 0.1 to fit the data and obtained consistent results. For TOI-2336 and TOI-2521, we also compared the transit depths by fitting the light curves of each individual sector independently, and the results are consistent. In addition, the contamination ratios of these 3 systems given by TICv8 on ExoFOP are all less than 0.05, indicating that the contamination is very small, so any errors in the contamination correction of the TESS PDCSAP light curves should be negligible. Therefore, we conclude that contamination or scattered light should not be an issue in the TESS light curves or affect our results. For TOI-2521, Since we detected a faint close companion (see Section \ref{subsect:Gemini}), though we believe that it would only result in a radius error of less than 1\%, we still conducted additional testing to determine its potential impact. The companion is 3.94 magnitudes fainter than the primary star at 832 nm, representing $\sim 2.7\%$ of its flux. Therefore, we fixed the dilution factor at 0.97 to fit the data, and the resulting radius ratio increased by less than $1\sigma$. Therefore, we fixed the dilution factor as 1 for all three targets in our reported results.

We set a Gaussian prior with a zero mean and a small variance for the mean out-of-transit flux M because the light curves have been detrended and normalized. We fitted an extra flux jitter term ($\sigma$) to account for additional noise and set a wide log-uniform prior based on the photometric uncertainties of the light curves. We adopted a quadratic limb-darkening law for the \tess \ light curves and fitted the limb-darkening parameters $q_1$ and $q_2$ \citep{Kipping2013_limb-darkening} with uninformative priors between 0 and 1. For the \lco \ light curves of TOI-2336, we set similar priors for the dilution factor, mean out-of-transit flux, extra flux jitter term and limb-darkening parameters.

For the RV data, we fitted for the RV semi-amplitude ($K$) for each system with a systemic velocity ($\mu$) for each instrument. We estimated rough values for these parameters from the data through visual inspection and set very wide and essentially uninformative uniform priors around these values. We fitted an extra RV jitter term to account for any additional systematics for each instrument. We set wide log-uniform priors for them according to the RV precision of each instrument. The RVs of NRES for TOI-1608 on 9th January 2022 (BJD=2459588.661) and of CORALIE for TOI-2336 on 9th May 2021 (BJD=2459343.687687) were observed in transit. However, based on the equation (1) in \cite{Triaud2018}, the predicted semi-amplitudes of the Rossiter-McLaughlin effect \citep{Rossiter1924_RMeffect1, McLaughlin1924_RMeffect2} of these systems are both less than 50 m/s, which is less than the uncertainties of our RVs. Therefore, these RVs were not highly biased due to the transit and should not influence our RV modeling for the orbits.

\begin{figure}
\centering
\includegraphics[width=0.49\textwidth]{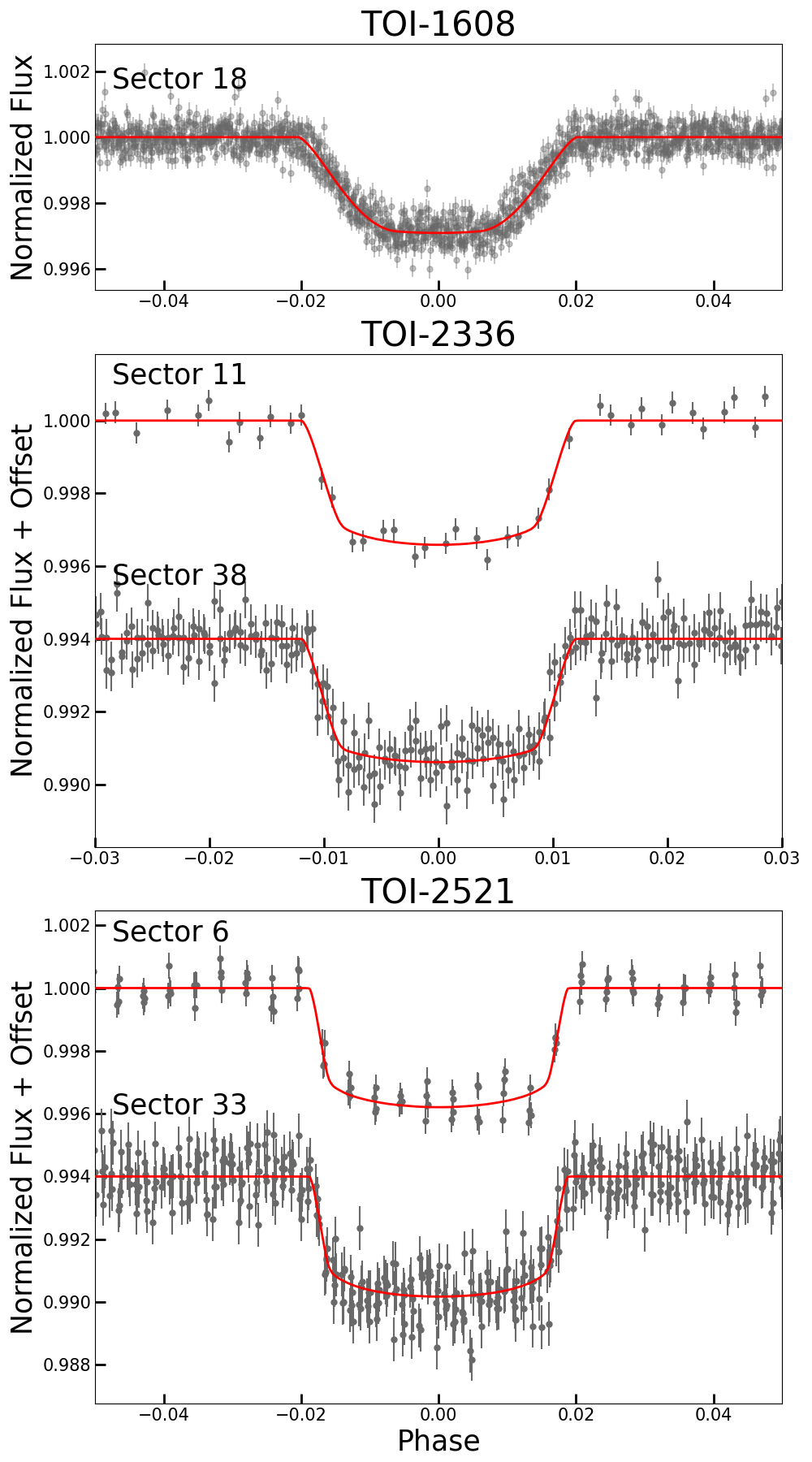}
\caption{The \tess \ light curves near the transit of TOI-1608, TOI-2336, and TOI-2521 with arbitrary offsets. The red line is the best-fit transit model obtained by \code{juliet}. The gray dots are the \tess \ data.} 
\label{fig_BD_lc}
\end{figure}

\begin{figure}
\centering
\includegraphics[width=0.49\textwidth]{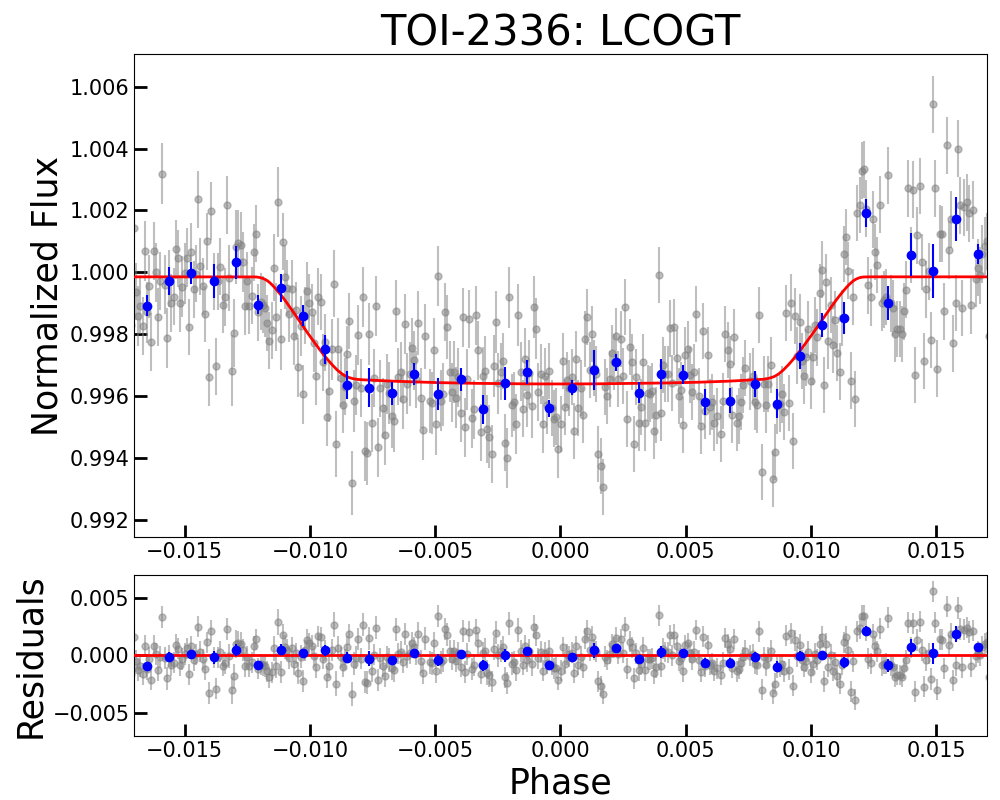}
\caption{The \lco \ light curves of TOI-2336 in the Pan-STARRS $z$-short ban ($z_S$). The red line is the best-fit transit model obtained by \code{juliet}. The gray dots are the \lco \ data and the blue dots are binned data.} 
\label{fig_2336_lco}
\end{figure}

\begin{figure*}
\centering
\includegraphics[width=\textwidth]{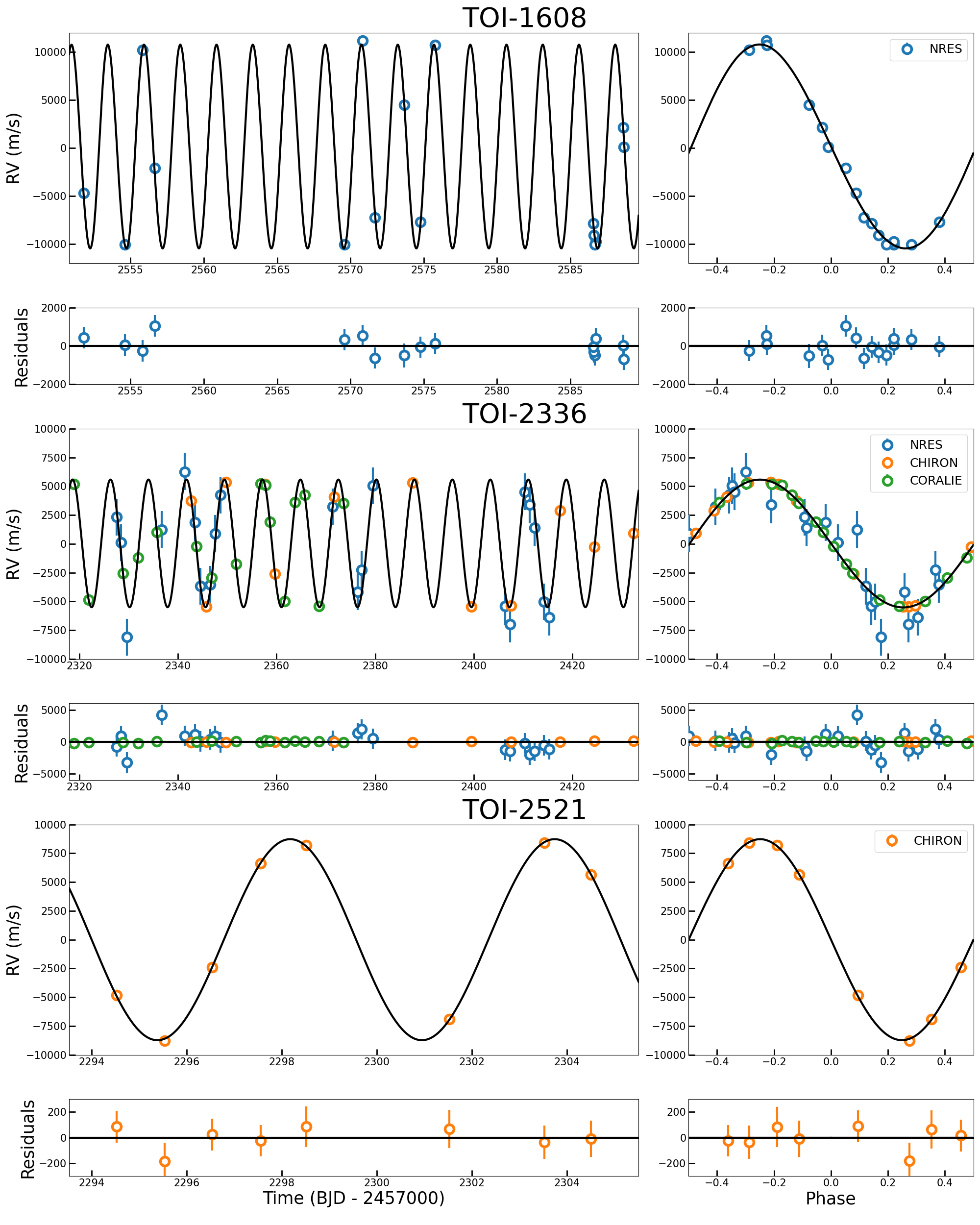}
\caption{The RVs of TOI-1608, TOI-2336 and TOI-2521. The black line is the best-fit model obtained by \code{juliet}. The error bars are the quadrature sum of the instrument jitter term and the measurement uncertainties for all RVs. The left row is the RVs as a function of time and the right row is the phase-folded RVs.} 
\label{fig_BD_rv}
\end{figure*}

We summarize the prior settings and the best-fit values for each parameter in our modeling in Table \ref{tab_1608_joint_model}, \ref{tab_2336_joint_model} and \ref{tab_2521_joint_model}. We show the light curves with the best-fit model in Figure \ref{fig_BD_lc} and \ref{fig_2336_lco}, and show the RV data with the best-fit model in Figure \ref{fig_BD_rv}.

We combine the parameters from the joint RV and transit modeling with the stellar parameters obtained in Section \ref{stellar_properties} to derive the physical parameters of the companions. The results are shown in Table \ref{tab:companion_param}. 

\begin{table*}
    \renewcommand{\arraystretch}{1.4}
    \caption{Prior settings and the best-fit values along with the 68\% credibility intervals in the final joint fit for TOI-1608. $\mathcal{N}$($\mu\ ,\ \sigma^{2}$) means a normal prior with mean $\mu$ and standard deviation $\sigma$. $\mathcal{U}$(a\ , \ b) stands for a uniform prior ranging from a to b. $\mathcal{J}$(a\ , \ b) stands for a Jeffrey's prior ranging from a to b.}.
    \begin{tabular}{lccr}
        \hline\hline
        Parameter       &Prior &Best-fit    &Description\\\hline
        \it{Companion's parameters}\\
        $P_{b}$ (days)  &$\mathcal{U}$ ($2.4$\ ,\ $2.6$)  &$2.47275^{+0.00004}_{-0.00004}$
        &Orbital period of TOI-1608b.\\
        $T_{0,b}$ (BJD)  &$\mathcal{U}$ ($2458791$\ ,\ $2458793$)   &$2458792.4599^{+0.0007}_{-0.0007}$
        &Mid-transit time of TOI-1608b.\\
        $r_{1,b}$  &$\mathcal{U}$ (0\ ,\ 1)   &$0.952^{+0.003}_{-0.004}$
        &Parametrisation for {\it p} and {\it b}.\\
        $r_{2,b}$  &$\mathcal{U}$ (0\ ,\ 1)   &$0.057^{+0.002}_{-0.002}$
        &Parametrisation for {\it p} and {\it b}.\\
        $e \sin \omega_{b}$                     &$\mathcal{U}$ (-0.7\ ,\ 0.7)  &$-0.02^{+0.03}_{-0.03}$ &Parametrisation for $e$ and $\omega_{b}$.\\
        $e \cos \omega_{b}$          &$\mathcal{U}$ (-0.7\ ,\ 0.7) &$0.02^{+0.03}_{-0.03}$  &Parametrisation for $e$ and $\omega_{b}$.\\

        \it{\tess\ photometry parameters}\\
        $D_{\rm TESS-S18}$  &Fixed    &$1$      &\tess\ photometric dilution factor.\\
        $M_{\rm TESS-S18}$  &$\mathcal{N}$ (0\ ,\ $0.1^{2}$)   &$0.0000004^{+0.000003}_{-0.000003}$      &Mean out-of-transit flux of \tess\ photometry.\\
        $\sigma_{\rm TESS-S18}$ (ppm) &$\mathcal{J}$ ($0.1$\ ,\ $1000$)  &$239^{+3}_{-3}$
          &\tess\ additive photometric jitter term.\\
        $q_{\rm 1, TESS-S18}$       &$\mathcal{U}$ (0\ ,\ 1)           &$0.12^{+0.12}_{-0.08}$   &Quadratic limb darkening coefficient.\\
        $q_{\rm 2, TESS-S18}$       &$\mathcal{U}$ (0\ ,\ 1)                 &$0.48^{+0.30}_{-0.32}$  &Quadratic limb darkening coefficient.\\
        
        \it{Stellar parameters}\\
        ${\rho}_{\ast}$ ($\rm kg\ m^{-3}$) &$\mathcal{J}$ ($10$\ ,\ $1000$)   &$224^{+27}_{-21}$
         &Stellar density.\\
         
        \it{RV parameters}\\
        $K_{b}$ ($\rm m\ s^{-1}$)  &$\mathcal{U}$ ($0$\ ,\ $20000$)      &$10649^{+232}_{-216}$  &RV semi-amplitude of TOI-1608b.\\
        
        \it{\nres RV parameters}\\
        $\rm \mu_{NRES}$ ($\rm m\ s^{-1}$) &$\mathcal{U}$ ($70000$\ ,\ $100000$)   &$82204^{+281}_{-283}$
        &Systemic velocity for \nres.\\
        $\rm \sigma_{NRES}$ ($\rm m\ s^{-1}$) &$\mathcal{J}$ ($1$\ ,\ $10000$)   &$542^{+139}_{-103}$
        &Extra jitter term for \nres.\\
        
        \hline\hline
    \label{tab_1608_joint_model}    
    \end{tabular}
\end{table*}

\begin{table*}
    \renewcommand{\arraystretch}{1.4}
    \caption{Similar to Table~\ref{tab_1608_joint_model} but for TOI-2336}.
    \begin{tabular}{lccr}
        \hline\hline
        Parameter       &Prior &Best-fit    &Description\\\hline
        \it{Companion's parameters}\\
        $P_{b}$ (days)  &$\mathcal{U}$ ($7.6$\ ,\ $7.8$)  &$7.711978^{+0.000013}_{-0.000013}$
        &Orbital period of TOI-2336b.\\
        $T_{0,b}$ (BJD)  &$\mathcal{U}$ ($2459358$\ ,\ $2459360$)   &$2459359.0493^{+0.0006}_{-0.0006}$
        &Mid-transit time of TOI-2336b.\\
        $r_{1,b}$  &$\mathcal{U}$ (0\ ,\ 1)   &$0.874^{+0.013}_{-0.014}$
        &Parametrisation for {\it p} and {\it b}.\\
        $r_{2,b}$  &$\mathcal{U}$ (0\ ,\ 1)   &$0.0592^{+0.0008}_{-0.0007}$
        &Parametrisation for {\it p} and {\it b}.\\
        $e \sin \omega_{b}$                     &$\mathcal{U}$ (-0.7\ ,\ 0.7)  &$-0.007^{+0.008}_{-0.007}$ &Parametrisation for $e$ and $\omega_{b}$.\\
        $e \cos \omega_{b}$          &$\mathcal{U}$ (-0.7\ ,\ 0.7) &$0.006^{+0.003}_{-0.003}$  &Parametrisation for $e$ and $\omega_{b}$.\\

        \it{\tess\ photometry parameters}\\
        $D_{\rm TESS-S11}$  &Fixed    &$1$      &\tess\ photometric dilution factor.\\
        $D_{\rm TESS-S38}$  &Fixed    &$1$  \\    
        $M_{\rm TESS-S11}$  &$\mathcal{N}$ (0\ ,\ $0.1^{2}$)   &$0.0000009^{+0.000009}_{-0.000009}$      &Mean out-of-transit flux of \tess\ photometry.\\
        $M_{\rm TESS-S38}$  &$\mathcal{N}$ (0\ ,\ $0.1^{2}$)   &$-0.000001^{+0.000008}_{-0.000008}$      \\
        $\sigma_{\rm TESS-S11}$ (ppm) &$\mathcal{J}$ ($1$\ ,\ $10000$)  &$12^{+35}_{-10}$
          &\tess\ additive photometric jitter term.\\
        $\sigma_{\rm TESS-S38}$ (ppm) &$\mathcal{J}$ ($1$\ ,\ $10000$)  &$117^{+28}_{-42}$ \\
        $q_{\rm 1, TESS-S11}$       &$\mathcal{U}$ (0\ ,\ 1)           &$0.19^{+0.12}_{-0.09}$   &Quadratic limb darkening coefficient.\\
        $q_{\rm 1, TESS-S38}$       &$\mathcal{U}$ (0\ ,\ 1)           &$0.12^{+0.07}_{-0.06}$   \\
        $q_{\rm 2, TESS-S11}$       &$\mathcal{U}$ (0\ ,\ 1)                 &$0.32^{+0.31}_{-0.20}$  &Quadratic limb darkening coefficient.\\
        $q_{\rm 2, TESS-S38}$       &$\mathcal{U}$ (0\ ,\ 1)                 &$0.56^{+0.27}_{-0.32}$  \\
        
        \it{\lco\ photometry parameters}\\
        $D_{\rm LCOGT}$  &Fixed    &$1$      &\lco\ photometric dilution factor.\\
        $M_{\rm LCOGT}$  &$\mathcal{N}$ (0\ ,\ $0.1^{2}$)   &$0.00015^{+0.00009}_{-0.00009}$      &Mean out-of-transit flux of \lco\ photometry.\\
        $\sigma_{\rm LCOGT}$ (ppm) &$\mathcal{J}$ ($1$\ ,\ $10000$)  &$1179^{+62}_{-62}$
          &\lco\ additive photometric jitter term.\\
        $q_{\rm 1, LCOGT}$       &$\mathcal{U}$ (0\ ,\ 1)           &$0.02^{+0.04}_{-0.02}$   &Quadratic limb darkening coefficient.\\
        $q_{\rm 2, LCOGT}$       &$\mathcal{U}$ (0\ ,\ 1)                 &$0.54^{+0.28}_{-0.32}$  &Quadratic limb darkening coefficient.\\
        
        \it{Stellar parameters}\\
        ${\rho}_{\ast}$ ($\rm kg\ m^{-3}$) &$\mathcal{J}$ ($10$\ ,\ $1000$)   &$246^{+38}_{-31}$
         &Stellar density.\\
         
        \it{RV parameters}\\
        $K_{b}$ ($\rm m\ s^{-1}$)  &$\mathcal{U}$ ($0$\ ,\ $20000$)      &$5549^{+37}_{-39}$  &RV semi-amplitude of TOI-2336b.\\
        
        \it{\nres RV parameters}\\
        $\rm \mu_{NRES}$ ($\rm m\ s^{-1}$) &$\mathcal{U}$ ($20000$\ ,\ $40000$)   &$33707^{+327}_{-314}$
        &Systemic velocity for \nres.\\
        $\rm \sigma_{NRES}$ ($\rm m\ s^{-1}$) &$\mathcal{J}$ ($0.1$\ ,\ $10000$)   &$1586^{+264}_{-205}$
        &Extra jitter term for \nres.\\
        
        \it{\chiron RV parameters}\\
        $\rm \mu_{CHIRON}$ ($\rm m\ s^{-1}$) &$\mathcal{U}$ ($20000$\ ,\ $40000$)   &$30288^{+52}_{-50}$
        &Systemic velocity for \chiron.\\
        $\rm \sigma_{CHIRON}$ ($\rm m\ s^{-1}$) &$\mathcal{J}$ ($0.1$\ ,\ $10000$)   &$4.7^{+27.3}_{-4.2}$
        &Extra jitter term for \chiron.\\
        
        \it{\coralie RV parameters}\\
        $\rm \mu_{CORALIE}$ ($\rm m\ s^{-1}$) &$\mathcal{U}$ ($20000$\ ,\ $40000$)   &$32432^{+36}_{-34}$
        &Systemic velocity for \coralie.\\
        $\rm \sigma_{CORALIE}$ ($\rm m\ s^{-1}$) &$\mathcal{J}$ ($0.1$\ ,\ $10000$)   &$104^{+40}_{-33}$
        &Extra jitter term for \coralie.\\
        
        \hline\hline
    \label{tab_2336_joint_model}    
    \end{tabular}
\end{table*}

\begin{table*}
    \renewcommand{\arraystretch}{1.4}
    \caption{Similar to Table~\ref{tab_1608_joint_model} but for TOI-2521}.
    \begin{tabular}{lccr}
        \hline\hline
        Parameter       &Prior &Best-fit    &Description\\\hline
        \it{Companion's parameters}\\
        $P_{b}$ (days)  &$\mathcal{U}$ ($5.5$\ ,\ $5.6$)  &$5.563060^{+0.000007}_{-0.000007}$
        &Orbital period of TOI-2521b.\\
        $T_{0,b}$ (BJD)  &$\mathcal{U}$ ($2459226$\ ,\ $2459228$)   &$2459227.2466^{+0.0007}_{-0.0007}$
        &Mid-transit time of TOI-2521b.\\
        $r_{1,b}$  &$\mathcal{U}$ (0\ ,\ 1)   &$0.71^{+0.07}_{-0.13}$
        &Parametrisation for {\it p} and {\it b}.\\
        $r_{2,b}$  &$\mathcal{U}$ (0\ ,\ 1)   &$0.0598^{+0.0011}_{-0.0013}$
        &Parametrisation for {\it p} and {\it b}.\\
        $e \sin \omega_{b}$                     &$\mathcal{U}$ (-0.7\ ,\ 0.7)  &$0.003^{+0.007}_{-0.007}$ &Parametrisation for $e$ and $\omega_{b}$.\\
        $e \cos \omega_{b}$          &$\mathcal{U}$ (-0.7\ ,\ 0.7) &$0.0003^{+0.004}_{-0.004}$  &Parametrisation for $e$ and $\omega_{b}$.\\

        \it{\tess\ photometry parameters}\\
        $D_{\rm TESS-S6}$  &Fixed    &$1$      &\tess\ photometric dilution factor.\\
        $D_{\rm TESS-S33}$  &Fixed    &$1$      \\
        $M_{\rm TESS-S6}$  &$\mathcal{N}$ (0\ ,\ $0.1^{2}$)   &$-0.000004^{+0.000012}_{-0.000012}$      &Mean out-of-transit flux of \tess\ photometry.\\
        $M_{\rm TESS-S33}$  &$\mathcal{N}$ (0\ ,\ $0.1^{2}$)   &$-0.000004^{+0.000011}_{-0.000011}$      \\
        $\sigma_{\rm TESS-S6}$ (ppm) &$\mathcal{J}$ ($0.001$\ ,\ $10$)  &$0.25^{+2.38}_{-0.24}$
          &\tess\ additive photometric jitter term.\\
        $\sigma_{\rm TESS-S33}$ (ppm) &$\mathcal{J}$ ($0.001$\ ,\ $10$)  &$0.12^{+1.99}_{-0.11}$\\
        $q_{\rm 1, TESS-S6}$       &$\mathcal{U}$ (0\ ,\ 1)           &$0.15^{+0.11}_{-0.07}$   &Quadratic limb darkening coefficient.\\
        $q_{\rm 1, TESS-S33}$       &$\mathcal{U}$ (0\ ,\ 1)           &$0.16^{+0.11}_{-0.08}$\\
        $q_{\rm 2, TESS-S6}$       &$\mathcal{U}$ (0\ ,\ 1)                 &$0.40^{+0.37}_{-0.28}$  &Quadratic limb darkening coefficient.\\
        $q_{\rm 2, TESS-S33}$       &$\mathcal{U}$ (0\ ,\ 1)                 &$0.31^{+0.27}_{-0.20}$  \\
        
        \it{Stellar parameters}\\
        ${\rho}_{\ast}$ ($\rm kg\ m^{-3}$) &$\mathcal{J}$ ($10$\ ,\ $1000$)   &$269^{+115}_{-77}$
         &Stellar density.\\
         
        \it{RV parameters}\\
        $K_{b}$ ($\rm m\ s^{-1}$)  &$\mathcal{U}$ ($0$\ ,\ $20000$)      &$8725^{+73}_{-59}$  &RV semi-amplitude of TOI-2521b.\\
        
        \it{\chiron RV parameters}\\
        $\rm \mu_{CHIRON}$ ($\rm m\ s^{-1}$) &$\mathcal{U}$ ($-40000$\ ,\ $-20000$)   &$-31334^{+49}_{-48}$
        &Systemic velocity for \chiron.\\
        $\rm \sigma_{CHIRON}$ ($\rm m\ s^{-1}$) &$\mathcal{J}$ ($0.1$\ ,\ $1000$)   &$113^{+96}_{-53}$
        &Extra jitter term for \chiron.\\
        
        \hline\hline
    \label{tab_2521_joint_model}    
    \end{tabular}
\end{table*}

\begin{table*}
    \renewcommand{\arraystretch}{1.4}
    \centering
    \caption{Physical parameters for TOI-1608b, TOI-2336b, and TOI-2521b.}
    \begin{tabular}{lcccr}
    \hline\hline
   Parameter &  TOI-1608   &   TOI-2336   &  TOI-2521   & Description\\ \hline
   $P$ (days)   &$2.47275^{+0.00004}_{-0.00004}$    &$7.711978^{+0.000013}_{-0.000013}$     &$5.563060^{+0.000007}_{-0.000007}$  &Orbital period.\\
   $R_b/R_{\ast}$   &$0.0574_{-0.0016}^{+0.0020}$  &$0.0592_{-0.0007}^{+0.0008}$  &$0.0598_{-0.0013}^{+0.0011}$    &Companion radius in units of stellar radius.\\
   $R_b$ ($R_J$)    &$1.21_{-0.06}^{+0.06}$     &$1.05_{-0.04}^{+0.04}$     &$1.01_{-0.04}^{+0.04}$  &Companion radius.\\
   $M_b$ ($M_J$)    &$90.7_{-3.7}^{+3.7}$   &$69.9_{-2.3}^{+2.3}$   &$77.5_{-3.3}^{+3.3}$   &Companion mass.\\
   $\rho_b$ ($\rm g~cm^{-3}$)   &$63.7_{-9.1}^{+10.3}$  &$75.0_{-7.9}^{+8.9}$  &$92.8_{-10.7}^{+12.3}$   &Companion density.\\
   $b$  &$0.929_{-0.006}^{+0.004}$  &$0.810_{-0.021}^{+0.019}$  &$0.56_{-0.20}^{+0.11}$  &Impact parameter.\\
   $a/R_{\ast}$     &$4.17_{-0.13}^{+0.16}$     &$9.18_{-0.40}^{+0.44}$     &$7.60_{-0.81}^{+0.96}$  &Semi-major axis in units of stellar radius.\\
   $a$ (au)     &$0.0419_{-0.0021}^{+0.0022}$   &$0.0777_{-0.0042}^{+0.0046}$   &$0.0615_{-0.0067}^{+0.0079}$    &Semi-major axis.\\
   $i$ (deg)    &$77.1_{-0.5}^{+0.5}$   &$84.9_{-0.4}^{+0.4}$   &$85.8_{-1.4}^{+1.8}$   &Inclination angle.\\
   $e$  &$0.041_{-0.019}^{+0.024}$  &$0.010_{-0.005}^{+0.006}$ &$<0.035^{[1]}$     &Eccentricity.\\
   $F$ ($F_{\oplus}$)   &$3155_{-278}^{+289}$   &$846_{-86}^{+92}$   &$721_{-153}^{+185}$   &Insolation flux relative to the Earth.\\
    \hline\hline 
    \end{tabular}
    \begin{tablenotes}
    \item[1]  [1] Here we report the upper limits at the 99\% confidence level.
    \end{tablenotes}
    \label{tab:companion_param}
\end{table*}

\subsection{Secondary eclipse} \label{subsect:secondary}

We conducted an analysis of the secondary eclipse signals in the \tess \ light curves. To prevent the removal of the secondary eclipse signal during detrending, we started by using \tess \ PDCSAP data and detrending the light curves. We masked out in-transit data as well as data predicted in the secondary eclipse based on the orbital parameters outlined in Section \ref{Joint_fit}. To account for the uncertainties in the orbital parameters, we masked out an additional period of time equal to the transit duration before and after the predicted secondary eclipse. We then employed the \code{PyTransit} package to fit the secondary eclipse. Our fitted parameters consisted of mid-transit time, orbital period, companion-to-star area ratio (equivalent to $R_b^2/R_{\ast}^2$), stellar density, impact parameter, companion-to-star flux ratio, $\sqrt{e} \sin \omega_{b}$, and $\sqrt{e} \cos \omega_{b}$ ($e$ and $\omega_{b}$ refer to the orbital eccentricity and the argument of periapsis). We set Gaussian prior with the mean and the variance derived from the results of Section \ref{Joint_fit} for all fitted parameters except the companion-to-star flux ratio, for which we assigned a uniform prior between 0 and 1.

\begin{figure}
\centering
\includegraphics[width=0.49\textwidth]{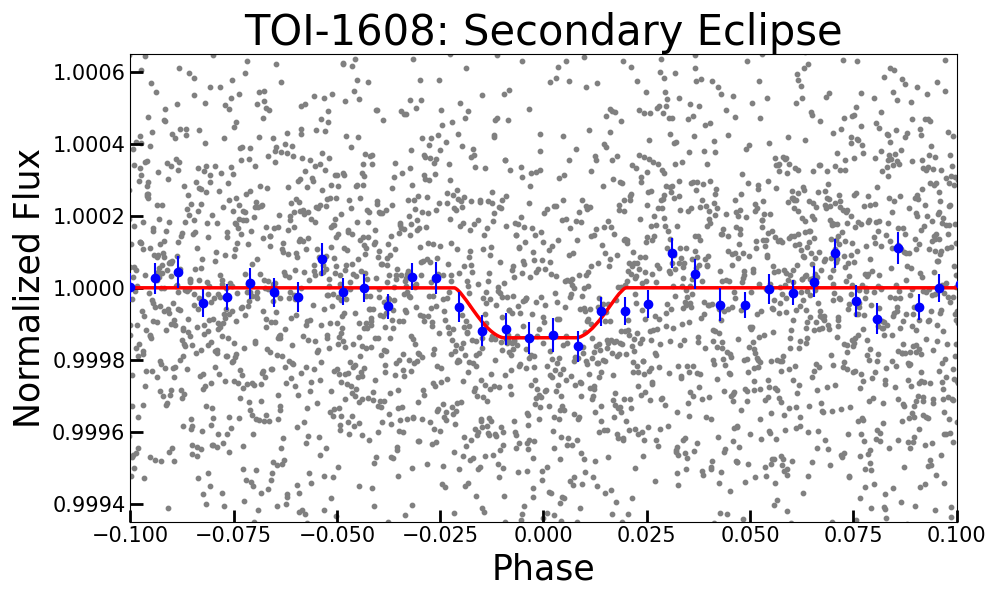}

\caption{The TESS light curves near the secondary eclipse of TOI-1608. The detrended TESS PDCSAP data, described in Section \ref{subsect:secondary}, are represented by gray dots and the binned data are in blue. The red line represents the best-fit secondary eclipse model obtained by \code{PyTransit}.} 
\label{fig_secondary_lc}
\end{figure}

We found that the secondary eclipse signal of TOI-1608 is significant, with a companion-to-star flux ratio of $0.043\pm 0.007$. The detrended light curve and the best-fit model for this result are shown in Figure \ref{fig_secondary_lc}. For TOI-2336 and TOI-2521, we found that their secondary eclipse signal is not significant and their flux ratio results are consistent with 0. Therefore, we only report the $3\sigma$ upper limit of the flux ratio, which is $<0.092$ for TOI-2336 and $<0.072$ for TOI-2521.

Then we used the obtained results to estimate the effective temperature of the companions. According to \cite{Charbonneau2005}, the flux ratio should be 
\begin{equation}
    \frac{F_b}{F_{\ast}}=\frac{\int F_b(\lambda)S(\lambda)\lambda d\lambda}{\int F_{\ast}(\lambda)S(\lambda)\lambda d\lambda},
\end{equation}
where $S(\lambda)$ is the \tess \ spectral response function, $F_b(\lambda)$ and $F_{\ast}(\lambda)$ represent the companion's and stellar surface fluxes as a function of the wavelength, respectively. For simplicity, we assumed blackbody spectra for both the companion and the star and neglected the reflected light, meaning that the albedo is 0. Combining the \tess \ spectral response function\footnote{\url{https://heasarc.gsfc.nasa.gov/docs/tess/the-tess-space-telescope.html}} and the stellar effective temperature from Table \ref{tab:stellar_param}, we estimated the effective temperature of the companions as follows: TOI-1608: $2983\pm 90$ K, TOI-2336: <3566 K, and TOI-2521: <3143 K. When assuming that the blackbody radiation of these companions is balanced with their incident flux, the corresponding equilibrium temperatures would be about 2081 K, 1509 K, and 1474 K, respectively. These values are all lower than the estimated effective temperature and upper limits, which is expected since these three targets are brown dwarfs or low-mass stars.

For TOI-1608, as its secondary eclipse signal is significant, we may use this signal to independently constrain its eccentricity. Although we used $e$ and $\omega_{b}$ from 
the transit to predict the time of the secondary eclipse when we detrended the light curves, the additional part we masked out ensures our results can remain independent with the $e$ and $\omega_{b}$ from transit. To constrain the eccentricity, we set a uniform prior from -1 to 1 for $\sqrt{e} \sin \omega_{b}$, and $\sqrt{e} \cos \omega_{b}$, kept other priors, and fitted the light curves. However, we found the secondary eclipse did not provide strong constraints on $e$ and $\omega_{b}$, as the posterior of $e$ was wide and consistent with a near zero $e$, consistent with our results from the joint fit listed in Table \ref{tab:companion_param}.

\section{Discussion}\label{discussion}

\subsection{Mass-radius relation and Inflation}

\begin{figure}
    \centering
    \includegraphics[width=0.49\textwidth]{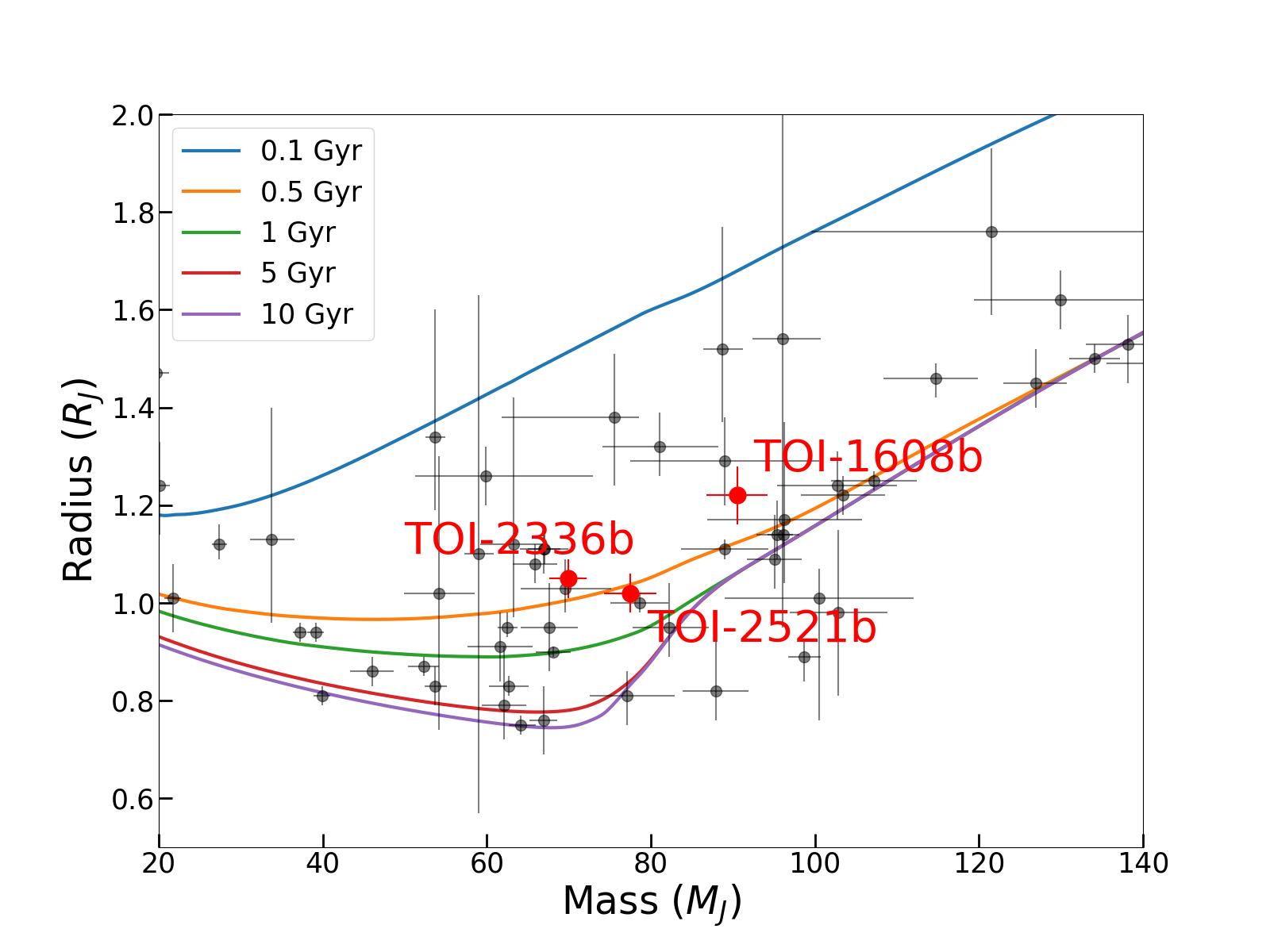}
    \caption{Radius-mass diagram for the transiting brown dwarfs and low-mass stars. The black dots are the `\sample', and the red dots are the three targets in this work. The isochrones of 0.1 Gyr, 0.5 Gry, 1 Gyr, 5 Gyr and 10 Gyr at solar metallicity from 
    \citet{Baraffe2015} and \citet{Phillips2020_ATMO2020} models are shown as lines with different colors. The sample and codes for generating this figure are available on \url{https://github.com/Ssealevel/Codes_TOI-1608_2336_2521}.} 
    \label{fig_BD_iso}
\end{figure}

We compare TOI-1608b, TOI-2336b and TOI-2521b with the models from \cite{Baraffe2015} and \cite{Phillips2020_ATMO2020} in the mass-radius diagram, which is shown in Figure \ref{fig_BD_iso}. For the following analysis, we also build a sample of transiting companions with a mass between $13\ M_J$ and $150\ M_J$. We use the sample compiled by \cite{Grieves2021} and add NGTS-19b \citep{Acton2021}, TOI-1278 B \citep{Artigau2021}, TOI-2119b \citep{Canas2022, Carmichael2022}, TIC-320687387 B \citep{Gill2022}, TOI-629b, TOI-1982b, and TOI-2543b \citep{Psaridi2022}. We call them the `\sample' in the following text, and we also plot this sample in Figure \ref{fig_BD_iso}. The sample and the codes for this figure are available on Github\footnote{\url{https://github.com/Ssealevel/Codes_TOI-1608_2336_2521}}. According to the plotted model isochrones, all of our three targets seem to be younger than $0.5\ Gyr$. This is much younger than the ages we obtained in Section \ref{stellar_properties} for the primary stars. For TOI-1608 and TOI-2336, if we adopt a very low age limit of $1\ Gyr$ for these systems, the predicted radii from the model isochrones would be $1.06\ R_J$ and $0.90\ R_J$, respectively, both of which are about $3\sigma$ smaller than the measured values from our joint transit and RV analyses. For TOI-2521, if we adopt a very low age limit of $5\ Gyr$ for the system, the predicted radius would be $0.84\ R_J$, which is $4\sigma$ smaller than our measured value. Therefore, TOI-1608b, TOI-2336b, and TOI-2521b should be all inflated.

The radius inflation of hot Jupiters has been observed and characterized in multiple previous studies \citep[e.g.,][]{Bouchy2011_BD_inflate1, Weiss2013}, and the stellar incident flux clearly plays an important role in shaping the sizes of these giant planets. \cite{Weiss2013} obtained an empirical relation for the inflation of planets heavier than $150\ M_{\oplus}$ (0.47 $M_J$)
\begin{equation} \label{eq_Jup_inflation}
    \frac{R_p}{R_{\oplus}} = 2.5\left( \frac{M_p}{M_{\oplus}} \right)^{-0.039} \left( {\rm \frac{F}{erg\ s^{-1}\ cm^{-2}}} \right)^{0.094}.
\end{equation}
However, for brown dwarf companions, previous research concluded that the stellar irradiation would not affect their radii as much as observed for the hot Jupiters \citep[e.g.,][]{Bouchy2011_BD_inflate1}. To compare the dependence on the incident flux for hot Jupiters, brown dwarfs, and low mass stars, we used the `\sample' and fitted a similar power-low relation to that of \cite{Weiss2013} as 
\begin{equation} \label{eq_fit_inflation}
    \frac{R_p}{R_J} = C\left( \frac{M_p}{M_J} \right)^{\alpha} \left( \frac{F}{F_{\oplus}} \right)^{\beta}.
\end{equation}
We fitted this relation for brown dwarfs and low mass stars separately with a cutoff at $80\ M_J$ \citep{Baraffe2002} to distinguish them. For both sets of samples, we calculated their time-averaged incident flux using the equation
\begin{equation}
    F = \sigma_{\rm sb} T_{\rm eff}^4 \left(\frac{R_*}{a} \right)^2 \sqrt{\frac{1}{1-e^2}},
\end{equation}
where $\sigma_{\rm sb}$ is the Stefan–Boltzmann constant. In the following analyses we excluded from the samples objects with precision on the radius or mass that is poorer than 30\%. We also excluded RIK 72b because it is a very young brown dwarf \citep{David2019_RIK72b} and probably still in the process of initial contraction after its formation. Finally, we have 34 brown dwarfs and 25 low mass stars in our fitting. We used \code{emcee} \citep{Foreman-Mackey2013_emcee} to perform a Markov chain Monte Carlo (MCMC) fitting for each sample. We set uniform priors between -5 and 5 for $\alpha$ and $\beta$ and a log-uniform prior between $10^{-5}$ and $10^5$ for $C$. The MCMC posteriors are shown in Figure \ref{fig_mcmc_post} and the best-fit values are listed in Table \ref{tab:inflation_fit}. The best-fit relation for the brown dwarfs ($13\ M_J\sim 80\ M_J$) is
\begin{equation} \label{eq_BD_inflation}
    \frac{R_p}{R_J} = 1.11\left( \frac{M_p}{M_J} \right)^{-0.052} \left( \frac{F}{F_{\oplus}} \right)^{0.009},
\end{equation}
and the relation for the low mass stars ($80\ M_J\sim 150\ M_J$) is
\begin{equation} \label{eq_star_inflation}
    \frac{R_p}{R_J} = 0.021\left( \frac{M_p}{M_J} \right)^{0.86} \left( \frac{F}{F_{\oplus}} \right)^{0.019}.
\end{equation}

\begin{figure}
\centering
\includegraphics[width=0.49\textwidth]{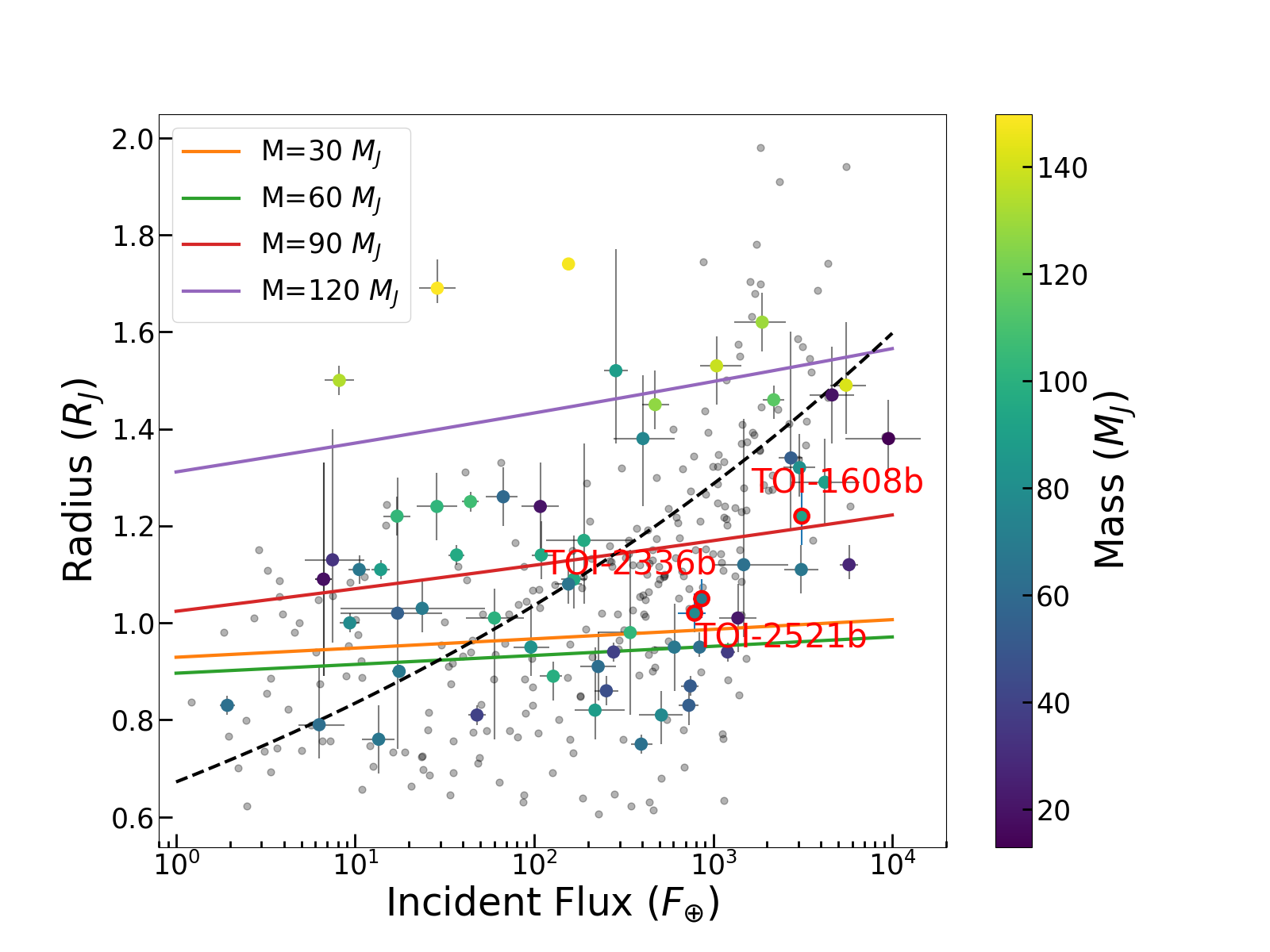}
\caption{The incident flux received by the companions vs.\ their radii. Dots with different colors are the `\sample', and the color scale represents the mass of these companions. The dots with red outlines are the three targets in this work and the gray dots in the background are hot Jupiters. Here we only show brown dwarfs and low mass stars with mass and radius measured with a precision better than 30\% and hot Jupiters with radius and incident flux with a precision better than 50\%. Equations \ref{eq_BD_inflation} and \ref{eq_star_inflation} are plotted in solid lines for two different selected masses in each mass range, and the dashed line presents the \citet{Weiss2013} relation for hot Jupiters with $1\ M_J$. The sample and codes for generating this figure are available on \url{https://github.com/Ssealevel/Codes_TOI-1608_2336_2521}.}
\label{fig_BD_inflation}
\end{figure}

\begin{figure}
\centering
\subfigure[Brown dwarfs]{\includegraphics[width=0.49\textwidth]{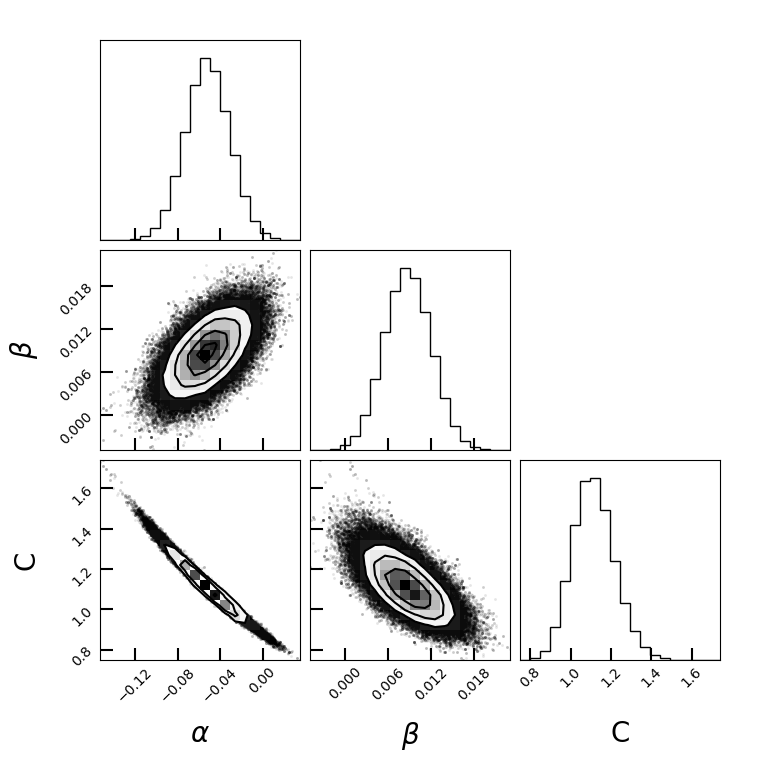}}

\subfigure[Low mass stars]{\includegraphics[width=0.49\textwidth]{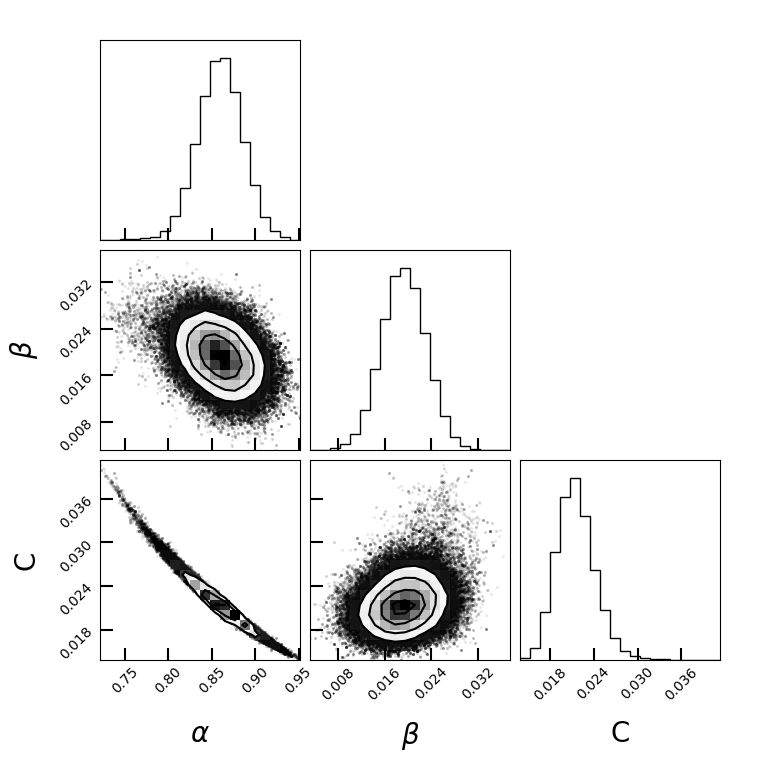}}
\caption{The posterior distribution of the MCMC fitting of Equation \ref{eq_fit_inflation} for brown dwarfs and low mass stars. $C$ is the normalization parameter, $\alpha$ is the power-law index for mass, and $\beta$ is the power-law index for incident flux. The corresponding results of these parameters are listed in Table \ref{tab:inflation_fit}.} 
\label{fig_mcmc_post}
\end{figure}

In Figure \ref{fig_BD_inflation}, we show the sample of transiting low-mass companions with mass around $13\ M_J\sim 150\ M_J$ as well as the sample of known hot Jupiters with the published values for the radius and the incident flux with a precision better than 50\%. The empirical relations for different types of objects are shown as different lines. Comparing these results, we find that the mass-radius relations for the three kinds of objects are different. The mass dependence for brown dwarfs and low mass stars from our fits is consistent with \cite{Chen2017_m_r_relation}. Besides, the power-law indices for incident flux in the mass-radius relation for brown dwarfs and low mass stars are significantly lower than that for hot Jupiters, which confirms that brown dwarfs are less affected by stellar irradiation \citep[e.g.,][]{Bouchy2011_BD_inflate1}. However, the best-fit indices for incident flux for brown dwarfs and low mass stars are not consistent with zero, which suggests that the effect on the size of these objects by the stellar irradiation is perhaps not totally negligible. Naively, one would speculate that objects with higher densities or higher surface gravity should be more difficult to inflate via external irradiation. Since brown dwarfs have higher densities than low mass stars on average \citep{Baraffe2003, Baraffe2008} \citep[also see Fig. 8 in][]{Persson2019}, it is consistent with the expectation that the brown dwarfs' radii have a weaker dependence on the incident flux than the low mass stars' (a smaller power-law index). Future studies with a larger and better characterized sample of low-mass objects may reveal more on this potential difference between these two populations. Finally, we caution that the radii of brown dwarfs and low mass stars can change considerably as they age (as can be seen in the isochrones in Figure \ref{fig_BD_iso}). For example, since the radii decrease with age, and irradiation is higher closer to the star where tidal effects are also stronger, if tides are strong enough to cause orbital decay and destruction of close-in objects within the main sequence lifetime of the host stars, then this may result in closer-in brown dwarfs being systematically younger and with larger radii than farther out objects. This may mimic a trend with irradiation. However, due to the lack of samples with accurate ages, we were unable to consider the dependence on age in this discussion. More samples with accurate age estimates and a more comprehensive analysis are needed in future works.

\begin{table}
    \renewcommand{\arraystretch}{1.5}
    \centering
    \caption{MCMC fitting results of Equation \ref{eq_fit_inflation} for brown dwarfs and low mass stars.}
    \begin{tabular}{ccc}
    \hline\hline
    Parameter   &  Brown dwarfs   &   Low mass stars\\ \hline
    $\alpha$    &$-0.052_{-0.021}^{+0.021}$     &$0.860_{-0.026}^{+0.025}$\\
    $\beta$     &$0.009_{-0.003}^{+0.003}$      &$0.019_{-0.004}^{+0.004}$ \\
    $C$     &$1.11_{-0.10}^{+0.11}$     &$0.021_{-0.002}^{+0.002}$\\
    \hline\hline 
    \end{tabular}
    \label{tab:inflation_fit}
\end{table}

\subsection{Tidal effects} \label{subsect_tidal}

Tidal effects are important in the evolution of close binary systems. There are several tidal effects to consider for binary systems with a relatively high mass ratio like the three cases here: tidal circularization, orbital decay, and spin-orbit synchronization. We discuss tidal circularization in Section \ref{subsubsect:circularization}. In Section \ref{subsubsect:Spin-orbit_synchronization}, we discuss spin-orbit synchronization as well as the fast rotation of TOI-1608 and TOI-2521. Finally, we discuss the orbital decay in Section \ref{subsubsect:orbital_decay}

\subsubsection{Tidal circularization} \label{subsubsect:circularization}

The tidal circularization effect is where the tidal force damps the eccentricity of the orbit. We examine the orbital circularization timescale
\begin{equation}
    \tau_e = \frac{e}{|de/dt|}
\end{equation}
of the three targets by using the equation from \cite{Jackson2008_Tidal}:
\begin{equation}
    \frac{1}{e}\frac{de}{dt} = -\left[\frac{63}{4}(GM_{\ast}^3)^{1/2}\frac{R_p^5}{Q_p M_p} + \frac{171}{16}(G/M_{\ast})^{1/2} \frac{R_{\ast}^5 M_p}{Q_{\ast}} \right]a^{-13/2},
\end{equation}
where $Q_p$ is the modified quality factor of the companion and $Q_*$ is the modified quality factor of the host star. The quality factor is impossible to measure precisely. For brown dwarf companions, this value is usually adopted as $Q_p=Q_{\ast}=10^6$ at face value. Under this assumption, the orbital circularization timescales are 2.13 Myr, 373 Myr, and 41.3 Myr for TOI-1608, TOI-2336, and TOI-2521, respectively. Compared to the stellar ages derived in Section \ref{stellar_properties}, the orbital circularization time-scales are much smaller, which means it is very likely that the orbits have been circularized. This is consistent with the measured low eccentricities of these three systems.

\subsubsection{Spin-orbit synchronization} \label{subsubsect:Spin-orbit_synchronization}

Dissipative tidal interactions alter both the orbital periods and the rotation periods of the host star and may lead to spin-orbit synchronization \citep{Goldreich1966_Tidal}. For TOI-1608 and TOI-2521, as discussed in Section \ref{subsect:stellar_age}, we have determined that they are old stars, but their fast rotations appear inconsistent with their age. However, since their rotation periods (see Table \ref{tab:stellar_param}) are very close to their companion orbit periods (see Table \ref{tab:companion_param}), the fast rotation might be explained by spin-orbit synchronization. To verify this scenario, we adopted the tidal model in \cite{Hut1981} and calculated the ratio of orbital and rotational angular momentum at the stable equilibrium configuration ($\alpha$) for these two systems. Here the equilibrium configuration means coplanarity (the spin axis of the two stars and the orbit are aligned), circularity (of the orbit), and spin-orbit synchronization of a binary system. In summary, \cite{Hut1981} suggested that such an equilibrium state is stable only when $\alpha>3$. When $\alpha<3$ and $\alpha-3 \ll 1$, the system will reach the circularity relatively quickly and take much more time to reach the spin-orbit synchronization. When $\alpha \sim 7$, the system will reach both circularity and synchronization roughly at the same time. When $\alpha \gg 7$, the system will reach synchronization much more quickly than it will reach circularity. Assuming the moment of inertia ${\rm I=0.07M_{\ast}R_{\ast}^2}$ as the sun\footnote{\url{https://nssdc.gsfc.nasa.gov/planetary/factsheet/sunfact.html}}, we obtained $\alpha=14$ for TOI-1608 and $\alpha=60$ for TOI-2521, which suggests that these systems can reach spin-orbit synchronization and that the spin-orbit synchronization will take less time than tidal circularization. In Section \ref{subsubsect:circularization}, we found these two systems are very likely to be circularized. Therefore, it is reasonable to suggest that these two systems have been spin-orbit synchronized, which led to the fast rotation of the host stars.

We also estimated the synchronization timescales for these two systems. Here we adopted the equations (2) and (3) in \cite{Albrecht2012}, which are based on the formulae from \cite{Zahn1977_tidal} and are calibrated with observations of binary stars:
\begin{equation} \label{sync_tCE}
    \frac{1}{\tau_{\rm CE}} = \frac{1}{\rm 10 \times 10^9 yr}q^2 \left(\frac{a/R_{\ast}}{40} \right)^{-6},
\end{equation}

\begin{equation} \label{sync_tRA}
    \frac{1}{\tau_{\rm RA}} = \frac{1}{\rm 0.25 \times 5 \times 10^9 yr}q^2 (1+q)^{5/6} \left(\frac{a/R_{\ast}}{6} \right)^{-17/2},
\end{equation}
where $\tau_{\rm CE}$ and $\tau_{\rm RA}$ are the synchronization timescales for cool stars with convective envelopes and hot stars with radiative envelopes, respectively, and q is the companion-to-star mass ratio. In \cite{Albrecht2012}, they adopted $T_{\rm eff}=6250$ K as the transition between stars with radiative and convective envelopes. As the effective temperatures of TOI-1608 and TOI-2521 are all smaller than 6250 K, we assumed convective envelopes and obtained $\tau_{\rm CE}=2.9$ Myr for TOI-1608 and $\tau_{\rm CE}=77$ Myr for TOI-2521. These timescales are much smaller than our estimated ages. Therefore, we conclude that TOI-1608 and TOI-2521 are very likely old stars, and their fast rotations result from spin-orbit synchronization.

For TOI-2336, the rotation period is $4.2\pm 0.5$ days, which is 7$\sigma$ smaller than the orbital period of 7.71 days. This significant difference makes spin-orbit synchronization unlikely. However, if the orbital period were the second harmonics of $8.4\pm 1.0$ days, synchronization would become a possibility. Nevertheless, considering that the $v\sin i$ derived from the spectra is faster than the velocity inferred by the rotation period of 4.2 days (as discussed in Section \ref{subsect_star_rotation}), the likelihood of an $\sim$8.4-day period is low. Therefore, it is unlikely for this system to have reached spin-orbit synchronization. In addition, we used Equation (\ref{sync_tRA}) to calculate its synchronization timescales and obtained $\tau_{\rm RA}=1.9\times 10^4$ Gyr, which is consistent with the notion that TOI-2336 has not yet reached synchronization.

\subsubsection{Orbital decay} \label{subsubsect:orbital_decay}

When the orbital period is shorter than the stellar rotation period, the semi-major axis of the companion's orbit will shrink and the companion will experience orbital tidal decay. According to the equation in \cite{Patra2017_Tidal_time1} derived from \cite{Goldreich1966_Tidal}, the tidal decay timescale is 
\begin{equation} \label{eq_tidal_decay}
    \tau_P = \frac{P}{|dP/dt|} = \frac{2Q_{\ast}}{27\pi} \left(\frac{M_{\ast}}{M_p} \right) \left(\frac{a}{R_{\ast}} \right)^5 P,
\end{equation}
where $P$ is the orbital period. For evolved host stars, the decay will become faster as the radius of the host star increases. For TOI-2336, since its orbital period is larger than the stellar rotation period, its orbit should not experience decay. For TOI-1608 and 2521, though we conclude in Section \ref{subsubsect:Spin-orbit_synchronization} that TOI-1608 and TOI-2521 have likely achieved spin-orbit synchronization, we still investigated their orbital decay timescales to gain insight into how rapidly their orbit would decay in case they had not achieved synchronization yet. Using equation \ref{eq_tidal_decay} and adopting $Q_{\ast}=10^6$, we obtained 3.05 Myr and 73.2 Myr for TOI-1608 and TOI-2521 respectively. The orbital decay timescale of TOI-1608b is similar to that of WASP-12b, whose orbital decay has been reported  \citep{Maciejewski2016_WASP1-2b_1, Patra2017_Tidal_time1, Yee2020_WASP-12b_3, Wong2022_WASP12}. Based on equation \ref{eq_tidal_decay}, we calculated the decay rate of the orbital period to be $\sim 70\ {\rm ms\ yr^{-1}}$, which only induces a difference of $\sim 4.6$ ms in our \tess\ observation. As the uncertainty on the orbital period in our analysis is 3.5 s, the orbital decay of TOI-1608b is not detectable in the \tess\ data gathered so far. 
Nevertheless, the orbital decay of WASP-12b was measured after being observed for a decade. Therefore, if TOI-1608 have not been spin-orbit synchronized, we might have a chance to observe their orbital decay in future observations (e.g., in \tess\ Extended Missions).  On the other hand, if TOI-1608 and 2521 have indeed reached tidal spin-orbit synchronization, magnetic braking could then cause further orbital decay with different time scales \citep[e.g.,][]{Barker2009_magnetic_braking1, Li2016_magnetic_braking2, Benbakoura2019_magnetic_braking3}.

\subsection{Eccentricity} \label{subsect:eccentricity}

\begin{figure}
\centering
\includegraphics[width=0.49\textwidth]{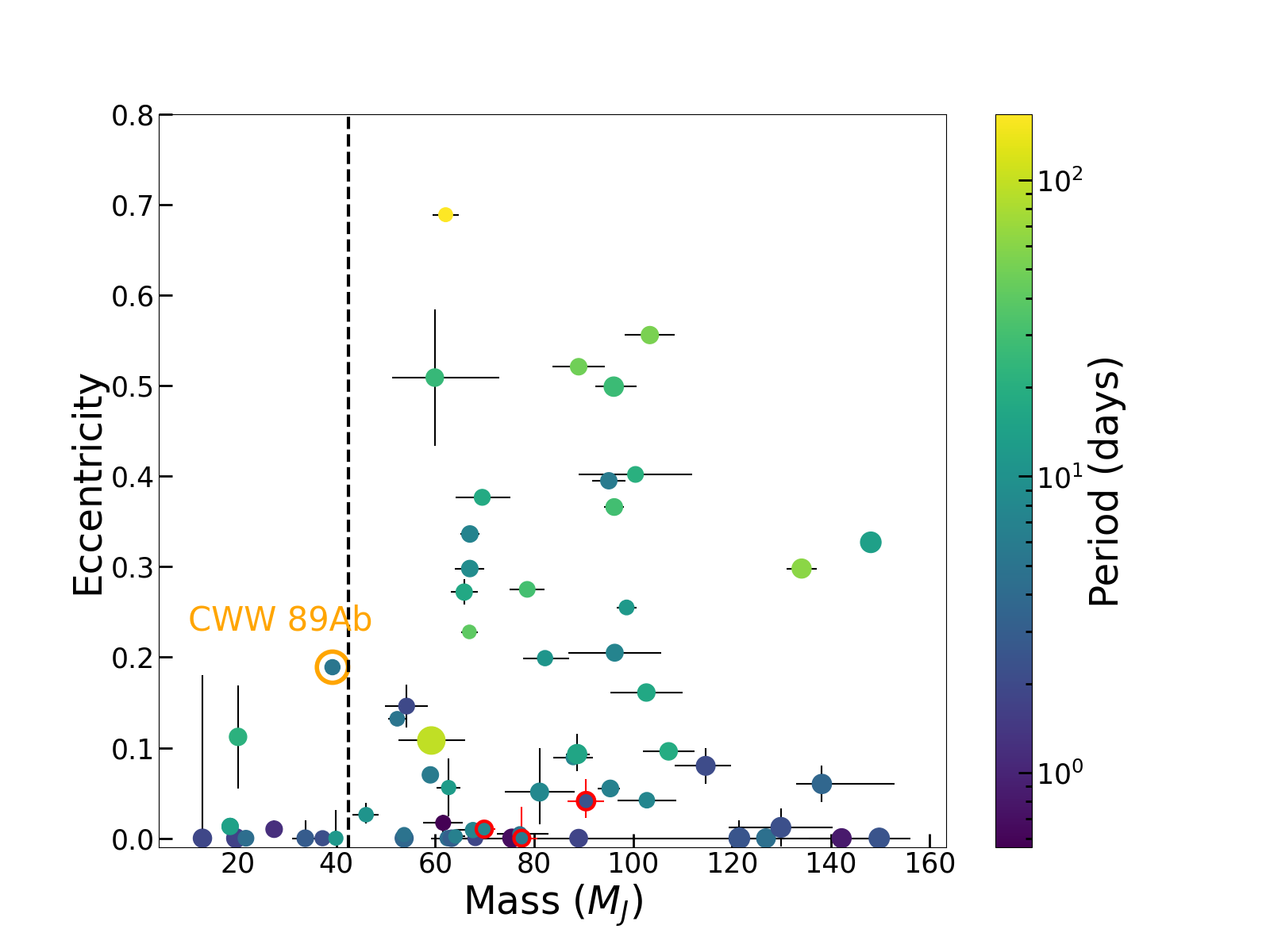}
\caption{The eccentricities of transiting companions with a mass between $13\ M_J$ and $150\ M_J$ as a function of mass. The dots with different colors are the `\sample' and our targets are circled in red. The dots are sized by the radii of the companions and the color shows their orbital periods. The dashed line is the mass threshold at $42.5\ M_J$ proposed by \citet{Ma2014_BDgap} separating two different brown dwarf populations. We denoted CWW 89Ab, a possible outlier in the low mass brown dwarf population, with the orange circle and presented a short discussion in Section \ref{subsect:eccentricity}. The sample and codes for generating this figure are available on \url{https://github.com/Ssealevel/Codes_TOI-1608_2336_2521}.}
\label{fig_BD_m_e}
\end{figure}

In section \ref{subsect_tidal}, we showed the low eccentricities of TOI-1608b, TOI-2336b and TOI-2521b are consistent with their relatively short tidal circularization timescales. Here we put these three system in a broader context by comparing their eccentricities with previous statistical results. In \cite{Ma2014_BDgap}, they analyzed brown dwarf companions and found different eccentricity distributions for brown dwarfs below and above $42.5\ M_J$. The eccentricity distribution of low mass brown dwarf companions is consistent with that of massive planets, while the high mass brown dwarf companions show more diversity in their eccentricity distribution, more consistent with binaries. We show the eccentricities of the `\sample' as well as our targets in Figure \ref{fig_BD_m_e}. More diverse eccentricities can be seen in the region above $42.5\ M_J$. All our targets are in the high mass region, but all have very low eccentricities in contrast to the majority of the high-mass brown dwarfs. We can see that companions with smaller periods tend to have lower eccentricities. This trend was also found by \cite{Ma2014_BDgap}, and they suggested that this is caused by tidal circularization. All of our targets have relatively small orbital periods and thus are consistent with their observed trend.

In addition, the possible outlier in the low mass region of Figure \ref{fig_BD_m_e} is CWW 89Ab \citep{Curtis2016_CWW89_1, Nowak2017_CWW89_2, Beatty2018_CWW89_3}, which is denoted by the orange circle. This system has an M dwarf companion on a wide orbit, which might help explain its architecture and provide insight into the formation and migration of brown dwarf companions.

As all our targets are companions orbiting aged stars, we also compared the eccentricities with the sample of hot Jupiters from \cite{Grunblatt2018}. They analyzed a sample of hot Jupiters around 419 dwarf stars and 136 giant stars with planetary radii $R_p > 0.4\ R_J$ and orbital periods between 4.5 and 30 days. They found that for the giant planets with short orbital periods, the ones orbiting evolved stars tend to have higher eccentricities (with a median eccentricity of $e\approx 0.152$) than the giant planets orbiting main-sequence dwarfs (with a median eccentricity of $e\approx 0.056$). They suggest that this may be caused by the increasing tidal effect of the evolved stars as their radii increase post-main-sequence: the planets with shorter tidal circularization timescales would also have relatively short tidal decay timescales and thus be engulfed. As a result, we would observe more planets with high eccentricities. All of our targets have eccentricities that are significantly lower than the median eccentricity of $e\approx 0.152$ of the evolved-star sample reported by \cite{Grunblatt2018}. This is not consistent with the trend in \cite{Grunblatt2018}. As we discussed in Section \ref{subsect_tidal}, orbital decay happens when the orbital period is shorter than the stellar rotation period and thus may not happen in our targets. Perhaps this mismatch shows the systematic difference between the post-main-sequence evolution of systems with brown dwarf or low-mass star companions and hot Jupiters. However, there is currently a small sample of brown dwarfs (or planets) around evolved stars, and more detections are needed to reveal the detailed statistical properties of this population in comparison with the planetary population.

\section{Summary} \label{summary}

In this paper, we confirmed one brown dwarf, TOI-2336b, and two objects near the hydrogen burning mass limit, TOI-1608b and TOI-2521b, transiting three aged stars. We analyzed space and ground photometry as well as RVs from ground-based high-resolution spectra. TOI-2336b has a radius of $1.05\pm 0.04\ R_J$, a mass of $69.9\pm 2.3\ M_J$ and an orbital period of 7.71 days. TOI-1608b has a radius of $1.21\pm 0.06\ R_J$, a mass of $90.7\pm 3.7\ M_J$ and an orbital period of 2.47 days. TOI-2521b has a radius of $1.01\pm 0.04\ R_J$, a mass of $77.5\pm 3.3\ M_J$ and an orbital period of 5.56 days. Their host stars are F (TOI-2336, TOI-1608) and G (TOI-2521) type subgiant stars. In addition, we detected the secondary eclipse signal in the light curve of TOI-1608, while no signal was detected for TOI-2336 and TOI-2521. The corresponding constraints on the effective temperatures are $2983\pm 90$ K for TOI-1608b, <3566 K for TOI-2336b, and <3143 K for TOI-2521b.

We found that all three companions have inflated radii via comparison to the evolution models for brown dwarfs and low mass stars. Combining data from the literature, we fitted a relationship between radius, mass, and incident flux for brown dwarfs and low mass stars, and we found that they are different between these two populations and also differ from the relationship of hot Jupiters. We found weak but statistically significant positive correlations between radius and flux for both brown dwarfs and low mass stars, which indicates that these companions may also be inflated by the irradiation from host stars as hot Jupiters, though to a much lesser degree.

We analyzed the tidal interactions in these three systems. We found that their relatively small eccentricities are consistent with their short orbital circularization timescales. We also compared their stellar rotation periods and orbital periods and found that TOI-1608 and TOI-2521 are very likely to have reached spin-orbit synchronization. This phenomenon adequately explains the unusually rapid rotation of their host stars, which appear to be at odds with gyrochronology given their evolutionary stages. We examined the angular momentum distribution of these two systems following \cite{Hut1981} and found that they can achieve and maintain stable spin-orbit synchronization. The estimated synchronization timescales are also significantly shorter than their ages, further supporting the synchronization scenario. Additionally, we found a short orbital decay timescale of 3.05\ Myr for TOI-1608, which may be detectable in future observations if it has not reached spin-orbit synchronization yet.

We compared the eccentricities of these three systems with previous statistical works. Compared with \cite{Ma2014_BDgap}, they do not follow the trend that high mass brown dwarf companions show more diversity in their eccentricity distribution than low mass brown dwarf. Instead, they are consistent with the trend that companions with smaller periods tend to have lower eccentricities. This is in contrast with the hot Jupiter sample in \cite{Grunblatt2018}, where short-period giant planets orbiting evolved stars tend to have larger eccentricities than the ones orbiting dwarfs. A larger sample of transiting brown dwarfs and low mass stars will reveal more on the similarities and differences between these populations and the gas giant planets and provide more insights about their formation and evolution.

\section*{Affiliations}
$^{1}$Department of Astronomy, Tsinghua University, Beijing 100084, China\\
$^{2}$Department of Physics and Kavli Institute for Astrophysics and Space Research, Massachusetts Institute of Technology, Cambridge, MA 02139, USA\\
$^{3}$Departamento de Matem\'atica y F\'isica Aplicadas, Facultad de Ingenier\'ia, Universidad Cat\'olica de la Sant\'isima Concepci\'on, Alonso de Rivera 2850, Concepci\'on, Chile\\
$^{4}$Centre for Astrophysics, University of Southern Queensland, West Street, Toowoomba, QLD 4350, Australia\\
$^{5}$Observatoire de Genève, Université de Genève, Chemin Pegasi, 51,1290 Versoix, Switzerland\\
$^{6}$Center for Astrophysics | Harvard \& Smithsonian, 60 Garden Street, Cambridge, MA 02138, USA\\
$^{7}$National Astronomical Observatories, Chinese Academy of Sciences, 20A Datun Road, Chaoyang District, Beijing 100012, China\\
$^{8}$Department of Physics and Astronomy, Vanderbilt University, Nashville, TN 37235, USA\\
$^{9}$Astrophysics Group, Keele University, Staffordshire, ST5 5BG, UK\\
$^{10}$NASA Ames Research Center, Moffett Field, CA 94035, USA\\
$^{11}$Department of Physics, Engineering and Astronomy, Stephen F. Austin State University, 1936 North St, Nacogdoches, TX 75962, USA\\
$^{12}$SETI Institute, Mountain View, CA 94043, USA\\
$^{13}$Jet Propulsion Laboratory, California Institute of Technology, Pasadena, CA 91109 USA\\
$^{14}$NASA Exoplanet Science Institute, IPAC, California Institute of Technology, Pasadena, CA 91125 USA\\
$^{15}$Center for Astrophysics | Harvard \& Smithsonian, 60 Garden Street, Cambridge, MA 02138, USA\\
$^{16}$Department of Astronomy, Columbia University, 550 West 120th Street, New York, NY, USA\\
$^{17}$Department of Astrophysics, American Museum of Natural History, Central Park West at 79th St., New York, NY 10024, USA\\
$^{18}$Department of Physics and Astronomy, Johns Hopkins University, 3400 N Charles St, Baltimore, MD 21218, USA\\
$^{19}$Department of Astronomy, The Ohio State University, 4055 McPherson Laboratory, 140 West 18th Avenue, Columbus, OH 43210 USA\\
$^{20}$Department of Physics and Astronomy, The University of North Carolina at Chapel Hill, Chapel Hill, NC 27599-3255, USA\\
$^{21}$Bay Area Environmental Research Institute, Moffett Field, CA 94035, USA\\
$^{22}$NASA Exoplanet Science Institute, Caltech/IPAC, Mail Code 100-22,1200 E. California Blvd., Pasadena, CA 91125, USA\\
$^{23}$Stellar Astrophysics Centre, Department of Physics and Astronomy, Aarhus University, Ny Munkegade 120, DK-8000 Aarhus C, Denmark\\
$^{24}$Proto-Logic LLC, 1718 Euclid Street NW, Washington, DC 20009, USA\\
$^{25}$Harvard-Smithsonian Center for Astrophysics, 60 Garden Street, Cambridge, MA 02138, USA\\
$^{26}$NASA Goddard Space Flight Center, 8800 Greenbelt Road, Greenbelt, MD 20771, USA\\
$^{27}$Department of Earth, Atmospheric, and Planetary Sciences, Massachusetts Institute of Technology, Cambridge, MA 02139, USA\\
$^{28}$Department of Aeronautics and Astronautics, Massachusetts Institute of Technology, Cambridge, MA 02139, USA\\
$^{29}$Kotizarovci Observatory, Sarsoni 90, 51216 Viskovo, Croatia\\
$^{30}$Planetary Discoveries in Fredericksburg, VA 22405, USA\\
$^{31}$Department of Astrophysical Sciences, Princeton University, Princeton, NJ 08544, USA\\
$^{32}$Max Planck Institute for Solar System Research, Justus-von-Liebig-Weg 3, D-37077 Gottingen, Germany\\

\section*{Acknowledgements}


We thank Doug Lin for his suggestions on this work, especially on the tidal effects. We also thank Theron W. Carmichael for his useful suggestions. We thank Johanna Teske for her comments.

T.G. and S.M. acknowledge the support from the National Science Foundation of China (Grant No. 12133005). We thank the Swiss National Science Foundation (SNSF) and the Geneva University for their continuous support to our planet search programs. This work has been carried out within the framework of the NCCR PlanetS supported by the Swiss National Science Foundation under grants 51NF40\_182901 and 51NF40\_205606. ML acknowledges support of the Swiss National Science Foundation under grant number PCEFP2\_194576.

C.A.C. acknowledges that this research was carried out at the Jet Propulsion Laboratory, California Institute of Technology, under a contract with the National Aeronautics and Space Administration (80NM0018D0004).

Funding for the TESS mission is provided by NASA's Science Mission Directorate. We acknowledge the use of public TESS data from pipelines at the TESS Science Office and at the TESS Science Processing Operations Center. This research has made use of the Exoplanet Follow-up Observation Program (ExoFOP; DOI: 10.26134/ExoFOP5) website, which is operated by the California Institute of Technology, under contract with the National Aeronautics and Space Administration under the Exoplanet Exploration Program. This paper includes data collected by the TESS mission that are publicly available from the Mikulski Archive for Space Telescopes (MAST). This research has made use of the NASA Exoplanet Archive, which is operated by the California Institute of Technology, under contract with the National Aeronautics and Space Administration under the Exoplanet Exploration Program. 

This work has made use of data from the European Space Agency (ESA) mission
{\it Gaia} (\url{https://www.cosmos.esa.int/gaia}), processed by the {\it Gaia}
Data Processing and Analysis Consortium (DPAC, \url{https://www.cosmos.esa.int/web/gaia/dpac/consortium}). Funding for the DPAC
has been provided by national institutions, in particular, the institutions
participating in the {\it Gaia} Multilateral Agreement. 

This work made use of Astropy:\footnote{http://www.astropy.org} a community-developed core Python package and an ecosystem of tools and resources for astronomy \citep{astropy:2013, astropy:2018, astropy:2022}. This research made use of Lightkurve, a Python package for Kepler and TESS data analysis \citep{Lightkurve2018}.

This work makes use of observations from the LCOGT network. Part of the LCOGT telescope time was granted by NOIRLab through the Mid-Scale Innovations Program (MSIP). MSIP is funded by NSF. Resources supporting this work were provided by the NASA High-End Computing (HEC) Program through the NASA Advanced Supercomputing (NAS) Division at Ames Research Center for the production of the SPOC data products.  

Some of the observations in this paper made use of the High-Resolution Imaging instrument Zorro and were obtained under Gemini LLP Proposal Number: GN/S-2021A-LP-105. Zorro was funded by the NASA Exoplanet Exploration Program and built at the NASA Ames Research Center by Steve B. Howell, Nic Scott, Elliott P. Horch, and Emmett Quigley. Zorro was mounted on the Gemini South telescope of the international Gemini Observatory, a program of NSF’s OIR Lab, which is managed by the Association of Universities for Research in Astronomy (AURA) under a cooperative agreement with the National Science Foundation. on behalf of the Gemini partnership: the National Science Foundation (United States), National Research Council (Canada), Agencia Nacional de Investigación y Desarrollo (Chile), Ministerio de Ciencia, Tecnología e Innovación (Argentina), Ministério da Ciência, Tecnologia, Inovações e Comunicações (Brazil), and Korea Astronomy and Space Science Institute (Republic of Korea).

\section*{Data Availability}
This paper includes photometric data collected by the \tess\ mission and ground instruments, which are publicly available in ExoFOP, at \url{https://exofop.ipac.caltech.edu/tess/target.php?id=138017750} for TOI-1608, \url{https://exofop.ipac.caltech.edu/tess/target.php?id=88902249} for TOI-2336, and \url{https://exofop.ipac.caltech.edu/tess/target.php?id=72556406} for TOI-2521. All spectroscopy data underlying this article are listed in the appendix. All of the high-resolution speckle imaging data is available at the NASA exoplanet Archive with no proprietary period. The \sample \ and the codes for generating figures in Section \ref{discussion} are available on \url{https://github.com/Ssealevel/Codes_TOI-1608_2336_2521}.



\bibliographystyle{mnras}
\bibliography{example} 

\begin{thebibliography}{}
\makeatletter
\relax
\def\mn@urlcharsother{\let\do\@makeother \do\$\do\&\do\#\do\^\do\_\do\%\do\~}
\def\mn@doi{\begingroup\mn@urlcharsother \@ifnextchar [ {\mn@doi@}
  {\mn@doi@[]}}
\def\mn@doi@[#1]#2{\def\@tempa{#1}\ifx\@tempa\@empty \href
  {http://dx.doi.org/#2} {doi:#2}\else \href {http://dx.doi.org/#2} {#1}\fi
  \endgroup}
\def\mn@eprint#1#2{\mn@eprint@#1:#2::\@nil}
\def\mn@eprint@arXiv#1{\href {http://arxiv.org/abs/#1} {{\tt arXiv:#1}}}
\def\mn@eprint@dblp#1{\href {http://dblp.uni-trier.de/rec/bibtex/#1.xml}
  {dblp:#1}}
\def\mn@eprint@#1:#2:#3:#4\@nil{\def\@tempa {#1}\def\@tempb {#2}\def\@tempc
  {#3}\ifx \@tempc \@empty \let \@tempc \@tempb \let \@tempb \@tempa \fi \ifx
  \@tempb \@empty \def\@tempb {arXiv}\fi \@ifundefined
  {mn@eprint@\@tempb}{\@tempb:\@tempc}{\expandafter \expandafter \csname
  mn@eprint@\@tempb\endcsname \expandafter{\@tempc}}}

\bibitem[\protect\citeauthoryear{{Acton} et~al.,}{{Acton}
  et~al.}{2021}]{Acton2021}
{Acton} J.~S.,  et~al., 2021, \mn@doi [\mnras] {10.1093/mnras/stab1459}, \href
  {https://ui.adsabs.harvard.edu/abs/2021MNRAS.505.2741A} {505, 2741}

\bibitem[\protect\citeauthoryear{{Albrecht} et~al.,}{{Albrecht}
  et~al.}{2012}]{Albrecht2012}
{Albrecht} S.,  et~al., 2012, \mn@doi [\apj] {10.1088/0004-637X/757/1/18},
  \href {https://ui.adsabs.harvard.edu/abs/2012ApJ...757...18A} {757, 18}

\bibitem[\protect\citeauthoryear{{Artigau} et~al.,}{{Artigau}
  et~al.}{2021}]{Artigau2021}
{Artigau} {\'E}.,  et~al., 2021, \mn@doi [\aj] {10.3847/1538-3881/ac096d},
  \href {https://ui.adsabs.harvard.edu/abs/2021AJ....162..144A} {162, 144}

\bibitem[\protect\citeauthoryear{{Astropy Collaboration} et~al.,}{{Astropy
  Collaboration} et~al.}{2013}]{astropy:2013}
{Astropy Collaboration} et~al., 2013, \mn@doi [\aap]
  {10.1051/0004-6361/201322068}, \href
  {http://adsabs.harvard.edu/abs/2013A%26A...558A..33A} {558, A33}

\bibitem[\protect\citeauthoryear{{Astropy Collaboration} et~al.,}{{Astropy
  Collaboration} et~al.}{2018}]{astropy:2018}
{Astropy Collaboration} et~al., 2018, \mn@doi [\aj] {10.3847/1538-3881/aabc4f},
  \href {https://ui.adsabs.harvard.edu/abs/2018AJ....156..123A} {156, 123}

\bibitem[\protect\citeauthoryear{{Astropy Collaboration} et~al.,}{{Astropy
  Collaboration} et~al.}{2022}]{astropy:2022}
{Astropy Collaboration} et~al., 2022, \mn@doi [apj] {10.3847/1538-4357/ac7c74},
  \href {https://ui.adsabs.harvard.edu/abs/2022ApJ...935..167A} {935, 167}

\bibitem[\protect\citeauthoryear{{Baraffe}, {Chabrier}, {Allard}  \&
  {Hauschildt}}{{Baraffe} et~al.}{2002}]{Baraffe2002}
{Baraffe} I.,  {Chabrier} G.,  {Allard} F.,   {Hauschildt} P.~H.,  2002,
  \mn@doi [\aap] {10.1051/0004-6361:20011638}, \href
  {https://ui.adsabs.harvard.edu/abs/2002A&A...382..563B} {382, 563}

\bibitem[\protect\citeauthoryear{{Baraffe}, {Chabrier}, {Barman}, {Allard}  \&
  {Hauschildt}}{{Baraffe} et~al.}{2003}]{Baraffe2003}
{Baraffe} I.,  {Chabrier} G.,  {Barman} T.~S.,  {Allard} F.,   {Hauschildt}
  P.~H.,  2003, \mn@doi [\aap] {10.1051/0004-6361:20030252}, \href
  {https://ui.adsabs.harvard.edu/abs/2003A&A...402..701B} {402, 701}

\bibitem[\protect\citeauthoryear{{Baraffe}, {Chabrier}  \& {Barman}}{{Baraffe}
  et~al.}{2008}]{Baraffe2008}
{Baraffe} I.,  {Chabrier} G.,   {Barman} T.,  2008, \mn@doi [\aap]
  {10.1051/0004-6361:20079321}, \href
  {https://ui.adsabs.harvard.edu/abs/2008A&A...482..315B} {482, 315}

\bibitem[\protect\citeauthoryear{{Baraffe}, {Homeier}, {Allard}  \&
  {Chabrier}}{{Baraffe} et~al.}{2015}]{Baraffe2015}
{Baraffe} I.,  {Homeier} D.,  {Allard} F.,   {Chabrier} G.,  2015, \mn@doi
  [\aap] {10.1051/0004-6361/201425481}, \href
  {http://adsabs.harvard.edu/abs/2015A%26A...577A..42B} {577, A42}

\bibitem[\protect\citeauthoryear{{Barker} \& {Ogilvie}}{{Barker} \&
  {Ogilvie}}{2009}]{Barker2009_magnetic_braking1}
{Barker} A.~J.,  {Ogilvie} G.~I.,  2009, \mn@doi [\mnras]
  {10.1111/j.1365-2966.2009.14694.x}, \href
  {https://ui.adsabs.harvard.edu/abs/2009MNRAS.395.2268B} {395, 2268}

\bibitem[\protect\citeauthoryear{{Beatty}, {Morley}, {Curtis}, {Burrows},
  {Davenport}  \& {Montet}}{{Beatty} et~al.}{2018}]{Beatty2018_CWW89_3}
{Beatty} T.~G.,  {Morley} C.~V.,  {Curtis} J.~L.,  {Burrows} A.,  {Davenport}
  J. R.~A.,   {Montet} B.~T.,  2018, \mn@doi [\aj] {10.3847/1538-3881/aad697},
  \href {https://ui.adsabs.harvard.edu/abs/2018AJ....156..168B} {156, 168}

\bibitem[\protect\citeauthoryear{{Benbakoura}, {R{\'e}ville}, {Brun}, {Le
  Poncin-Lafitte}  \& {Mathis}}{{Benbakoura}
  et~al.}{2019}]{Benbakoura2019_magnetic_braking3}
{Benbakoura} M.,  {R{\'e}ville} V.,  {Brun} A.~S.,  {Le Poncin-Lafitte} C.,
  {Mathis} S.,  2019, \mn@doi [\aap] {10.1051/0004-6361/201833314}, \href
  {https://ui.adsabs.harvard.edu/abs/2019A&A...621A.124B} {621, A124}

\bibitem[\protect\citeauthoryear{{Bouchy} et~al.,}{{Bouchy}
  et~al.}{2011}]{Bouchy2011_BD_inflate1}
{Bouchy} F.,  et~al., 2011, \mn@doi [\aap] {10.1051/0004-6361/201015276}, \href
  {https://ui.adsabs.harvard.edu/abs/2011A&A...525A..68B} {525, A68}

\bibitem[\protect\citeauthoryear{{Brahm}, {Jord{\'a}n}  \& {Espinoza}}{{Brahm}
  et~al.}{2017}]{Brahm2017}
{Brahm} R.,  {Jord{\'a}n} A.,   {Espinoza} N.,  2017, \mn@doi [\pasp]
  {10.1088/1538-3873/aa5455}, \href
  {https://ui.adsabs.harvard.edu/abs/2017PASP..129c4002B} {129, 034002}

\bibitem[\protect\citeauthoryear{{Brown}, {Gilliland}, {Noyes}  \&
  {Ramsey}}{{Brown} et~al.}{1991}]{brown1991}
{Brown} T.~M.,  {Gilliland} R.~L.,  {Noyes} R.~W.,   {Ramsey} L.~W.,  1991,
  \mn@doi [\apj] {10.1086/169725}, \href
  {https://ui.adsabs.harvard.edu/abs/1991ApJ...368..599B} {368, 599}

\bibitem[\protect\citeauthoryear{{Brown} et~al.,}{{Brown}
  et~al.}{2013}]{Brown2013}
{Brown} T.~M.,  et~al., 2013, \mn@doi [\pasp] {10.1086/673168}, \href
  {http://adsabs.harvard.edu/abs/2013PASP..125.1031B} {125, 1031}

\bibitem[\protect\citeauthoryear{{Buchhave} et~al.,}{{Buchhave}
  et~al.}{2010}]{buchhave2010}
{Buchhave} L.~A.,  et~al., 2010, \mn@doi [\apj] {10.1088/0004-637X/720/2/1118},
  \href {https://ui.adsabs.harvard.edu/abs/2010ApJ...720.1118B} {720, 1118}

\bibitem[\protect\citeauthoryear{{Buchhave} et~al.,}{{Buchhave}
  et~al.}{2012}]{Buchhave2012}
{Buchhave} L.~A.,  et~al., 2012, \mn@doi [\nat] {10.1038/nature11121}, \href
  {https://ui.adsabs.harvard.edu/abs/2012Natur.486..375B} {486, 375}

\bibitem[\protect\citeauthoryear{{Burrows} et~al.,}{{Burrows}
  et~al.}{1997}]{Burrows1997_BD_low_mass_range_2}
{Burrows} A.,  et~al., 1997, \mn@doi [\apj] {10.1086/305002}, \href
  {https://ui.adsabs.harvard.edu/abs/1997ApJ...491..856B} {491, 856}

\bibitem[\protect\citeauthoryear{{Ca{\~n}as} et~al.,}{{Ca{\~n}as}
  et~al.}{2022}]{Canas2022}
{Ca{\~n}as} C.~I.,  et~al., 2022, \mn@doi [\aj] {10.3847/1538-3881/ac415f},
  \href {https://ui.adsabs.harvard.edu/abs/2022AJ....163...89C} {163, 89}

\bibitem[\protect\citeauthoryear{{Caldwell} et~al.,}{{Caldwell}
  et~al.}{2020}]{Caldwell2020}
{Caldwell} D.~A.,  et~al., 2020, \mn@doi [Research Notes of the American
  Astronomical Society] {10.3847/2515-5172/abc9b3}, \href
  {https://ui.adsabs.harvard.edu/abs/2020RNAAS...4..201C} {4, 201}

\bibitem[\protect\citeauthoryear{{Campante} et~al.,}{{Campante}
  et~al.}{2016}]{Campante2016}
{Campante} T.~L.,  et~al., 2016, \mn@doi [\apj] {10.3847/0004-637X/830/2/138},
  \href {https://ui.adsabs.harvard.edu/abs/2016ApJ...830..138C} {830, 138}

\bibitem[\protect\citeauthoryear{{Carmichael} et~al.,}{{Carmichael}
  et~al.}{2021}]{Carmichael2021}
{Carmichael} T.~W.,  et~al., 2021, \mn@doi [\aj] {10.3847/1538-3881/abd4e1},
  \href {https://ui.adsabs.harvard.edu/abs/2021AJ....161...97C} {161, 97}

\bibitem[\protect\citeauthoryear{{Carmichael} et~al.,}{{Carmichael}
  et~al.}{2022}]{Carmichael2022}
{Carmichael} T.~W.,  et~al., 2022, \mn@doi [\mnras] {10.1093/mnras/stac1666},
  \href {https://ui.adsabs.harvard.edu/abs/2022MNRAS.514.4944C} {514, 4944}

\bibitem[\protect\citeauthoryear{{Castelli} \& {Hubrig}}{{Castelli} \&
  {Hubrig}}{2004}]{Castelli2004}
{Castelli} F.,  {Hubrig} S.,  2004, \mn@doi [\aap]
  {10.1051/0004-6361:20041011}, \href
  {https://ui.adsabs.harvard.edu/abs/2004A&A...425..263C} {425, 263}

\bibitem[\protect\citeauthoryear{{Chaplin} et~al.,}{{Chaplin}
  et~al.}{2011}]{Chaplin2011}
{Chaplin} W.~J.,  et~al., 2011, \mn@doi [\apjl] {10.1088/2041-8205/732/1/L5},
  \href {https://ui.adsabs.harvard.edu/abs/2011ApJ...732L...5C} {732, L5}

\bibitem[\protect\citeauthoryear{{Charbonneau} et~al.,}{{Charbonneau}
  et~al.}{2005}]{Charbonneau2005}
{Charbonneau} D.,  et~al., 2005, \mn@doi [\apj] {10.1086/429991}, \href
  {https://ui.adsabs.harvard.edu/abs/2005ApJ...626..523C} {626, 523}

\bibitem[\protect\citeauthoryear{{Chen} \& {Kipping}}{{Chen} \&
  {Kipping}}{2017}]{Chen2017_m_r_relation}
{Chen} J.,  {Kipping} D.,  2017, \mn@doi [\apj] {10.3847/1538-4357/834/1/17},
  \href {https://ui.adsabs.harvard.edu/abs/2017ApJ...834...17C} {834, 17}

\bibitem[\protect\citeauthoryear{{Ciardi}, {Beichman}, {Horch}  \&
  {Howell}}{{Ciardi} et~al.}{2015}]{Ciardi2015}
{Ciardi} D.~R.,  {Beichman} C.~A.,  {Horch} E.~P.,   {Howell} S.~B.,  2015,
  \mn@doi [\apj] {10.1088/0004-637X/805/1/16}, \href
  {https://ui.adsabs.harvard.edu/abs/2015ApJ...805...16C} {805, 16}

\bibitem[\protect\citeauthoryear{{Collier Cameron} et~al.,}{{Collier Cameron}
  et~al.}{2007}]{2007MNRAS.380.1230C}
{Collier Cameron} A.,  et~al., 2007, \mn@doi [\mnras]
  {10.1111/j.1365-2966.2007.12195.x}, \href
  {https://ui.adsabs.harvard.edu/abs/2007MNRAS.380.1230C} {380, 1230}

\bibitem[\protect\citeauthoryear{{Collins}, {Kielkopf}, {Stassun}  \&
  {Hessman}}{{Collins} et~al.}{2017}]{Collins2017}
{Collins} K.~A.,  {Kielkopf} J.~F.,  {Stassun} K.~G.,   {Hessman} F.~V.,  2017,
  \mn@doi [\aj] {10.3847/1538-3881/153/2/77}, \href
  {http://adsabs.harvard.edu/abs/2017AJ....153...77C} {153, 77}

\bibitem[\protect\citeauthoryear{{Curtis} et~al.,}{{Curtis}
  et~al.}{2016}]{Curtis2016_CWW89_1}
{Curtis} J.,  et~al., 2016, in 19th Cambridge Workshop on Cool Stars, Stellar
  Systems, and the Sun (CS19). Cambridge Workshop on Cool Stars, Stellar
  Systems, and the Sun.
p.~95, \mn@doi{10.5281/zenodo.58758}

\bibitem[\protect\citeauthoryear{{Curtis} et~al.,}{{Curtis}
  et~al.}{2020}]{Curtis2020_Gyrochronology1}
{Curtis} J.~L.,  et~al., 2020, \mn@doi [\apj] {10.3847/1538-4357/abbf58}, \href
  {https://ui.adsabs.harvard.edu/abs/2020ApJ...904..140C} {904, 140}

\bibitem[\protect\citeauthoryear{{Cutri} et~al.,}{{Cutri}
  et~al.}{2003}]{Cutri2003_2MASS_JKH1}
{Cutri} R.~M.,  et~al., 2003, {2MASS All Sky Catalog of point sources.}

\bibitem[\protect\citeauthoryear{{David}, {Hillenbrand}, {Gillen}, {Cody},
  {Howell}, {Isaacson}  \& {Livingston}}{{David}
  et~al.}{2019}]{David2019_RIK72b}
{David} T.~J.,  {Hillenbrand} L.~A.,  {Gillen} E.,  {Cody} A.~M.,  {Howell}
  S.~B.,  {Isaacson} H.~T.,   {Livingston} J.~H.,  2019, \mn@doi [\apj]
  {10.3847/1538-4357/aafe09}, \href
  {https://ui.adsabs.harvard.edu/abs/2019ApJ...872..161D} {872, 161}

\bibitem[\protect\citeauthoryear{{Donati} et~al.,}{{Donati}
  et~al.}{2008}]{Donati2008_tau_Boo}
{Donati} J.~F.,  et~al., 2008, \mn@doi [\mnras]
  {10.1111/j.1365-2966.2008.12946.x}, \href
  {https://ui.adsabs.harvard.edu/abs/2008MNRAS.385.1179D} {385, 1179}

\bibitem[\protect\citeauthoryear{{Dotter}}{{Dotter}}{2016}]{Dotter2016_MIST}
{Dotter} A.,  2016, \mn@doi [\apjs] {10.3847/0067-0049/222/1/8}, \href
  {https://ui.adsabs.harvard.edu/abs/2016ApJS..222....8D} {222, 8}

\bibitem[\protect\citeauthoryear{{Espinoza}}{{Espinoza}}{2018}]{Espinoza2018_bp_sample}
{Espinoza} N.,  2018, \mn@doi [Research Notes of the American Astronomical
  Society] {10.3847/2515-5172/aaef38}, \href
  {https://ui.adsabs.harvard.edu/abs/2018RNAAS...2..209E} {2, 209}

\bibitem[\protect\citeauthoryear{{Espinoza}, {Kossakowski}  \&
  {Brahm}}{{Espinoza} et~al.}{2019}]{Espinoza2019_Juliet}
{Espinoza} N.,  {Kossakowski} D.,   {Brahm} R.,  2019, \mn@doi [\mnras]
  {10.1093/mnras/stz2688}, \href
  {https://ui.adsabs.harvard.edu/abs/2019MNRAS.490.2262E} {490, 2262}

\bibitem[\protect\citeauthoryear{F\H{u}r\'esz}{F\H{u}r\'esz}{2008}]{tres}
F\H{u}r\'esz G.,  2008, PhD thesis, University of Szeged, Hungary

\bibitem[\protect\citeauthoryear{{Findeisen}, {Hillenbrand}  \&
  {Soderblom}}{{Findeisen} et~al.}{2011}]{Findeisen:2011}
{Findeisen} K.,  {Hillenbrand} L.,   {Soderblom} D.,  2011, \mn@doi [\aj]
  {10.1088/0004-6256/142/1/23}, \href
  {https://ui.adsabs.harvard.edu/abs/2011AJ....142...23F} {142, 23}

\bibitem[\protect\citeauthoryear{{Foreman-Mackey}, {Hogg}, {Lang}  \&
  {Goodman}}{{Foreman-Mackey} et~al.}{2013}]{Foreman-Mackey2013_emcee}
{Foreman-Mackey} D.,  {Hogg} D.~W.,  {Lang} D.,   {Goodman} J.,  2013, \mn@doi
  [\pasp] {10.1086/670067}, \href
  {https://ui.adsabs.harvard.edu/abs/2013PASP..125..306F} {125, 306}

\bibitem[\protect\citeauthoryear{{Foreman-Mackey}, {Agol}, {Angus}  \&
  {Ambikasaran}}{{Foreman-Mackey} et~al.}{2017}]{Foreman-Mackey2017_celerite}
{Foreman-Mackey} D.,  {Agol} E.,  {Angus} R.,   {Ambikasaran} S.,  2017,
  \mn@doi [AJ] {10.3847/1538-3881/aa9332}, 154, 220

\bibitem[\protect\citeauthoryear{{Fulton}, {Petigura}, {Blunt}  \&
  {Sinukoff}}{{Fulton} et~al.}{2018}]{Fulton2018_radvel}
{Fulton} B.~J.,  {Petigura} E.~A.,  {Blunt} S.,   {Sinukoff} E.,  2018, \mn@doi
  [\pasp] {10.1088/1538-3873/aaaaa8}, \href
  {http://adsabs.harvard.edu/abs/2018PASP..130d4504F} {130, 044504}

\bibitem[\protect\citeauthoryear{{Furlan} \& {Howell}}{{Furlan} \&
  {Howell}}{2017}]{FH1}
{Furlan} E.,  {Howell} S.~B.,  2017, \mn@doi [\aj] {10.3847/1538-3881/aa7b70},
  \href {https://ui.adsabs.harvard.edu/abs/2017AJ....154...66F} {154, 66}

\bibitem[\protect\citeauthoryear{{Furlan} \& {Howell}}{{Furlan} \&
  {Howell}}{2020}]{FH2}
{Furlan} E.,  {Howell} S.~B.,  2020, \mn@doi [\apj] {10.3847/1538-4357/ab9c9c},
  \href {https://ui.adsabs.harvard.edu/abs/2020ApJ...898...47F} {898, 47}

\bibitem[\protect\citeauthoryear{{Gaia Collaboration} et~al.,}{{Gaia
  Collaboration} et~al.}{2016}]{Gaia2016_Gaia_1}
{Gaia Collaboration} et~al., 2016, \mn@doi [\aap]
  {10.1051/0004-6361/201629272}, \href
  {https://ui.adsabs.harvard.edu/abs/2016A&A...595A...1G} {595, A1}

\bibitem[\protect\citeauthoryear{{Gaia Collaboration} et~al.,}{{Gaia
  Collaboration} et~al.}{2021}]{Gaia2021_DR3}
{Gaia Collaboration} et~al., 2021, \mn@doi [\aap]
  {10.1051/0004-6361/202039657}, \href
  {https://ui.adsabs.harvard.edu/abs/2021A&A...649A...1G} {649, A1}

\bibitem[\protect\citeauthoryear{{Gill} et~al.,}{{Gill}
  et~al.}{2022}]{Gill2022}
{Gill} S.,  et~al., 2022, \mn@doi [\mnras] {10.1093/mnras/stac798}, \href
  {https://ui.adsabs.harvard.edu/abs/2022MNRAS.513.1785G} {513, 1785}

\bibitem[\protect\citeauthoryear{{Giuricin}, {Mardirossian}  \&
  {Mezzetti}}{{Giuricin} et~al.}{1984}]{Giuricin1984}
{Giuricin} G.,  {Mardirossian} F.,   {Mezzetti} M.,  1984, \aap, \href
  {https://ui.adsabs.harvard.edu/abs/1984A&A...131..152G} {131, 152}

\bibitem[\protect\citeauthoryear{{Goldreich} \& {Soter}}{{Goldreich} \&
  {Soter}}{1966}]{Goldreich1966_Tidal}
{Goldreich} P.,  {Soter} S.,  1966, \mn@doi [\icarus]
  {10.1016/0019-1035(66)90051-0}, \href
  {https://ui.adsabs.harvard.edu/abs/1966Icar....5..375G} {5, 375}

\bibitem[\protect\citeauthoryear{{Grether} \& {Lineweaver}}{{Grether} \&
  {Lineweaver}}{2006}]{Grether2006}
{Grether} D.,  {Lineweaver} C.~H.,  2006, \mn@doi [\apj] {10.1086/500161},
  \href {https://ui.adsabs.harvard.edu/abs/2006ApJ...640.1051G} {640, 1051}

\bibitem[\protect\citeauthoryear{{Grieves} et~al.,}{{Grieves}
  et~al.}{2021}]{Grieves2021}
{Grieves} N.,  et~al., 2021, \mn@doi [\aap] {10.1051/0004-6361/202141145},
  \href {https://ui.adsabs.harvard.edu/abs/2021A&A...652A.127G} {652, A127}

\bibitem[\protect\citeauthoryear{{Grunblatt} et~al.,}{{Grunblatt}
  et~al.}{2016}]{Grunblatt2016_Jup_reinflate}
{Grunblatt} S.~K.,  et~al., 2016, \mn@doi [\aj] {10.3847/0004-6256/152/6/185},
  \href {https://ui.adsabs.harvard.edu/abs/2016AJ....152..185G} {152, 185}

\bibitem[\protect\citeauthoryear{{Grunblatt} et~al.,}{{Grunblatt}
  et~al.}{2018}]{Grunblatt2018}
{Grunblatt} S.~K.,  et~al., 2018, \mn@doi [\apjl] {10.3847/2041-8213/aacc67},
  \href {https://ui.adsabs.harvard.edu/abs/2018ApJ...861L...5G} {861, L5}

\bibitem[\protect\citeauthoryear{{Grunblatt} et~al.,}{{Grunblatt}
  et~al.}{2022}]{Grunblatt2022}
{Grunblatt} S.~K.,  et~al., 2022, \mn@doi [\aj] {10.3847/1538-3881/ac4972},
  \href {https://ui.adsabs.harvard.edu/abs/2022AJ....163..120G} {163, 120}

\bibitem[\protect\citeauthoryear{{Henry} et~al.,}{{Henry}
  et~al.}{2018}]{Henry2018}
{Henry} T.~J.,  et~al., 2018, \mn@doi [\aj] {10.3847/1538-3881/aac262}, \href
  {https://ui.adsabs.harvard.edu/abs/2018AJ....155..265H} {155, 265}

\bibitem[\protect\citeauthoryear{{Howell} \& {Furlan}}{{Howell} \&
  {Furlan}}{2022}]{HF}
{Howell} S.~B.,  {Furlan} E.,  2022, \mn@doi [Frontiers in Astronomy and Space
  Sciences] {10.3389/fspas.2022.871163}, \href
  {https://ui.adsabs.harvard.edu/abs/2022FrASS...9.1163H} {9, 871163}

\bibitem[\protect\citeauthoryear{{Howell}, {Everett}, {Sherry}, {Horch}  \&
  {Ciardi}}{{Howell} et~al.}{2011}]{Howell2011}
{Howell} S.~B.,  {Everett} M.~E.,  {Sherry} W.,  {Horch} E.,   {Ciardi} D.~R.,
  2011, \mn@doi [\aj] {10.1088/0004-6256/142/1/19}, \href
  {https://ui.adsabs.harvard.edu/abs/2011AJ....142...19H} {142, 19}

\bibitem[\protect\citeauthoryear{{Howell}, {Everett}, {Horch}, {Winters},
  {Hirsch}, {Nusdeo}  \& {Scott}}{{Howell} et~al.}{2016}]{Howell2016}
{Howell} S.~B.,  {Everett} M.~E.,  {Horch} E.~P.,  {Winters} J.~G.,  {Hirsch}
  L.,  {Nusdeo} D.,   {Scott} N.~J.,  2016, \mn@doi [\apjl]
  {10.3847/2041-8205/829/1/L2}, \href
  {https://ui.adsabs.harvard.edu/abs/2016ApJ...829L...2H} {829, L2}

\bibitem[\protect\citeauthoryear{{Howell}, {Scott}, {Matson}, {Everett},
  {Furlan}, {Gnilka}, {Ciardi}  \& {Lester}}{{Howell}
  et~al.}{2021}]{Howell2021}
{Howell} S.~B.,  {Scott} N.~J.,  {Matson} R.~A.,  {Everett} M.~E.,  {Furlan}
  E.,  {Gnilka} C.~L.,  {Ciardi} D.~R.,   {Lester} K.~V.,  2021, \mn@doi
  [Frontiers in Astronomy and Space Sciences] {10.3389/fspas.2021.635864},
  \href {https://ui.adsabs.harvard.edu/abs/2021FrASS...8...10H} {8, 10}

\bibitem[\protect\citeauthoryear{{Hut}}{{Hut}}{1980}]{Hut1980}
{Hut} P.,  1980, \aap, \href
  {https://ui.adsabs.harvard.edu/abs/1980A&A....92..167H} {92, 167}

\bibitem[\protect\citeauthoryear{{Hut}}{{Hut}}{1981}]{Hut1981}
{Hut} P.,  1981, \aap, \href
  {https://ui.adsabs.harvard.edu/abs/1981A&A....99..126H} {99, 126}

\bibitem[\protect\citeauthoryear{{Jackson}, {Greenberg}  \& {Barnes}}{{Jackson}
  et~al.}{2008}]{Jackson2008_Tidal}
{Jackson} B.,  {Greenberg} R.,   {Barnes} R.,  2008, \mn@doi [\apj]
  {10.1086/529187}, \href
  {https://ui.adsabs.harvard.edu/abs/2008ApJ...678.1396J} {678, 1396}

\bibitem[\protect\citeauthoryear{{Jenkins} et~al.,}{{Jenkins}
  et~al.}{2016}]{Jenkins2016_SPOC}
{Jenkins} J.~M.,  et~al., 2016, in {Chiozzi} G.,  {Guzman} J.~C.,  eds,
  Society of Photo-Optical Instrumentation Engineers (SPIE) Conference Series
  Vol. 9913, Software and Cyberinfrastructure for Astronomy IV. p. 99133E,
  \mn@doi{10.1117/12.2233418}

\bibitem[\protect\citeauthoryear{{Jensen}}{{Jensen}}{2013}]{Jensen2013}
{Jensen} E.,  2013, {Tapir: A web interface for transit/eclipse observability}
  (\mn@eprint {ascl} {1306.007})

\bibitem[\protect\citeauthoryear{{Kipping}}{{Kipping}}{2013}]{Kipping2013_limb-darkening}
{Kipping} D.~M.,  2013, \mn@doi [\mnras] {10.1093/mnras/stt1435}, \href
  {https://ui.adsabs.harvard.edu/abs/2013MNRAS.435.2152K} {435, 2152}

\bibitem[\protect\citeauthoryear{{Kjeldsen} \& {Bedding}}{{Kjeldsen} \&
  {Bedding}}{1995}]{kb1995}
{Kjeldsen} H.,  {Bedding} T.~R.,  1995, \aap, \href
  {https://ui.adsabs.harvard.edu/abs/1995A&A...293...87K} {293, 87}

\bibitem[\protect\citeauthoryear{{Kreidberg}}{{Kreidberg}}{2015}]{Kreidberg2015_batman}
{Kreidberg} L.,  2015, \mn@doi [Publications of the Astronomical Society of the
  Pacific] {10.1086/683602}, \href
  {https://ui.adsabs.harvard.edu/\#abs/2015PASP..127.1161K} {127, 1161}

\bibitem[\protect\citeauthoryear{Kurucz}{Kurucz}{1992}]{kurucz1992}
Kurucz R.~L.,  1992, in , The Stellar Populations of Galaxies.
Springer, pp 225--232

\bibitem[\protect\citeauthoryear{{Lester} et~al.,}{{Lester}
  et~al.}{2021}]{Lester}
{Lester} K.~V.,  et~al., 2021, \mn@doi [\aj] {10.3847/1538-3881/ac0d06}, \href
  {https://ui.adsabs.harvard.edu/abs/2021AJ....162...75L} {162, 75}

\bibitem[\protect\citeauthoryear{{Levato}}{{Levato}}{1974}]{Levato1974}
{Levato} H.,  1974, \aap, \href
  {https://ui.adsabs.harvard.edu/abs/1974A&A....35..259L} {35, 259}

\bibitem[\protect\citeauthoryear{{Li} \& {Winn}}{{Li} \&
  {Winn}}{2016}]{Li2016_magnetic_braking2}
{Li} G.,  {Winn} J.~N.,  2016, \mn@doi [\apj] {10.3847/0004-637X/818/1/5},
  \href {https://ui.adsabs.harvard.edu/abs/2016ApJ...818....5L} {818, 5}

\bibitem[\protect\citeauthoryear{{Li}, {Tenenbaum}, {Twicken}, {Burke},
  {Jenkins}, {Quintana}, {Rowe}  \& {Seader}}{{Li}
  et~al.}{2019}]{Li:DVmodelFit2019}
{Li} J.,  {Tenenbaum} P.,  {Twicken} J.~D.,  {Burke} C.~J.,  {Jenkins} J.~M.,
  {Quintana} E.~V.,  {Rowe} J.~F.,   {Seader} S.~E.,  2019, \mn@doi [\pasp]
  {10.1088/1538-3873/aaf44d}, \href
  {https://ui.adsabs.harvard.edu/abs/2019PASP..131b4506L} {131, 024506}

\bibitem[\protect\citeauthoryear{{Lightkurve Collaboration}
  et~al.,}{{Lightkurve Collaboration} et~al.}{2018}]{Lightkurve2018}
{Lightkurve Collaboration} et~al., 2018, {Lightkurve: Kepler and TESS time
  series analysis in Python}, Astrophysics Source Code Library (\mn@eprint
  {ascl} {1812.013})

\bibitem[\protect\citeauthoryear{{Lund} et~al.,}{{Lund}
  et~al.}{2016}]{lund2016}
{Lund} M.~N.,  et~al., 2016, \mn@doi [\pasp]
  {10.1088/1538-3873/128/970/124204}, \href
  {https://ui.adsabs.harvard.edu/abs/2016PASP..128l4204L} {128, 124204}

\bibitem[\protect\citeauthoryear{{Ma} \& {Ge}}{{Ma} \&
  {Ge}}{2014}]{Ma2014_BDgap}
{Ma} B.,  {Ge} J.,  2014, \mn@doi [\mnras] {10.1093/mnras/stu134}, \href
  {https://ui.adsabs.harvard.edu/abs/2014MNRAS.439.2781M} {439, 2781}

\bibitem[\protect\citeauthoryear{{Maciejewski} et~al.,}{{Maciejewski}
  et~al.}{2016}]{Maciejewski2016_WASP1-2b_1}
{Maciejewski} G.,  et~al., 2016, \mn@doi [\aap] {10.1051/0004-6361/201628312},
  \href {https://ui.adsabs.harvard.edu/abs/2016A&A...588L...6M} {588, L6}

\bibitem[\protect\citeauthoryear{{Marcy} \& {Butler}}{{Marcy} \&
  {Butler}}{2000}]{Marcy2000}
{Marcy} G.~W.,  {Butler} R.~P.,  2000, \mn@doi [\pasp] {10.1086/316516}, \href
  {https://ui.adsabs.harvard.edu/abs/2000PASP..112..137M} {112, 137}

\bibitem[\protect\citeauthoryear{{Marley} et~al.,}{{Marley}
  et~al.}{2021}]{Marley2021}
{Marley} M.~S.,  et~al., 2021, \mn@doi [\apj] {10.3847/1538-4357/ac141d}, \href
  {https://ui.adsabs.harvard.edu/abs/2021ApJ...920...85M} {920, 85}

\bibitem[\protect\citeauthoryear{{Masuda} \& {Winn}}{{Masuda} \&
  {Winn}}{2020}]{Masuda2020}
{Masuda} K.,  {Winn} J.~N.,  2020, \mn@doi [\aj] {10.3847/1538-3881/ab65be},
  \href {https://ui.adsabs.harvard.edu/abs/2020AJ....159...81M} {159, 81}

\bibitem[\protect\citeauthoryear{{Matson}, {Howell}, {Horch}  \&
  {Everett}}{{Matson} et~al.}{2018}]{Matson2018}
{Matson} R.~A.,  {Howell} S.~B.,  {Horch} E.~P.,   {Everett} M.~E.,  2018,
  \mn@doi [\aj] {10.3847/1538-3881/aac778}, \href
  {https://ui.adsabs.harvard.edu/abs/2018AJ....156...31M} {156, 31}

\bibitem[\protect\citeauthoryear{{Matson}, {Howell}  \& {Ciardi}}{{Matson}
  et~al.}{2019}]{Matson}
{Matson} R.~A.,  {Howell} S.~B.,   {Ciardi} D.~R.,  2019, \mn@doi [\aj]
  {10.3847/1538-3881/ab1755}, \href
  {https://ui.adsabs.harvard.edu/abs/2019AJ....157..211M} {157, 211}

\bibitem[\protect\citeauthoryear{{Maxted} et~al.,}{{Maxted}
  et~al.}{2011}]{2011PASP..123..547M}
{Maxted} P.~F.~L.,  et~al., 2011, \mn@doi [\pasp] {10.1086/660007}, \href
  {https://ui.adsabs.harvard.edu/abs/2011PASP.123..547M} {123, 547}

\bibitem[\protect\citeauthoryear{{McLaughlin}}{{McLaughlin}}{1924}]{McLaughlin1924_RMeffect2}
{McLaughlin} D.~B.,  1924, \mn@doi [\apj] {10.1086/142826}, \href
  {https://ui.adsabs.harvard.edu/abs/1924ApJ....60...22M} {60, 22}

\bibitem[\protect\citeauthoryear{{Morton}}{{Morton}}{2015}]{Morton2015_isochrones}
{Morton} T.~D.,  2015, {isochrones: Stellar model grid package}, Astrophysics
  Source Code Library, record ascl:1503.010 (\mn@eprint {ascl} {1503.010})

\bibitem[\protect\citeauthoryear{{Nakajima}, {Oppenheimer}, {Kulkarni},
  {Golimowski}, {Matthews}  \& {Durrance}}{{Nakajima}
  et~al.}{1995}]{Nakajima1995_Gl229b1}
{Nakajima} T.,  {Oppenheimer} B.~R.,  {Kulkarni} S.~R.,  {Golimowski} D.~A.,
  {Matthews} K.,   {Durrance} S.~T.,  1995, \mn@doi [\nat] {10.1038/378463a0},
  \href {https://ui.adsabs.harvard.edu/abs/1995Natur.378..463N} {378, 463}

\bibitem[\protect\citeauthoryear{{Nowak} et~al.,}{{Nowak}
  et~al.}{2017}]{Nowak2017_CWW89_2}
{Nowak} G.,  et~al., 2017, \mn@doi [\aj] {10.3847/1538-3881/aa5cb6}, \href
  {https://ui.adsabs.harvard.edu/abs/2017AJ....153..131N} {153, 131}

\bibitem[\protect\citeauthoryear{{Oppenheimer}, {Kulkarni}, {Matthews}  \&
  {Nakajima}}{{Oppenheimer} et~al.}{1995}]{Oppenheimer1995_Gl229b2}
{Oppenheimer} B.~R.,  {Kulkarni} S.~R.,  {Matthews} K.,   {Nakajima} T.,  1995,
  \mn@doi [Science] {10.1126/science.270.5241.1478}, \href
  {https://ui.adsabs.harvard.edu/abs/1995Sci...270.1478O} {270, 1478}

\bibitem[\protect\citeauthoryear{{Palle} et~al.,}{{Palle}
  et~al.}{2021}]{Palle2021}
{Palle} E.,  et~al., 2021, \mn@doi [\aap] {10.1051/0004-6361/202039937}, \href
  {https://ui.adsabs.harvard.edu/abs/2021A&A...650A..55P} {650, A55}

\bibitem[\protect\citeauthoryear{{Paredes}, {Henry}, {Quinn}, {Gies},
  {Hinojosa-Go{\~n}i}, {James}, {Jao}  \& {White}}{{Paredes}
  et~al.}{2021}]{Paredes2021}
{Paredes} L.~A.,  {Henry} T.~J.,  {Quinn} S.~N.,  {Gies} D.~R.,
  {Hinojosa-Go{\~n}i} R.,  {James} H.-S.,  {Jao} W.-C.,   {White} R.~J.,  2021,
  \mn@doi [\aj] {10.3847/1538-3881/ac082a}, \href
  {https://ui.adsabs.harvard.edu/abs/2021AJ....162..176P} {162, 176}

\bibitem[\protect\citeauthoryear{{Patra}, {Winn}, {Holman}, {Yu}, {Deming}  \&
  {Dai}}{{Patra} et~al.}{2017}]{Patra2017_Tidal_time1}
{Patra} K.~C.,  {Winn} J.~N.,  {Holman} M.~J.,  {Yu} L.,  {Deming} D.,   {Dai}
  F.,  2017, \mn@doi [\aj] {10.3847/1538-3881/aa6d75}, \href
  {https://ui.adsabs.harvard.edu/abs/2017AJ....154....4P} {154, 4}

\bibitem[\protect\citeauthoryear{{Paunzen}}{{Paunzen}}{2015}]{Paunzen:2015}
{Paunzen} E.,  2015, \mn@doi [\aap] {10.1051/0004-6361/201526413}, \href
  {https://ui.adsabs.harvard.edu/abs/2015A&A...580A..23P} {580, A23}

\bibitem[\protect\citeauthoryear{{Pecaut} \& {Mamajek}}{{Pecaut} \&
  {Mamajek}}{2013}]{Pecaut2013_Mamajek_Table2}
{Pecaut} M.~J.,  {Mamajek} E.~E.,  2013, \mn@doi [\apjs]
  {10.1088/0067-0049/208/1/9}, \href
  {https://ui.adsabs.harvard.edu/abs/2013ApJS..208....9P} {208, 9}

\bibitem[\protect\citeauthoryear{{Pecaut}, {Mamajek}  \& {Bubar}}{{Pecaut}
  et~al.}{2012}]{Pecaut2012_Mamajek_Table1}
{Pecaut} M.~J.,  {Mamajek} E.~E.,   {Bubar} E.~J.,  2012, \mn@doi [\apj]
  {10.1088/0004-637X/746/2/154}, \href
  {https://ui.adsabs.harvard.edu/abs/2012ApJ...746..154P} {746, 154}

\bibitem[\protect\citeauthoryear{{Pepe} et~al.,}{{Pepe}
  et~al.}{2002}]{Pepe2002}
{Pepe} F.,  et~al., 2002, The Messenger, \href
  {https://ui.adsabs.harvard.edu/abs/2002Msngr.110....9P} {110, 9}

\bibitem[\protect\citeauthoryear{{Persson} et~al.,}{{Persson}
  et~al.}{2019}]{Persson2019}
{Persson} C.~M.,  et~al., 2019, \mn@doi [\aap] {10.1051/0004-6361/201935505},
  \href {https://ui.adsabs.harvard.edu/abs/2019A&A...628A..64P} {628, A64}

\bibitem[\protect\citeauthoryear{{Phillips} et~al.,}{{Phillips}
  et~al.}{2020}]{Phillips2020_ATMO2020}
{Phillips} M.~W.,  et~al., 2020, \mn@doi [\aap] {10.1051/0004-6361/201937381},
  \href {https://ui.adsabs.harvard.edu/abs/2020A&A...637A..38P} {637, A38}

\bibitem[\protect\citeauthoryear{{Pollacco} et~al.,}{{Pollacco}
  et~al.}{2006}]{2006PASP..118.1407P}
{Pollacco} D.~L.,  et~al., 2006, \mn@doi [\pasp] {10.1086/508556}, \href
  {http://adsabs.harvard.edu/abs/2006PASP..118.1407P} {118, 1407}

\bibitem[\protect\citeauthoryear{{Psaridi} et~al.,}{{Psaridi}
  et~al.}{2022}]{Psaridi2022}
{Psaridi} A.,  et~al., 2022, \mn@doi [\aap] {10.1051/0004-6361/202243454},
  \href {https://ui.adsabs.harvard.edu/abs/2022A&A...664A..94P} {664, A94}

\bibitem[\protect\citeauthoryear{{Queloz} et~al.,}{{Queloz}
  et~al.}{2001}]{Queloz2001}
{Queloz} D.,  et~al., 2001, The Messenger, \href
  {https://ui.adsabs.harvard.edu/abs/2001Msngr.105....1Q} {105, 1}

\bibitem[\protect\citeauthoryear{{Rebolo}, {Zapatero Osorio}  \&
  {Mart{\'\i}n}}{{Rebolo} et~al.}{1995}]{Rebolo1995_Teide1_1}
{Rebolo} R.,  {Zapatero Osorio} M.~R.,   {Mart{\'\i}n} E.~L.,  1995, \mn@doi
  [\nat] {10.1038/377129a0}, \href
  {https://ui.adsabs.harvard.edu/abs/1995Natur.377..129R} {377, 129}

\bibitem[\protect\citeauthoryear{{Rebolo}, {Martin}, {Basri}, {Marcy}  \&
  {Zapatero-Osorio}}{{Rebolo} et~al.}{1996}]{Rebolo1996_Teide1_2}
{Rebolo} R.,  {Martin} E.~L.,  {Basri} G.,  {Marcy} G.~W.,   {Zapatero-Osorio}
  M.~R.,  1996, \mn@doi [\apjl] {10.1086/310263}, \href
  {https://ui.adsabs.harvard.edu/abs/1996ApJ...469L..53R} {469, L53}

\bibitem[\protect\citeauthoryear{{Ricker} et~al.,}{{Ricker}
  et~al.}{2015}]{Ricker2015}
{Ricker} G.~R.,  et~al., 2015, \mn@doi [Journal of Astronomical Telescopes,
  Instruments, and Systems] {10.1117/1.JATIS.1.1.014003}, \href
  {https://ui.adsabs.harvard.edu/abs/2015JATIS...1a4003R} {1, 014003}

\bibitem[\protect\citeauthoryear{{Rossiter}}{{Rossiter}}{1924}]{Rossiter1924_RMeffect1}
{Rossiter} R.~A.,  1924, \mn@doi [\apj] {10.1086/142825}, \href
  {https://ui.adsabs.harvard.edu/abs/1924ApJ....60...15R} {60, 15}

\bibitem[\protect\citeauthoryear{{Saunders} et~al.,}{{Saunders}
  et~al.}{2022}]{Saunders2022_Jup_inflate}
{Saunders} N.,  et~al., 2022, \mn@doi [\aj] {10.3847/1538-3881/ac38a1}, \href
  {https://ui.adsabs.harvard.edu/abs/2022AJ....163...53S} {163, 53}

\bibitem[\protect\citeauthoryear{{Schlegel}, {Finkbeiner}  \&
  {Davis}}{{Schlegel} et~al.}{1998}]{Schlegel:1998}
{Schlegel} D.~J.,  {Finkbeiner} D.~P.,   {Davis} M.,  1998, \mn@doi [\apj]
  {10.1086/305772}, \href
  {https://ui.adsabs.harvard.edu/abs/1998ApJ...500..525S} {500, 525}

\bibitem[\protect\citeauthoryear{{Scott} et~al.,}{{Scott} et~al.}{2021}]{SH}
{Scott} N.~J.,  et~al., 2021, \mn@doi [Frontiers in Astronomy and Space
  Sciences] {10.3389/fspas.2021.716560}, \href
  {https://ui.adsabs.harvard.edu/abs/2021FrASS...8..138S} {8, 138}

\bibitem[\protect\citeauthoryear{{Siverd} et~al.,}{{Siverd}
  et~al.}{2018}]{Siverd2018}
{Siverd} R.~J.,  et~al., 2018, in {Evans} C.~J.,  {Simard} L.,   {Takami} H.,
  eds,  Society of Photo-Optical Instrumentation Engineers (SPIE) Conference
  Series Vol. 10702, Ground-based and Airborne Instrumentation for Astronomy
  VII. p. 107026C, \mn@doi{10.1117/12.2312800}

\bibitem[\protect\citeauthoryear{{Smith} et~al.,}{{Smith}
  et~al.}{2012}]{Smith2012_PDC2}
{Smith} J.~C.,  et~al., 2012, \mn@doi [\pasp] {10.1086/667697}, \href
  {https://ui.adsabs.harvard.edu/abs/2012PASP..124.1000S} {124, 1000}

\bibitem[\protect\citeauthoryear{{Spada} \& {Lanzafame}}{{Spada} \&
  {Lanzafame}}{2020}]{Spada2020_Gyrochronology2}
{Spada} F.,  {Lanzafame} A.~C.,  2020, \mn@doi [\aap]
  {10.1051/0004-6361/201936384}, \href
  {https://ui.adsabs.harvard.edu/abs/2020A&A...636A..76S} {636, A76}

\bibitem[\protect\citeauthoryear{{Speagle}}{{Speagle}}{2020}]{Speagle2020_DYNESTY}
{Speagle} J.~S.,  2020, \mn@doi [\mnras] {10.1093/mnras/staa278}, \href
  {https://ui.adsabs.harvard.edu/abs/2020MNRAS.493.3132S} {493, 3132}

\bibitem[\protect\citeauthoryear{{Spiegel}, {Burrows}  \& {Milsom}}{{Spiegel}
  et~al.}{2011}]{Spiegel2011}
{Spiegel} D.~S.,  {Burrows} A.,   {Milsom} J.~A.,  2011, \mn@doi [\apj]
  {10.1088/0004-637X/727/1/57}, \href
  {https://ui.adsabs.harvard.edu/abs/2011ApJ...727...57S} {727, 57}

\bibitem[\protect\citeauthoryear{{Stassun} \& {Torres}}{{Stassun} \&
  {Torres}}{2016}]{Stassun:2016}
{Stassun} K.~G.,  {Torres} G.,  2016, \mn@doi [\aj]
  {10.3847/0004-6256/152/6/180}, \href
  {https://ui.adsabs.harvard.edu/abs/2016AJ....152..180S} {152, 180}

\bibitem[\protect\citeauthoryear{{Stassun} \& {Torres}}{{Stassun} \&
  {Torres}}{2021}]{StassunTorres:2021}
{Stassun} K.~G.,  {Torres} G.,  2021, \mn@doi [\apjl]
  {10.3847/2041-8213/abdaad}, \href
  {https://ui.adsabs.harvard.edu/abs/2021ApJ...907L..33S} {907, L33}

\bibitem[\protect\citeauthoryear{{Stassun}, {Collins}  \& {Gaudi}}{{Stassun}
  et~al.}{2017}]{Stassun:2017}
{Stassun} K.~G.,  {Collins} K.~A.,   {Gaudi} B.~S.,  2017, \mn@doi [\aj]
  {10.3847/1538-3881/aa5df3}, \href
  {https://ui.adsabs.harvard.edu/abs/2017AJ....153..136S} {153, 136}

\bibitem[\protect\citeauthoryear{{Stassun}, {Corsaro}, {Pepper}  \&
  {Gaudi}}{{Stassun} et~al.}{2018a}]{Stassun:2018}
{Stassun} K.~G.,  {Corsaro} E.,  {Pepper} J.~A.,   {Gaudi} B.~S.,  2018a,
  \mn@doi [\aj] {10.3847/1538-3881/aa998a}, \href
  {https://ui.adsabs.harvard.edu/abs/2018AJ....155...22S} {155, 22}

\bibitem[\protect\citeauthoryear{{Stassun} et~al.,}{{Stassun}
  et~al.}{2018b}]{Stassun2017_TIC}
{Stassun} K.~G.,  et~al., 2018b, \mn@doi [\aj] {10.3847/1538-3881/aad050},
  \href {https://ui.adsabs.harvard.edu/abs/2018AJ....156..102S} {156, 102}

\bibitem[\protect\citeauthoryear{{Stassun} et~al.,}{{Stassun}
  et~al.}{2019}]{Stassun2019_TICv8}
{Stassun} K.~G.,  et~al., 2019, \mn@doi [\aj] {10.3847/1538-3881/ab3467}, \href
  {https://ui.adsabs.harvard.edu/abs/2019AJ....158..138S} {158, 138}

\bibitem[\protect\citeauthoryear{{Stumpe} et~al.,}{{Stumpe}
  et~al.}{2012}]{Stumpe2012_PDC1}
{Stumpe} M.~C.,  et~al., 2012, \mn@doi [\pasp] {10.1086/667698}, \href
  {https://ui.adsabs.harvard.edu/abs/2012PASP..124..985S} {124, 985}

\bibitem[\protect\citeauthoryear{{Stumpe}, {Smith}, {Catanzarite}, {Van Cleve},
  {Jenkins}, {Twicken}  \& {Girouard}}{{Stumpe} et~al.}{2014}]{Stumpe2014_PDC3}
{Stumpe} M.~C.,  {Smith} J.~C.,  {Catanzarite} J.~H.,  {Van Cleve} J.~E.,
  {Jenkins} J.~M.,  {Twicken} J.~D.,   {Girouard} F.~R.,  2014, \mn@doi [\pasp]
  {10.1086/674989}, \href
  {https://ui.adsabs.harvard.edu/abs/2014PASP..126..100S} {126, 100}

\bibitem[\protect\citeauthoryear{{Tarter}}{{Tarter}}{1975}]{Tarter1975}
{Tarter} J.~C.,  1975, PhD thesis, University of California, Berkeley

\bibitem[\protect\citeauthoryear{{Tayar}, {Stassun}  \& {Corsaro}}{{Tayar}
  et~al.}{2019}]{Tayar2019}
{Tayar} J.,  {Stassun} K.~G.,   {Corsaro} E.,  2019, \mn@doi [\apj]
  {10.3847/1538-4357/ab3db1}, \href
  {https://ui.adsabs.harvard.edu/abs/2019ApJ...883..195T} {883, 195}

\bibitem[\protect\citeauthoryear{{Tokovinin}}{{Tokovinin}}{2018}]{Tokovinin2018}
{Tokovinin} A.,  2018, \mn@doi [\pasp] {10.1088/1538-3873/aaa7d9}, \href
  {https://ui.adsabs.harvard.edu/abs/2018PASP..130c5002T} {130, 035002}

\bibitem[\protect\citeauthoryear{{Tokovinin}, {Fischer}, {Bonati}, {Giguere},
  {Moore}, {Schwab}, {Spronck}  \& {Szymkowiak}}{{Tokovinin}
  et~al.}{2013}]{Tokovinin2013}
{Tokovinin} A.,  {Fischer} D.~A.,  {Bonati} M.,  {Giguere} M.~J.,  {Moore} P.,
  {Schwab} C.,  {Spronck} J. F.~P.,   {Szymkowiak} A.,  2013, \mn@doi [\pasp]
  {10.1086/674012}, \href
  {https://ui.adsabs.harvard.edu/abs/2013PASP..125.1336T} {125, 1336}

\bibitem[\protect\citeauthoryear{{Torres}, {Andersen}  \&
  {Gim{\'e}nez}}{{Torres} et~al.}{2010}]{Torres:2010}
{Torres} G.,  {Andersen} J.,   {Gim{\'e}nez} A.,  2010, \mn@doi [\aapr]
  {10.1007/s00159-009-0025-1}, \href
  {https://ui.adsabs.harvard.edu/abs/2010A&ARv..18...67T} {18, 67}

\bibitem[\protect\citeauthoryear{{Triaud}}{{Triaud}}{2018}]{Triaud2018}
{Triaud} A. H.~M.~J.,  2018, in {Deeg} H.~J.,  {Belmonte} J.~A.,  eds, ,
  Handbook of Exoplanets.
p.~2, \mn@doi{10.1007/978-3-319-55333-7_2}

\bibitem[\protect\citeauthoryear{{Twicken} et~al.,}{{Twicken}
  et~al.}{2018}]{Twicken:DVdiagnostics2018}
{Twicken} J.~D.,  et~al., 2018, \mn@doi [\pasp] {10.1088/1538-3873/aab694},
  \href {http://adsabs.harvard.edu/abs/2018PASP..130f4502T} {130, 064502}

\bibitem[\protect\citeauthoryear{{Veras}}{{Veras}}{2016}]{Veras2016}
{Veras} D.,  2016, \mn@doi [Royal Society Open Science] {10.1098/rsos.150571},
  \href {https://ui.adsabs.harvard.edu/abs/2016RSOS....350571V} {3, 150571}

\bibitem[\protect\citeauthoryear{{Villaver}, {Livio}, {Mustill}  \&
  {Siess}}{{Villaver} et~al.}{2014}]{Villaver2014}
{Villaver} E.,  {Livio} M.,  {Mustill} A.~J.,   {Siess} L.,  2014, \mn@doi
  [\apj] {10.1088/0004-637X/794/1/3}, \href
  {https://ui.adsabs.harvard.edu/abs/2014ApJ...794....3V} {794, 3}

\bibitem[\protect\citeauthoryear{{Weiss} et~al.,}{{Weiss}
  et~al.}{2013}]{Weiss2013}
{Weiss} L.~M.,  et~al., 2013, \mn@doi [\apj] {10.1088/0004-637X/768/1/14},
  \href {https://ui.adsabs.harvard.edu/abs/2013ApJ...768...14W} {768, 14}

\bibitem[\protect\citeauthoryear{{Wong}, {Shporer}, {Vissapragada},
  {Greklek-McKeon}, {Knutson}, {Winn}  \& {Benneke}}{{Wong}
  et~al.}{2022}]{Wong2022_WASP12}
{Wong} I.,  {Shporer} A.,  {Vissapragada} S.,  {Greklek-McKeon} M.,  {Knutson}
  H.~A.,  {Winn} J.~N.,   {Benneke} B.,  2022, \mn@doi [\aj]
  {10.3847/1538-3881/ac5680}, \href
  {https://ui.adsabs.harvard.edu/abs/2022AJ....163..175W} {163, 175}

\bibitem[\protect\citeauthoryear{{Wright} et~al.,}{{Wright}
  et~al.}{2010}]{Wright2010_WISE}
{Wright} E.~L.,  et~al., 2010, \mn@doi [\aj] {10.1088/0004-6256/140/6/1868},
  \href {https://ui.adsabs.harvard.edu/abs/2010AJ....140.1868W} {140, 1868}

\bibitem[\protect\citeauthoryear{{Yee} et~al.,}{{Yee}
  et~al.}{2020}]{Yee2020_WASP-12b_3}
{Yee} S.~W.,  et~al., 2020, \mn@doi [\apjl] {10.3847/2041-8213/ab5c16}, \href
  {https://ui.adsabs.harvard.edu/abs/2020ApJ...888L...5Y} {888, L5}

\bibitem[\protect\citeauthoryear{{Yu}, {Huber}, {Bedding}  \& {Stello}}{{Yu}
  et~al.}{2018}]{Yu2018}
{Yu} J.,  {Huber} D.,  {Bedding} T.~R.,   {Stello} D.,  2018, \mn@doi [\mnras]
  {10.1093/mnrasl/sly123}, \href
  {https://ui.adsabs.harvard.edu/abs/2018MNRAS.480L..48Y} {480, L48}

\bibitem[\protect\citeauthoryear{{Zahn}}{{Zahn}}{1977}]{Zahn1977_tidal}
{Zahn} J.~P.,  1977, \aap, \href
  {https://ui.adsabs.harvard.edu/abs/1977A&A....57..383Z} {57, 383}

\bibitem[\protect\citeauthoryear{{Ziegler}, {Tokovinin}, {Brice{\~n}o}, {Mang},
  {Law}  \& {Mann}}{{Ziegler} et~al.}{2020}]{Ziegler2020}
{Ziegler} C.,  {Tokovinin} A.,  {Brice{\~n}o} C.,  {Mang} J.,  {Law} N.,
  {Mann} A.~W.,  2020, \mn@doi [\aj] {10.3847/1538-3881/ab55e9}, \href
  {https://ui.adsabs.harvard.edu/abs/2020AJ....159...19Z} {159, 19}

\makeatother
\end{thebibliography}




\appendix

\section{The posterior distribution of isochrone fitting for TOI-1608}

\begin{figure*}
\centering
\includegraphics[width=\textwidth]{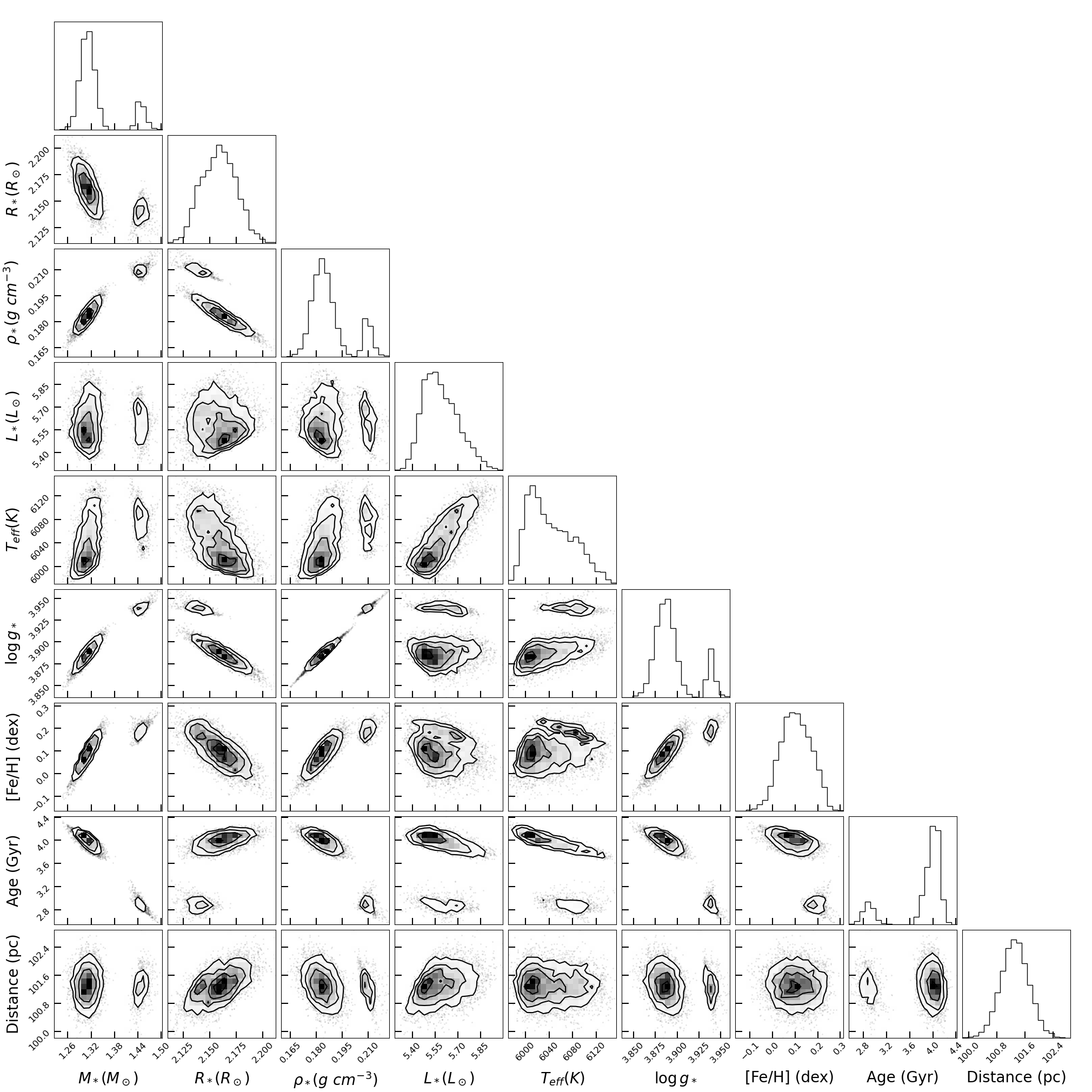}
\caption{The posterior distribution of isochrone fitting for TOI-1608. The bimodal solution can be seen in the distribution of $M_{\ast}$, $\rho_\ast$, $\log g_{\ast}$ and age.} 
\label{fig_1608_iso_post}
\end{figure*}

\section{RV measurements}

\begin{table}
    \centering
    \caption{RV measurements of TOI-1608.}
    \begin{tabular}{cccc}
    \hline\hline
    BJD$_{\rm TDB}$   &RV ($\rm m\ s^{-1}$)   &$\rm \sigma_{\rm RV}$ ($\rm m\ s^{-1}$)  &Instrument\\ \hline
    2459551.809   &77505.8   &111.0   &NRES \\
    2459554.611   &72125.0   &115.4   &NRES \\
    2459555.828   &92426.6   &113.9   &NRES \\
    2459556.669   &80122.9   &117.2   &NRES \\
    2459569.599   &72143.9   &117.0   &NRES \\
    2459570.815   &93377.1   &136.1   &NRES \\
    2459571.662   &74946.0   &112.5   &NRES \\
    2459573.657   &86689.7   &317.6   &NRES \\
    2459574.790   &74478.1   &111.6   &NRES \\
    2459575.762   &92945.6   &119.0   &NRES \\
    2459586.564   &74329.9   &128.5   &NRES \\
    2459586.624   &73127.7   &135.1   &NRES \\
    2459586.691   &72150.8   &99.8   &NRES \\
    2459586.757   &72454.5   &140.4   &NRES \\
    2459588.610   &84335.5   &147.5   &NRES \\
    2459588.661   &82298.2   &101.6   &NRES \\
    \hline\hline 
    \end{tabular}
    \label{tab:1608_RV_data}
\end{table}

\begin{table}
    \centering
    \caption{RV measurements of TOI-2336.}
    \begin{tabular}{cccc}
    \hline\hline
    BJD$_{\rm TDB}$   &RV ($\rm m\ s^{-1}$)   &$\rm \sigma_{\rm RV}$ ($\rm m\ s^{-1}$)  &Instrument\\ \hline
    2459327.48570   &36031.7   &146.3   &NRES \\
    2459328.38208   &33814.8   &130.3   &NRES \\
    2459329.55779   &25601.2   &205.3   &NRES \\
    2459336.62242   &34946.6   &169.7   &NRES \\
    2459341.32044   &39963.4   &127.5   &NRES \\
    2459343.48120   &35564.8   &113.3   &NRES \\
    2459344.56582   &30027.7   &129.6   &NRES \\
    2459346.54750   &30188.4   &132.2   &NRES \\
    2459347.49433   &34602.6   &102.0   &NRES \\
    2459348.57917   &37950.9   &112.0   &NRES \\
    2459371.34618   &36932.0   &97.0   &NRES \\
    2459376.46575   &29548.0   &119.3   &NRES \\
    2459377.29886   &31467.1   &142.8   &NRES \\
    2459379.50367   &38767.5   &134.0   &NRES \\
    2459406.40547   &28286.2   &134.2   &NRES \\
    2459407.41335   &26729.2   &134.1   &NRES \\
    2459410.41467   &38226.1   &132.9   &NRES \\
    2459411.40852   &37089.3   &110.6   &NRES \\
    2459412.37610   &35113.3   &128.6   &NRES \\
    2459414.22659   &28683.6   &113.1   &NRES \\
    2459415.38816   &27329.2   &113.1   &NRES \\
    2459342.68168   &34035   &219   &CHIRON \\
    2459345.72623   &24826   &164   &CHIRON \\
    2459349.70953   &35637   &180   &CHIRON \\
    2459357.62661   &35468   &206   &CHIRON \\
    2459359.65638   &27702   &198   &CHIRON \\
    2459371.64418   &34388   &128   &CHIRON \\
    2459387.65515   &35618   &178   &CHIRON \\
    2459399.56554   &24837   &215   &CHIRON \\
    2459407.61081   &24916   &186   &CHIRON \\
    2459417.57493   &33182   &191   &CHIRON \\
    2459424.53740   &30008   &214   &CHIRON \\
    2459432.50942   &31233   &205   &CHIRON \\
    2459318.877767   &37598.76   &121.12   &CORALIE \\
    2459321.804744   &27564.77   &111.03   &CORALIE \\
    2459328.784617   &29880.55   &119.24   &CORALIE \\
    2459331.872147   &31218.65   &99.95   &CORALIE \\
    2459335.705030   &33463.34   &109.05   &CORALIE \\
    2459343.687687   &32204.50   &125.93   &CORALIE \\
    2459346.774496   &29472.14   &99.30   &CORALIE \\
    2459351.752823   &30681.14   &120.70   &CORALIE \\
    2459356.753344   &37656.42   &123.56   &CORALIE \\
    2459357.723789   &37554.35   &77.79   &CORALIE \\
    2459358.643619   &34352.13   &99.49   &CORALIE \\
    2459361.597369   &27438.35   &65.13   &CORALIE \\
    2459363.736155   &36053.07   &107.73   &CORALIE \\
    2459365.700203   &36685.84   &69.86   &CORALIE \\
    2459368.608494   &27031.54   &78.39   &CORALIE \\
    2459373.599021   &35976.49   &64.58   &CORALIE \\
    \hline\hline 
    \end{tabular}
    \label{tab:2336_RV_data}
\end{table}

\begin{table}
    \centering
    \caption{RV measurements of TOI-2521.}
    \begin{tabular}{cccc}
    \hline\hline
    BJD$_{\rm TDB}$   &RV ($\rm m\ s^{-1}$)   &$\rm \sigma_{\rm RV}$ ($\rm m\ s^{-1}$)  &Instrument\\ \hline
    2459294.52983   &-36161   &53   &CHIRON \\
    2459295.53896   &-40116   &87   &CHIRON \\
    2459296.53688   &-33715   &53   &CHIRON \\
    2459297.55627   &-24716   &49   &CHIRON \\
    2459298.51469   &-23135   &112   &CHIRON \\
    2459301.52687   &-38223   &97   &CHIRON \\
    2459303.52923   &-22904   &66   &CHIRON \\
    2459304.50997   &-25694   &86   &CHIRON \\
    \hline\hline 
    \end{tabular}
    \label{tab:2521_RV_data}
\end{table}




\bsp	
\label{lastpage}
\end{document}